\documentclass[fleqn,usenatbib,useAMS]{mnras}
\usepackage{pifont}
\usepackage{amsmath}
\usepackage{graphicx,epstopdf}
\DeclareGraphicsExtensions{.ps}
\usepackage[utf8]{inputenc}
\usepackage{url}
\usepackage{lscape}
\usepackage{float}
\usepackage[section]{placeins}
\usepackage{color}

\usepackage{newtxtext,newtxmath}
\usepackage[T1]{fontenc}
\usepackage{ae,aecompl}
\usepackage{amssymb}	% Extra maths symbols
\usepackage{longtable}
\usepackage{pdflscape}
\usepackage{ulem}

\newcommand{\teff}{$T_\textrm{eff}$}
\newcommand{\logg}{$\log g$}

\newcommand{\mh}{[M/H]}
\newcommand{\luminosity}{$\log (L_\star/{\rm L_\odot})$}
\newcommand{\kms}{km\,s$^{-1}$}
\newcommand{\vsini}{$v\sin i$}
\newcommand{\vrot}{$v_{\rm rot}$}
\def\spose#1{\hbox to 0pt{#1\hss}}
\def\lta{\mathrel{\spose{\lower 3pt\hbox{$\mathchar"218$}}
     \raise 2.0pt\hbox{$\mathchar"13C$}}}
\def\gta{\mathrel{\spose{\lower 3pt\hbox{$\mathchar"218$}}
     \raise 2.0pt\hbox{$\mathchar"13E$}}}

\title[Study of Chemically Peculiar Stars]{Study of Chemically Peculiar Stars-I : High-resolution Spectroscopy and {\it K2} Photometry of Am Stars in the Region of M44}

\author[Joshi et al.]{Santosh Joshi$^{1}$\thanks{E-mail: santosh@aries.res.in},
Otto Trust$^{2}$\thanks{E-mail: otrust@must.ac.ug},
E. Semenko$^{3,4}$,
P.~E. Williams$^{5} $,
P. Lampens $^{6}$,
P. De Cat$^{6}$,
\newauthor L. Vermeylen$^{6}$,
D.~L. Holdsworth$^{7}$,
R.~A.\ Garc\'\i a$^{8,9}$,
S. Mathur$^{10,11}$,
A.~R.~G. Santos$^{12,13}$,
\newauthor 
D. Mkrtichian$^{3} $,
A. Goswami$^{14} $,
M. Cuntz$^{15} $,
A.~P. Yadav$^{16}$,
M. Sarkar$^{1}$,
 B.~C. Bhatt$^{14}$,
\newauthor  F. Kahraman Ali\c{c}avu\c{s}$^{17,18}$,
 M.~D. Nhlapo$^{19}$,
 M.~N. Lund$^{20}$,
P.~P. Goswami$^{14}$,
I. Savanov$^{21}$,
\newauthor
A. Jorissen$^{22}$,
E. Jurua$^{2}$,
E. Avvakumova$^{23}$,
E.~S. Dmitrienko$^{24} $,
N.~K. Chakradhari$^{25}$,
\newauthor
M.~K. Das$^{26}$,
S. Chowdhury$^{17}$,
O.~P. Abedigamba$^{19,27}$,
I. Yakunin$^{4}$,
B. Letarte$^{19}$,  
and D. Karinkuzhi$^{28}$ \\
Affiliations are listed at the end of the paper}

\date{Accepted XXX. Received YYY; in original form ZZZ}

\pubyear{2021}

\begin{document}
\label{firstpage}
\pagerange{\pageref{firstpage}--\pageref{lastpage}}
\maketitle

\begin{abstract}
We present a study based on the high-resolution spectroscopy and {\it K2} space photometry of five chemically peculiar stars in the region of the open cluster M44. The analysis of the high-precision photometric {\it K2} data reveals that the light variations in HD\,73045 and HD\,76310 are rotational in nature and caused by spots or cloud-like co-rotating structures, which are non-stationary and short-lived. The time-resolved radial velocity measurements, in combination with the {\it K2} photometry, confirm that HD\,73045 does not show any periodic variability on timescales shorter than 1.3\,d, contrary to previous reports in the literature. In addition to these new rotational variables, we discovered a new heartbeat system, HD\,73619, where no pulsational signatures are seen. The spectroscopic and spectropolarimetric analyses indicate that HD\,73619 belongs to the  peculiar Am class, with either a weak or no magnetic field considering the 200\,G detection limit of our study. The Least-Squares Deconvolution (LSD) profiles for HD\,76310 indicate a complex structure in its spectra suggesting that this star is either part of a binary system or surrounded by a cloud shell. When placed in the Hertzsprung-Russell diagram, all studied stars  are evolved from main-sequence and situated in the $\delta$\,Scuti instability strip. The present work is relevant for further detailed studies of CP stars, such as inhomogeneities (including spots) in the absence of magnetic fields and the origin of the pulsational variability in heartbeat systems.
\end{abstract}

% asteroseismology -- stars: chemically peculiar -- stars: star spots – stars
% keyword1 -- keyword2 -- keyword3

\section{Introduction}
\label{section:intro}

The chemically peculiar (CP) stars are a group of  main sequence (MS) B-, A-, and F-type stars having peculiar surface elemental abundances; they are characterized by abnormal spectral lines strengths \citep{1974ARA&A..12..257P}. The chemical anomalies in these stars are thought to be confined to the outer stellar layers and to arise from gravitational settling and radiative levitation of certain elements, a process known as atomic diffusion \citep{1970ApJ...160..641M, 1981A&A...103..244M}. The present study is confined to one subset of CP stars, the metallic-lined A (Am) stars, which are generally non-magnetic in nature and   characterized by under-abundances of some light elements such as Ca and Sc, but slight/moderate over-abundances of iron peak elements, e.g., Zn, Sr, Y, Zr, and Ba. The projected rotational velocities of these stars are generally smaller than for ordinary A stars (\vsini\ typically $<$ 120 \kms), with the majority of the Am stars being members of close binary systems. Rotational braking through tidal interaction is regarded as a possible cause of the low rotational velocities.  

Using four years of high-precision photometry from the nominal {\it Kepler} mission and the {\it K2} campaigns, \cite{2015MNRAS.448.1378B} investigated the light variations in 29 Am stars and found that most of the Am stars in the {\it Kepler} field have light curves with the characteristics of rotational modulation arising from star spots or co-rotating structures. The origin of spots in Am stars seems to be different from solar-like spots as these stars do not show any signs of intense magnetic fields able to produce such magnetic features. Magnetic fields on the order of sub-Gauss strengths have been reported in some Am stars, e.g., Sirius A \citep{2011A&A...532L..13P}, Vega \citep{2015A&A...577A..64B}, $\beta$\,UMa, and $\theta$\,Leo \citep{2015IAUS..305...67B}, Alhena \citep{2016MNRAS.459L..81B} and $\rho$\,Pup \citep{2017MNRAS.468L..46N}. It is thought that for the majority of these stars that convective flows in the atmospheres may disrupt any spot-like features \citep{2003ASPC..305..190K}. Hence, the rotational modulation in some Am stars indicates that either a weak magnetic field may lead to surface inhomogeneities in the form of spots across the stellar surface or, alternatively, there is some unknown mechanism(s) producing those spots. If a weak magnetic field is indeed present, then the basic processes operating in Am stars would need to be revisited since magnetic fields have been omitted from the diffusion models attempting to explain their unusual chemical abundances \citep{10.3389/fspas.2021.653558}.
 
Stars showing variable amplitudes or harmonics in their light curves provide the best cases for identifying rotational modulation. However, it is extremely difficult to distinguish between rotation and binarity at low frequencies, noting that the amplitude of the periodic variation due to co-rotating star spots changes in a relatively short time-scale (perhaps a few months).
 
Short-term variability, similar to that seen in $\delta$\,Scuti stars, has been reported in many Am stars and is understood to arise from pulsations driven by the $\kappa$-mechanism operating in the He\,\textsc{ii} ionization zone \citep{2000ASPC..210..215P} combined with turbulent pressure in the H/He\,\textsc{i} ionization zone \citep{2017MNRAS.465.2662S,10.1093/mnras/stz2787}. In the context of the present study, \cite{2015IAUS..307..218J} suspected a short-term pulsational variability in HD\,73045; hitherto awaiting confirmation. 
 
In close binary systems, stellar pulsations may also be tidally induced \citep{1995ApJ...449..294K, 2012ApJ...753...86T}; e.g., KOI-54 \citep{2011ApJS..197....4W,2012MNRAS.420.3126F, 2012MNRAS.421..983B,2014MNRAS.440.3036O} and KIC\,3230227 \citep{2017ApJ...834...59G}. Highly eccentric ($e>0.3$) binary systems with orbital periods between a fraction of a day and tens of days showing a sudden  increase in brightness at periastron passage, on the order of several parts-per-thousand (ppt), are known as heartbeat (HB) stars \citep{2002MNRAS.333..262H,2009A&A...508.1375M,2013MNRAS.434..925H,2014A&A...564A..36B, 2016MNRAS.463.1199H,2018MNRAS.473.5165H, 2020arXiv201211559K}. The name stems from the resemblance of the light curve to a heartbeat in an electrocardiogram. About 180 heartbeat stars have been discovered to date. Due to the small-amplitude light variations and their short orbital periods, these targets were mainly detected using space missions such as {\it Kepler} \citep{2013MNRAS.434..925H, 2016AJ....151...68K,  2016MNRAS.463.1199H,2017MNRAS.472.1538F,2018MNRAS.473.5165H,  2020ApJ...888...95G}. The Am stars are often part of binary systems \citep{2007MNRAS.380.1064C}, though Am stars in heartbeat systems have not been reported in the literature yet, thus the discovery of a new heartbeat beat system in one of the CP stars HD\,73619 would be significant.

In the present work, we revisit all stars previously observed as part of the Nainital-Cape survey project \citep{2000BASI...28..251A, 2001A&A...371.1048M, 2003MNRAS.344..431J,2006A&A...455..303J, 2009A&A...507.1763J,  2010MNRAS.401.1299J, 2012MNRAS.424.2002J, 2016A&A...590A.116J, 2017MNRAS.467..633J} with data in the {\it Kepler} and {\it K2} archives.  It is our aim to search for low frequency, high-precision, and photometric variability. No stars from the Nainital-Cape survey were observed in the nominal {\it Kepler} field, whilst eight stars were found in the {\it K2} archive. In the current paper, we present the results on extrinsic (rotational and heartbeat) variables, as well as the results for stars where variability could not be ascertained. 

\section{Observations and Data Reduction}
\label{obs}

 The photometric observations were obtained using both ground- and space-based telescopes while spectroscopy and spectropolarimetry were performed through ground-based telescopes. The specifications of the telescopes and state-of-art instruments used for the observations are listed in Table\,\ref{table:tele}.
 
\begin{table*}
\begin{center}
\caption{The specifications of the telescopes and back-end instruments used for the photometric and spectroscopic observations.}
% \smallskip
\label{table:tele}
%\fontsize{7.50}{9.0}\selectfont
\begin{tabular}{lllllr}  % l = left, c = centered
\hline
\hline
\noalign{\smallskip}
Category& Name of & Diameter & Location & Detector/ & FoV/  \\
     &  Telescope & (m)     &          &  Spectrograph  & Resolution   \\
\noalign{\smallskip}
\hline
\noalign{\smallskip}
Ground-based & DFOT  & 1.30  & Devasthal (India)& Andor’s DZ436 & 18$^{'}\times18^{'}$\\
Photometry  &       &      &           & Andor’s iXon EM+ DU-897   & 4.7$^{'}\times4.7^{'}$\\
            & ST    & 1.04  & Nainital (India)& 2k Wright CCD&  13 $^{'}\times13^{'}$\\
%    &              &       &      &     & PMT      & & && 0.5$^{'}\times0.5^{'}$\\
% 3. & SAAO & 0.50-m++++++++++++++++++++++++++++++++++++++++++++++++------------------------  & Sutherland (South Africa) & PMT & 0.5$^{'}\times0.5^{'}$\\
 & MASTER-II-URAL & 0.40 & Ural Federal University (Russia)&  U16M CCD& 2$^\circ\times 4^\circ$\\
 & PROMPT-8 & 0.60  & CTIO (Chile) &Apogee, F42  & 26.6$^{'}\times22.6^{'}$ \\
% 6. & SuperWasp & 20-cm & SAAO \& Canary Islands & 2048$\times$2048 & 7.8$^\circ\times7.8^\circ$ \\
\hline
\noalign{\smallskip}
Space-based & {\it K2}   & 0.95 & Space& e2v CCD90s  & 10.25 $^\circ\times10.25^\circ $ \\
Photometry&&&&&\\\\
\noalign{\smallskip}
\hline
\noalign{\smallskip}
 Spectroscopy& BTA   & 6.00  & Nizhniy Arkhyz (Russia)& NES,MSS&    39\,000, 15\,000 \\ 
 & TNO  &  2.40 &Chiang Mai (Thailand) & MRES & 17\,000  \\
 & HCT  & 2.00 & Hanle (India)& HESP  & 30\,000  \\
 & Mercator & 1.20 & La Palma (Spain)& HERMES & 85\,000 \\
\hline
%\tableline\
\end{tabular}
\end{center}
\end{table*}

\subsection{Ground-Based Differential Photometry of \texorpdfstring{HD\,73045}{}}
\label{diff}

To investigate the variability of the single-lined spectroscopic binary (SB1) HD\,73045, as suspected by \cite{2015IAUS..307..218J}, this object was extensively monitored photometrically with the 1.3-m Devasthal Fast Optical Telescope (DFOT; \citealt{2011CSci..101.1020S}), the 1.04-m Sampurnanand Telescope (ST; \citealt{1972oams.conf...20S}), the 2$\times$0.40-m MASTER-II-URAL telescope \citep{2010AdAst2010E..30L}, and the 0.6-m Panchromatic Robotic Optical Monitoring and Polarimetry Telescope (PROMPT-8; \citealt{2014AstBu..69..368B}). These instruments are introduced in Table\,\ref{table:tele}.

\begin{figure}
\centering
 \includegraphics[width=\columnwidth]{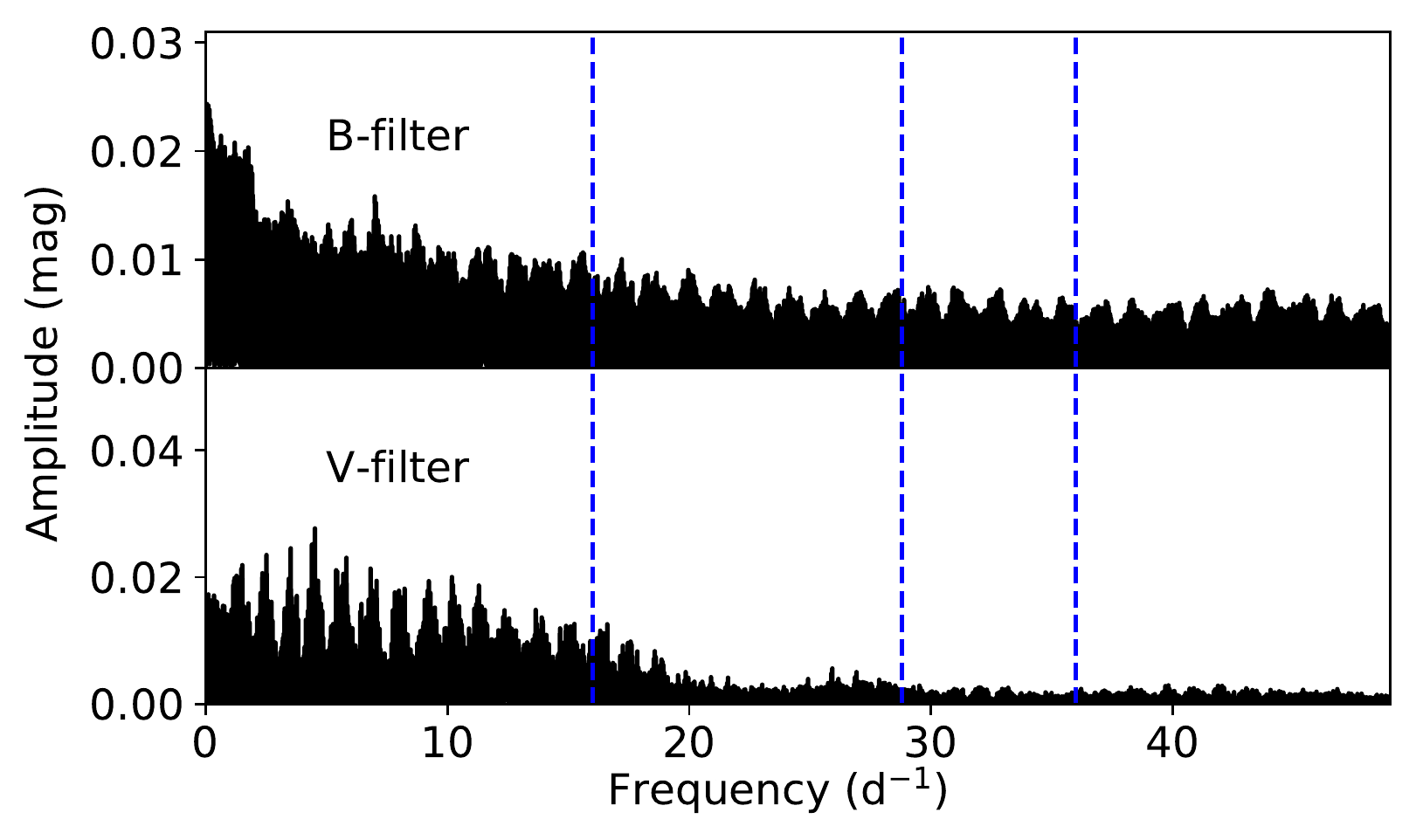}
\caption{The amplitude spectra of HD\,73045 using the combined data from ground-based observations in B (top panel) and V (bottom panel) filters. There are no significant frequencies with signal-to-noise ratio (SNR) above 4 (0.048\,mag for B-filter and 0.051\,mag for V-filter), hence the star is classified as non-variable. The blue dashed vertical lines represent the short-term periodicity as suspected by \citet{2015IAUS..307..218J}.}
 \label{ground_ob}
\end{figure}

All observations were carried out through a Johnson B and/or V filter. Calibration frames were acquired during each night for the correction of background and pixel sensitivity. The science frames were processed by subtracting master bias and dark frames before normalising them with a median flat-field frame from which the cosmic rays were removed.
The instrumental magnitudes of HD\,73045 and the comparison stars (BD\,$+19^{\circ}2046$ and TYC\,1395-855-1) were obtained through aperture photometry using the in-built \textsc{daophot} package of IRAF \citep{Davis94areference}. The characteristics of the ground based photometric observations are given in Table\,\ref{table:groundlog} and the individual light curves and power spectra are  presented in Figs.\,\ref{lc:B} and \ref{lc:V}. The top and bottom panels of Fig.\,\ref{ground_ob} show the combined amplitude spectra of HD\,73045 in the B and V filters, respectively. 

\subsection{{\it K2} Space Photometry}
\label{space}

Our target stars were observed by the {\it K2} mission that operated in a long cadence (LC; 29.4\,min) and a short cadence (SC; 58.8\,s) mode 
\citep{2014PASP..126..398H}. {\it K2} pointed towards the Praesepe (Beehive) cluster (M44) during the campaigns C05 (2015/04/27--2015/07/10), C16 (2017/12/07--2018/02/25), and C18 (2018/05/12--2018/07/02). The {\it K2} observations of our targets are summarized in Table\,\ref{table:campaigns}.

\begin{table}
\begin{center}
\caption{The {\it K2} campaigns concerning our programme stars.}
\smallskip
\label{table:campaigns}
%\fontsize{7.2}{9.0}\selectfont
\begin{tabular}{cccccccc}  % l = left, c = centered
\hline
\hline   
\noalign{\smallskip}
Star Name & EPIC Number & \multicolumn{6}{c}{{\it K2} Observation Campaigns} \\
   &          &       &       &       &       &       &      \\    
   &          & \multicolumn{2}{c}{C05} &  \multicolumn{2}{c}{C16} & \multicolumn{2}{c}{C18} \\    
   &          & \multicolumn{2}{c}{$\sim$79 d} &  \multicolumn{2}{c}{$\sim$80 d} & \multicolumn{2}{c}{$\sim$50 d} \\    
\noalign{\smallskip}
   &          & LC & SC &  LC & SC & LC & SC \\    
\noalign{\smallskip}
\hline
\noalign{\smallskip}
HD\,73045 & 211910450 & \checkmark & - & - & - & \checkmark & \checkmark \\
HD\,73574 & 211989558 & \checkmark & - & - & - & \checkmark & - \\
HD\,73618 & 211955190 & \checkmark & - & \checkmark & - & \checkmark & - \\
HD\,73619 & 211954496 & \checkmark & - & \checkmark & - & \checkmark & - \\
HD\,76310 & 212082764 & \checkmark & - & \checkmark & - & \checkmark & - \\
\noalign{\smallskip}
\hline
%\tableline\
\end{tabular}

\end{center}
\end{table}

\begin{figure}
\begin{center}
\includegraphics[trim={100 90 120 100}, clip, width=\columnwidth]{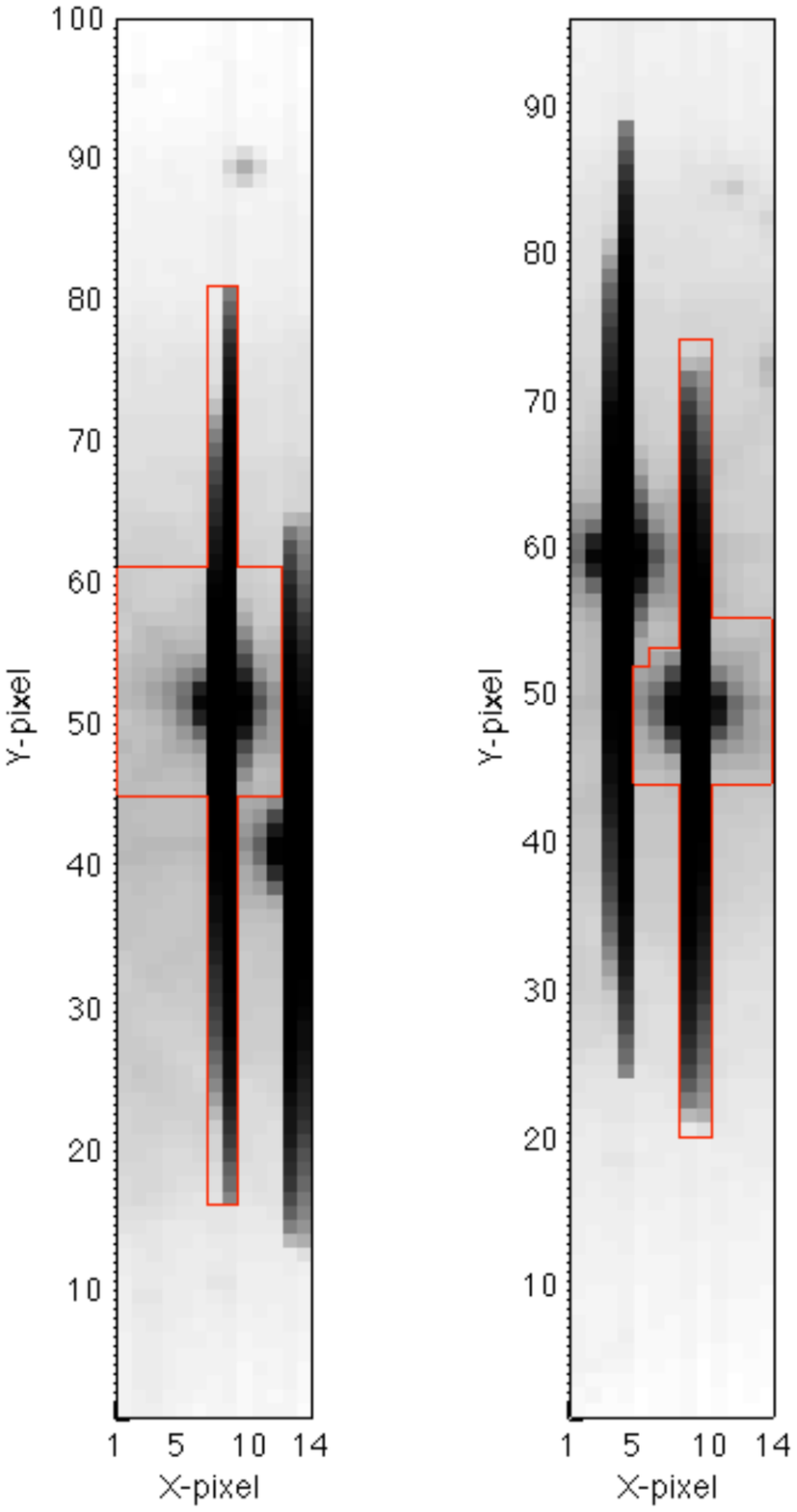}
\caption{Target pixel images from the C18 time-series of HD\,73618 (left) and HD\,73619 (right). Note that each star appears in both images. The red border signifies the boundary of a custom mask within which the pixels are isolated and summed to produce a single simple aperture photometry flux data point.}
\label{K2_HD73618_HD73619_C18_mask}
\end{center}
\end{figure}

\begin{figure*}
\begin{center}
\includegraphics[width=0.7\textwidth]{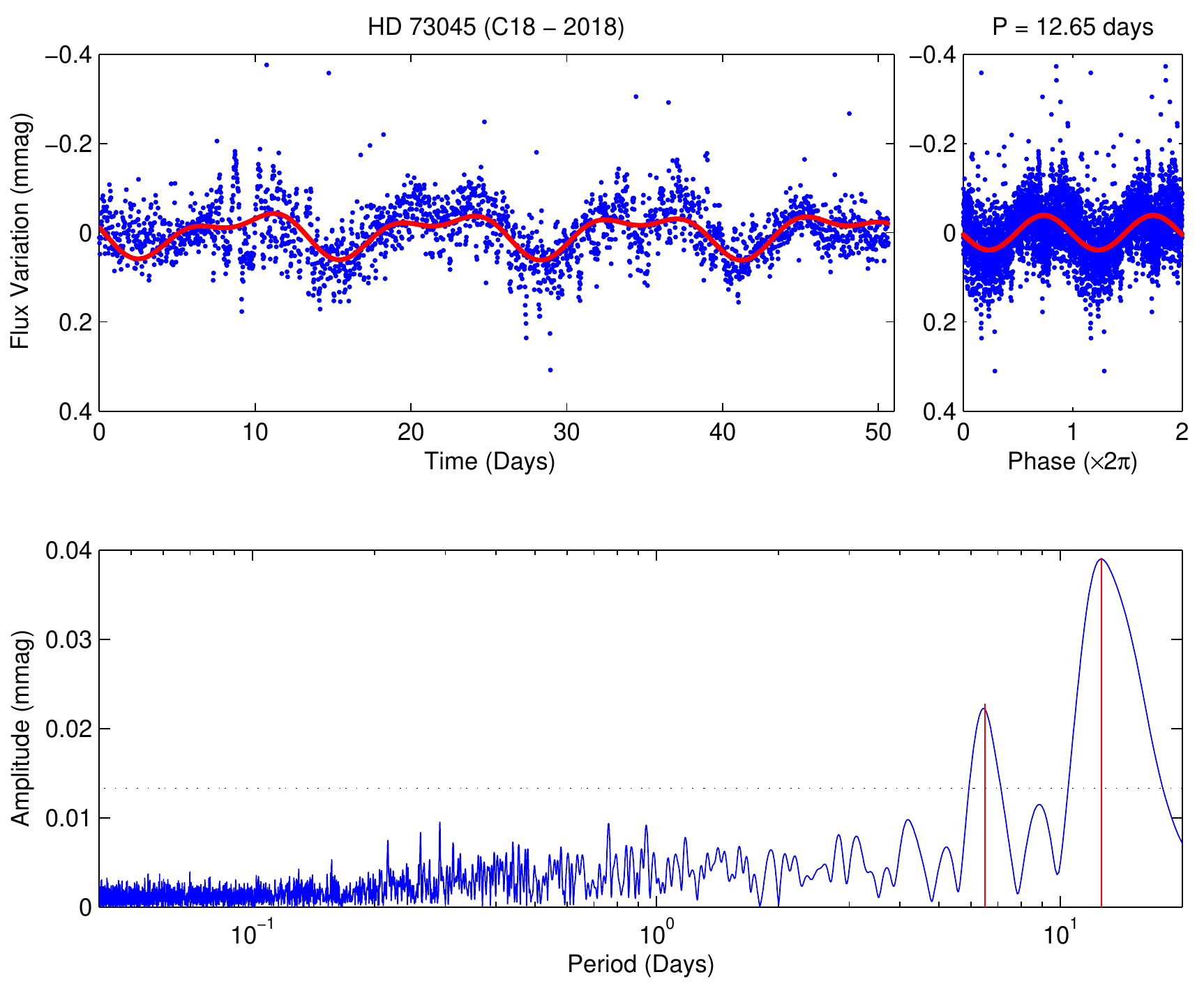}
\caption{Top left: Time-series illustrating photometric flux variation (in mmag) of HD\,73045 for the LC 2018 {\it K2} Campaign 18. The data (blue points) include intrinsic signals as well as systemic and noise artifacts. The sinusoidal model, with parameters given in Table\,\ref{table:KepSineComps}, is shown as the red line. Top right: The photometric data (blue points) phase folded on the dominant variation period of 12.2\,d. The red line represents the best-fit to this data. The data is repeated over two periods for clarity of presentation. Bottom: Linear-log periodogram of spectral amplitudes (blue line), in mmag. The components of the sinusoidal model are shown by the red stems. The black dotted line corresponds to a False Alarm Probability of $10^{-8}$, and the periodogram peaks above this line are considered to be confirmed signals.}
\label{K2_HD73045_C18}
\end{center}
\end{figure*}

These photometric data sets were downloaded from the Barbara A. Mikulski Archive for Space Telescopes (MAST) database{\footnote{https://archive.stsci.edu/k2/epic/search.php}}. Each target pixel file (TPF) was checked for the presence of additional stars.
As HD\,73045, HD\,76310, and HD\,73574 were the sole stars in their TPF images, their respective {\it K2} FITS light curve files were directly downloaded in the PDC\_SAP format \citep{2012PASP..124..985S, 2012PASP..124.1000S}, having undergone pre-search data conditioning (PDC) processing of their simple aperture photometry (SAP) flux data. The underlying position-dependent artifacts were removed by processing the data through the K2SC algorithm \citep{2015MNRAS.447.2880A}. The TPF images of HD\,73618 and HD\,73619 showed the presence of nearby stars (Fig.\, \ref{K2_HD73618_HD73619_C18_mask}), which had to be masked  out to isolate the flux of these target stars. The remaining images were averaged and a custom mask was produced to isolate the target star. This mask was applied to each image of the campaign and the total flux for each image was deduced, producing a light curve similar to the available SAP flux data. These data sets were further processed with the K2SC algorithm. Fig.\,\ref{K2_HD73045_C18} (top-left, in blue) shows an example light curve for the C18 observation of HD\,73045.  A detailed discussion of our analysis is given in Appendix\,\ref{k2_appendix}. A periodogram produced by the Lomb-Scargle algorithm is shown in Fig.\,\ref{K2_HD73045_C18} (bottom, in blue). The frequency-varying artifact in the light-curve is due to the roll-angle variation \citep{2016MNRAS.459.2408A}, reversing twice near 10 and 40 days respectively. This artifact is not prevalent in the periodogram, but is distributed between 0.25\,--\,2\,days (0.5\,--\,4\,d$^{-1}$).

Using a multi-component sinusoidal model, frequencies, periods, amplitudes, and phases, relating to the variability detected in the K2 light curves, are derived, and are tabulated  in Table\,\ref{table:KepSineComps}. Fig.\,\ref{K2_HD73045_C18} (top-left, in red) illustrates such a model of the HD 73045 C18 light curve, for which the derived amplitudes and frequencies of the components are included in the periodogram (Fig.\,\ref{K2_HD73045_C18}, bottom) as red stems. The light curve data is phase-folded on the period with the largest amplitude and overplotted with that period's sinusoidal component (Fig.\,\ref{K2_HD73045_C18}, top-right). Similar figures for the remaining campaigns of all targets are given in Appendix\,\ref{appendixC} (Figs.\,\ref{K2_HD73045_C05}--\ref{K2_HD76310_C18}).

\begin{table*}
\centering
\caption{Parameters of the sinusoidal component (ordered by amplitude), related to the variations observed within the analysed flux data, as derived from the fitting process. For each data set, these components have been summed to produce a model of their respective light curves.}
\fontsize{7.2}{9.0}\selectfont
\label{table:KepSineComps}
\begin{tabular}{lcccccc}  % l = left, c = centered
\hline
\hline 
%\noalign{\smallskip}
Star Name & Campaign & Frequency &   Period   &  Amplitude   &     Phase   \\
                  &  & (d$^{-1}$)&     (d)    &  (mmag)      &     (rad)   \\
\hline
%\noalign{\smallskip}
HD\,73045 & C05  & 0.0783  $\pm$ 0.0002 & 12.77 $\pm$ 0.03  & 0.113     $\pm$ 0.003   &  1.44     $\pm$ 0.03   \\
                  &          & 0.1555  $\pm$ 0.0005 & 6.43   $\pm$ 0.02  & 0.035     $\pm$ 0.002   &  -0.47   $\pm$ 0.06   \\
                  &          & 0.0640  $\pm$ 0.0005 & 15.6   $\pm$ 0.1    & 0.030     $\pm$ 0.002   &  -2.38   $\pm$ 0.07  \\
                  &          & 0.1395  $\pm$ 0.0006 & 7.17   $\pm$ 0.03  & 0.026     $\pm$ 0.002   &  -1.31   $\pm$ 0.08  \\
                  &          & 0.1668  $\pm$ 0.0008 & 6.00   $\pm$ 0.03  & 0.020     $\pm$ 0.002   &  -0.9   $\pm$ 0.1  \\
\noalign{\smallskip}
                  & C18  & 0.0791  $\pm$ 0.0005 & 12.64  $\pm$ 0.08  & 0.039     $\pm$ 0.002   &  -0.008  $\pm$ 0.05 \\
                  &          & 0.1535  $\pm$ 0.0008& 6.51    $\pm$ 0.03  & 0.023     $\pm$ 0.002  &  -0.54   $\pm$ 0.07 \\
\hline
%\noalign{\smallskip}
HD\,76310 & C05  & 0.2165  $\pm$ 0.0002 & 4.618 $\pm$ 0.004  & 0.50      $\pm$ 0.01    &  2.81    $\pm$ 0.02 \\
                  &          & 0.2039  $\pm$ 0.0002 & 4.905 $\pm$ 0.005  & 0.320    $\pm$ 0.008  &  -1.39    $\pm$ 0.03 \\
                  &          & 0.2290  $\pm$ 0.0002 & 4.367 $\pm$ 0.005  & 0.197    $\pm$ 0.007  &  0.19   $\pm$ 0.03  \\
                  &          & 0.1863  $\pm$ 0.0003 & 5.366 $\pm$ 0.008  & 0.149    $\pm$ 0.006  &  -0.84   $\pm$ 0.04 \\
                  &          & 0.4198  $\pm$ 0.0004  & 2.382 $\pm$ 0.002  & 0.106    $\pm$ 0.005  &  -2.56    $\pm$ 0.05  \\
%                  &          & 0.0481  $\pm$ 0.0004  & 20.8   $\pm$ 0.2      & 0.101    $\pm$ 0.005  &  -1.76   $\pm$ 0.05  \\
                  &          & 0.1668  $\pm$ 0.0004  & 5.99   $\pm$ 0.01    & 0.099    $\pm$ 0.004  &  -2.75   $\pm$ 0.05 \\
\noalign{\smallskip}
                  & C16  & 0.2089   $\pm$ 0.0002  & 4.786   $\pm$ 0.004   & 0.40   $\pm$ 0.01   &  -2.22   $\pm$ 0.02 \\
                  &          & 0.1856   $\pm$ 0.0002  & 5.388   $\pm$ 0.005   & 0.315 $\pm$ 0.008 &  -0.65   $\pm$ 0.02  \\
                  &          & 0.0756   $\pm$ 0.0004  & 13.23   $\pm$ 0.07     & 0.099 $\pm$ 0.006 &  0.90    $\pm$ 0.06  \\
                  &          & 0.4302   $\pm$ 0.0004  & 2.325   $\pm$ 0.002   & 0.084 $\pm$ 0.005 &  -2.51   $\pm$ 0.06  \\
                  &          & 0.9699   $\pm$ 0.0004  & 1.0310 $\pm$ 0.0005 & 0.080 $\pm$ 0.005 &  -2.25    $\pm$ 0.06  \\
                  &          & 0.0573   $\pm$ 0.0005  & 17.4 $\pm$ 0.2 & 0.065 $\pm$ 0.005 &  2.95    $\pm$ 0.08  \\
\noalign{\smallskip}
                  & C18  & 0.2065   $\pm$ 0.0002  & 4.843  $\pm$ 0.006    & 1.09   $\pm$ 0.02  &  -1.05    $\pm$ 0.02  \\
                  &          & 0.1841   $\pm$ 0.0005  & 5.43    $\pm$ 0.02     & 0.25   $\pm$ 0.01   &  -2.47    $\pm$ 0.04 \\
                  &          & 0.4270   $\pm$ 0.0006  & 2.342  $\pm$ 0.003   & 0.18   $\pm$ 0.01   &  -2.4   $\pm$ 0.06 \\
\hline
%\noalign{\smallskip}
HD\,73574 & C05  & 0.0700   $\pm$ 0.0004  & 14.28  $\pm$ 0.08     & 0.85   $\pm$ 0.04   &  2.87    $\pm$ 0.05  \\
          &      & 0.0925   $\pm$ 0.0008  & 10.8    $\pm$ 0.1      & 0.36   $\pm$ 0.04   &  1.7     $\pm$ 0.1   \\
\noalign{\smallskip}
                  & C18  & 0.092   $\pm$ 0.002  & 10.8  $\pm$ 0.2    & 0.047   $\pm$ 0.007  &  0.4    $\pm$ 0.1  \\
\hline
%\noalign{\smallskip}
 HD\,73618 & C05 &0.2644 $\pm$ 0.0002  &  3.782 $\pm$ 0.003  &  0.40  $\pm$ 0.01  &   1.89 $\pm$ 0.03 \\
& &0.2055 $\pm$ 0.0002  &  4.867 $\pm$ 0.006   & 0.248 $\pm$ 0.008  &  0.90 $\pm$ 0.03  \\
 & &0.4100 $\pm$ 0.0004  &  2.439 $\pm$ 0.002   & 0.136 $\pm$ 0.007 &  2.98 $\pm$ 0.05  \\
 & &0.2531 $\pm$ 0.0005  & 3.951 $\pm$  0.008  & 0.094 $\pm$ 0.006 &  2.78 $\pm$ 0.07 \\
 & &0.4287 $\pm$ 0.0005  & 2.333 $\pm$ 0.003   & 0.092 $\pm$ 0.006 &  1.00 $\pm$ 0.07  \\
 & &0.1766 $\pm$ 0.0006  & 5.66  $\pm$ 0.02    & 0.077 $\pm$ 0.006 & -1.81 $\pm$ 0.08  \\
\noalign{\smallskip}
                  & C16  & 0.2609 $\pm$ 0.0002 &   3.833 $\pm$ 0.003  &  0.32 $\pm$ 0.009  &  1.28 $\pm$ 0.03 \\
 &&0.2052 $\pm$ 0.0002 &  4.874 $\pm$ 0.004 & 0.268 $\pm$ 0.007  &  2.74 $\pm$ 0.03 \\
 &&0.4013 $\pm$ 0.0003 & 2.493 $\pm$ 0.002 & 0.151 $\pm$ 0.005  & -2.03 $\pm$ 0.04 \\
 &&0.2244 $\pm$ 0.0003 &  4.456 $\pm$ 0.005 & 0.125 $\pm$ 0.005 &   2.86 $\pm$ 0.04 \\
 &&0.4167 $\pm$ 0.0003 & 2.399 $\pm$ 0.002 & 0.099 $\pm$ 0.004  &  2.51 $\pm$ 0.05 \\
 &&0.2365 $\pm$ 0.0003 & 4.228 $\pm$ 0.006 & 0.085 $\pm$ 0.004  & -2.90 $\pm$ 0.05 \\
 &&0.2877 $\pm$ 0.0004 & 3.476 $\pm$ 0.005 & 0.066 $\pm$ 0.004  & -0.23 $\pm$ 0.06 \\
 &&0.4471 $\pm$ 0.0005 &2.237 $\pm$ 0.002 & 0.062 $\pm$ 0.004  &  1.49 $\pm$ 0.06 \\
\noalign{\smallskip}
                  & C18  & 0.2656 $\pm$ 0.0003  &  3.765 $\pm$ 0.004  &  0.60  $\pm$ 0.01 &    -1.54 $\pm$ 0.02 \\
 &&0.2120 $\pm$ 0.0004 &    4.717 $\pm$  0.009  &   0.224 $\pm$ 0.008  & 1.47 $\pm$ 0.04 \\
 &&0.1951 $\pm$ 0.0005 &  5.13 $\pm$ 0.01  &  0.174 $\pm$ 0.007  &  -0.80 $\pm$ 0.04 \\
 &&0.4180 $\pm$ 0.0004 &  2.392 $\pm$ 0.002  &  0.173 $\pm$ 0.007 & 2.89 $\pm$ 0.04 \\
 &&0.2373 $\pm$ 0.0005 &  4.22	$\pm$ 0.01  &   0.113 $\pm$ 0.006   &  2.11 $\pm$ 0.05 \\
 &&0.2941 $\pm$ 0.0005 &  3.401	$\pm$ 0.006  &   0.111 $\pm$ 0.005   &  -0.38 $\pm$ 0.05 \\
\hline
%\noalign{\smallskip}
HD\,73619 & C05  & 0.0771 $\pm$ 0.0002 & 12.97 $\pm$  0.03   &  0.394 $\pm$ 0.009 &   -2.01 $\pm$ 0.02  \\
 && 0.1548 $\pm$ 0.0002 &  6.46 $\pm$ 0.01  &  0.208 $\pm$ 0.007  &   -2.91 $\pm$ 0.03  \\
 && 0.2326 $\pm$ 0.0003 &  4.300 $\pm$ 0.005  &  0.144 $\pm$ 0.006  &   2.52 $\pm$ 0.04 \\
 && 0.3104 $\pm$ 0.0005 &   3.221 $\pm$ 0.005  &  0.085 $\pm$ 0.005  &   1.45 $\pm$ 0.06  \\
 && 0.3882 $\pm$ 0.0006 &   2.576 $\pm$ 0.004  &  0.059 $\pm$ 0.005  &  0.77 $\pm$ 0.09 \\
 && 0.465 $\pm$  0.001  &   2.152 $\pm$ 0.005  &  0.033 $\pm$ 0.005  &  0.3 $\pm$  0.1 \\
 && 0.545 $\pm$  0.001  &   1.834 $\pm$ 0.005  &  0.025 $\pm$ 0.005  &  -2.1 $\pm$  0.2 \\
\noalign{\smallskip}
                  & C18  & 0.0775 $\pm$ 0.0003 &  12.92 $\pm$  0.05   &   0.344 $\pm$ 0.01   &  -1.27 $\pm$ 0.03  \\
 && 0.1525 $\pm$ 0.0004 &   6.46 $\pm$  0.02  &    0.179 $\pm$ 0.007  &   -0.9 $\pm$ 0.04  \\
 && 0.2294 $\pm$ 0.0006 & 4.36 $\pm$ 0.01 &    0.113 $\pm$ 0.006  &  -0.81 $\pm$ 0.05  \\
 && 0.3063 $\pm$  0.006 &   3.265 $\pm$   0.008   &  0.074 $\pm$ 0.005  &  -0.77 $\pm$  0.07 \\
 && 0.384 $\pm$ 0.001 &   2.607 $\pm$ 0.007 &    0.049 $\pm$ 0.005  &   -0.6 $\pm$ 0.1  \\
 && 0.462  $\pm$ 0.002  &   2.16  $\pm$ 0.01  &    0.023 $\pm$ 0.005  &    -0.5 $\pm$  0.2 \\
 && 0.540 $\pm$ 0.003 &   1.85 $\pm$ 0.01 &    0.017 $\pm$ 0.005  &   -0.2 $\pm$ 0.3 \\
%\noalign{\smallskip}
\hline
%\tableline\
\end{tabular}
\end{table*}

\subsection{High-Resolution Spectroscopic and Spectropolarimetric Observations}
\label{hrs}

High-resolution spectroscopy provides an opportunity of deriving a wide range of information on the target stars, such as 
%their
basic stellar parameters 
%%% PDC:.
like the effective temperature (\teff), surface gravity (\logg), and metallicity (\mh). The high-resolution spectra of our targets were gathered with four spectrographs: (1) the Hanle \'{E}chelle Spectrograph (HESP; \citealt{2015ExA....39..423C, 2018SPIE10702E..6KS}) at the 2.01-m Himalayan Chandra Telescope (HCT) of IIA, Bengaluru, (2) the Middle Resolution \'{E}chelle Spectrograph (MRES) at the 2.4-m telescope of the TNO (Thailand), (3) the High Efficiency and Resolution Mercator \'{E}chelle Spectrograph (HERMES; \citealt{2011A&A...526A..69R}) at the 1.2-m Mercator telescope, and (4) the Nasmyth \'{E}chelle Spectrometer\,(NES; \citealt{Panchuk2017}) at the 6-m Bolshoi Teleskop Alt-azimutalnyi (BTA) of the Special Astrophysical Observatory (SAO; Russia).  

For HD\,73619, we also used the Main Stellar Spectrograph (MSS; \citealt{2004mast.conf..286C, 2014AstBu..69..339P}) at the 6-m BTA for measuring stellar magnetic field strengths. Our target was observed along with two standard stars, i.e., HD\,52711 (non-magnetic) and 53\,Cam (magnetic). These spectropolarimetric observations were carried out in a series of paired exposures obtained in two orthogonal orientations of a retarder. 

Data reduction was performed using the pipeline developed for each particular spectrograph that generally involves overscan correction, averaged bias subtraction, flat-field correction, and wavelength calibration. The spectropolarimetric data reduction was performed using the \textsc{zeeman} package developed within the ESO MIDAS environment for the specific format of CCD spectra taken with an image slicer \citep{2006MNRAS.372.1804K}. 

\section{Radial Velocity Measurements}
\label{rv}

We measured the radial velocities (RVs) of all target stars using the Least-Squares Deconvolution (LSD) technique \citep{kochukhov2010}.
For HD\,73045, the obtained RV time-series data were analysed using the standard Discrete Fourier Transform (DFT) technique.
The results of the RV analysis for HD\,73045 obtained from HERMES and MRES are shown in Figs.\,\ref{rv:fig_mres} and \ref{rv:fig_hermes}, respectively. The frequency spectra at different epochs do not show any significant peak with SNR\,$>$\,4. Therefore, we conclude that in the observational limit ($3\sigma\sim300$\,m\,s$^{-1}$), HD\,73045 does not vary spectroscopically on a short time scale. The RV of HD\,73045 was also measured from the single spectrum from NES. All RV measurements are plotted with the orbital parameters given by \cite{2007MNRAS.380.1064C}, see Fig.\,\ref{hd73045orb}. Based on the residuals shown in the bottom panel of Fig.\,\ref{hd73045orb}, our results are consistent with the known orbital solution (with an orbital period of 435.57 days). 

\begin{table}
\centering
\caption{Summary of the high-resolution spectroscopic observations. Spectropolarimetric  observations are listed in the last row (MSS  {\@} BTA).} 
\label{table:spectro}
% \begin{center}
\fontsize{7.10}{9.0}\selectfont
\begin{tabular}{ccccc}  % l = left, c = centered
\hline
\hline
%\noalign{\smallskip}
 \multicolumn{5}{c}{Spectrograph \@ Telescope}\\
 \noalign{\smallskip}
 \hline
 %\noalign{\smallskip}
 Star&  HJD (2\,450\,000+) & Total No. & Spectral   & SNR \\
 & (day)         & Spectra  & Range (nm) &     \\
\hline
%\noalign{\smallskip}
\multicolumn{5}{c}{MRES  @ TNT} \\
%\noalign{\smallskip}
HD\,73045 &7007.391\,--\,7007.467 & 38 & 406\,--\,879 & 33\,--\,310 \\
          & 7009.408\,--\,7009.470 & 29  & & (@ 572 nm) \\
          & 7011.365\,--\,7011.470 & 31  &  &  \\ 
HD\,76310  &8954.116&1&&\\
%\noalign{\smallskip}
\hline
%\noalign{\smallskip}
 \multicolumn{5}{c}{ NES @ BTA}\\
 %\noalign{\smallskip}
  HD\,73045 & 7350.462 & 1 & 409\,--\,686  & 300 \\
     &  &  && (@ 555 nm) \\
%\noalign{\smallskip}
\hline
%\noalign{\smallskip} 
\multicolumn{5}{c}{HERME @ Mercator}\\
%\noalign{\smallskip}
HD\,73045&  7490.375--7490.496 & 30  & 390\,--\,893 & 40\,--\,45  \\
  & 7491.377\,--\,7491.425 & 10   & & (@ 650 nm) \\
                                & 7512.375\,--\,7512.460 & 22    &   \\
%\noalign{\smallskip}
\hline
%\noalign{\smallskip}
\multicolumn{5}{c}{ HESP @ HCT}\\
%\noalign{\smallskip}
HD\,73574&  8473.186&  1 & 350\,--\,1000  & 143 \\
HD\,73619&  8473.223& 1 &                &96 \\
HD\,76310& 8473.261& 1 &                 &100 \\
HD\,73618&  8555.074& 1 &                  &220 \\
& & & & (@ 550 nm) \\
%\noalign{\smallskip}
\hline
%\noalign{\smallskip}
 \multicolumn{5}{c}{ MSS  @ BTA }\\
 %\noalign{\smallskip}
 HD\,73619&8578.300\,--\,8578.330 & 3&444\,--\,498 &150\,--\,200  \\ 
    & 8579.293\,--\,8579.314 &     3& &100\,--\,250 \\
    & & & & (@ 455 nm) \\
%\noalign{\smallskip}
\hline
\end{tabular}
% \end{center}
\end{table}

\begin{figure}
	\begin{center}
		\includegraphics[width=\columnwidth]{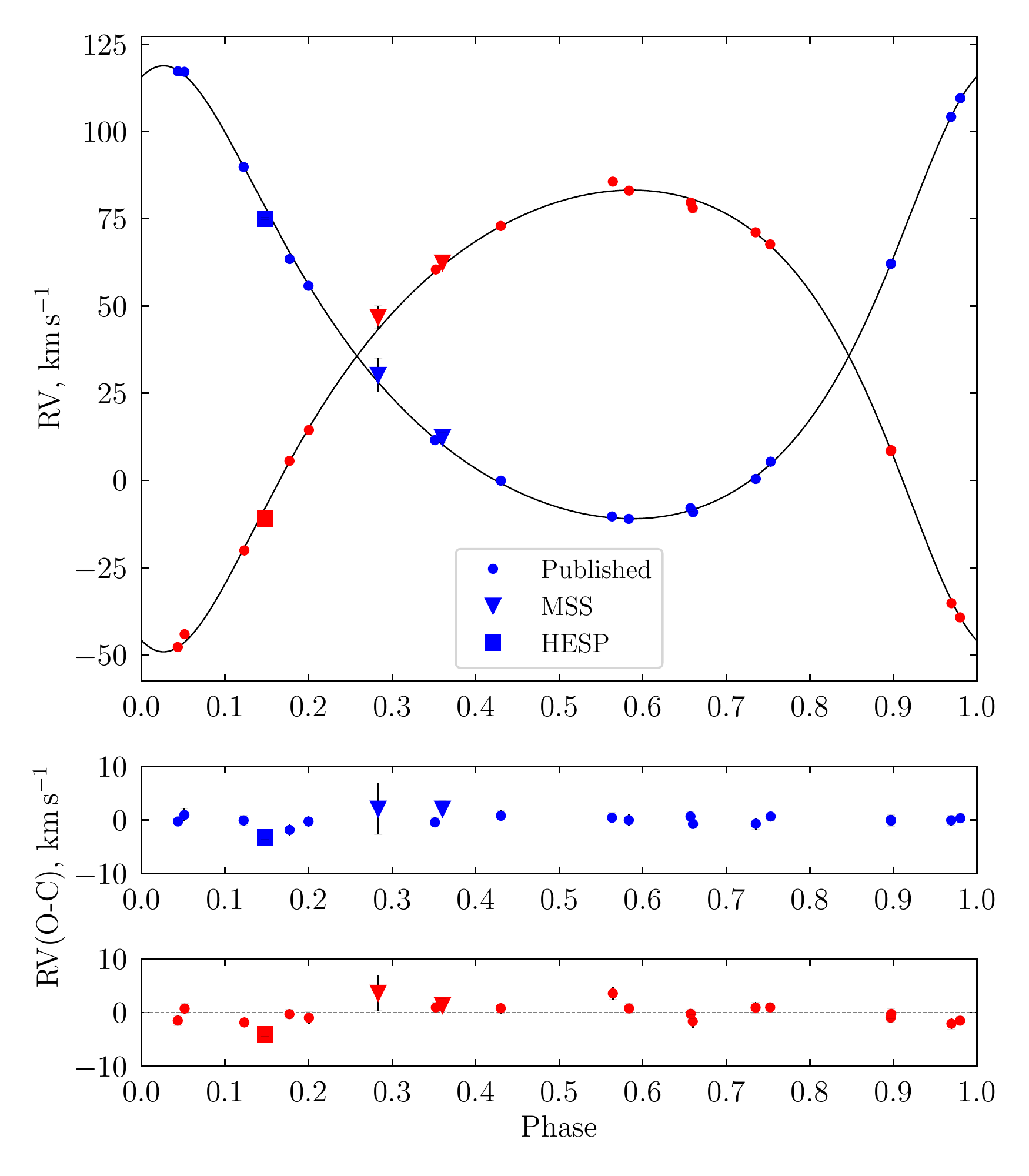}
		\caption{Radial velocity measurements of HD\,73619 derived from both the HESP and MSS spectra are plotted together with the orbital solution given by \citet{2000A&A...354..881D}. Blue and red symbols represent the measured RV of the primary and secondary component, respectively.}
		\label{orb_hd73619}
	\end{center}
\end{figure}

For HD\,73619, the RV measurements based on the MSS and HESP spectra (see Table\,\ref{table:hd76319-binary}) are fitted well using the orbital parameters (with an orbital period of 12.91 days) derived by \cite{2000A&A...354..881D}. Fig.\,\ref{orb_hd73619} shows the orbital solution of HD\,73619 where the observations are compared to the model. Our RV observations for both HD\,73045 and HD\,73619 agree with the known orbital parameters; hence, we did not re-derive them. 

 \begin{table*}
 \begin{center}
\caption{Basic parameters of the components of HD\,73619 and HD\,76310 obtained through high-resolution spectroscopy. The numbers 1 and 2 are assigned to the primary and secondary component, respectively. The spectral coverage was 430\,--\,680\,nm. We assumed equal light factors for both components of HD\,73619, whereas for HD\,76310, a light ratio ($l_1$) of 0.887 was used.  The stars HD\,73045, HD\,73574, and HD\,73618 are assumed to be single stars, as either they are truly single, or the secondary star does not contribute significantly to the lines of the spectrum.}

\label{table:hd76319-binary}
%\fontsize{7.2}{9.0}\selectfont 
\begin{tabular}{cccccccccc}
\hline
\hline
\noalign{\smallskip}
Star&Spectrograph & \multicolumn{2}{c}{RV} & \multicolumn{2}{c}{$T\rm_{eff}$}&\multicolumn{2}{c}{\logg}&\multicolumn{2}{c}{\vsini} \\
\noalign{\smallskip}
 &&\multicolumn{2}{c}{(\kms)}&\multicolumn{2}{c}{(K)}&\multicolumn{2}{c}{(cgs)}&\multicolumn{2}{c}{(\kms)}\\
% \cline{3-10}
&& 1 & 2 & 1 & 2 & 1  & 2 & 1 & 2 \\
 %\noalign{\smallskip}
\hline
\noalign{\smallskip}

     HD\,73619&HESP& -8.6\,$\pm$\,0.1&77.5\,$\pm$\,0.1&7150\,$\pm$\,90&7220\,$\pm$\,100&3.28\,$\pm$\,0.60&3.02\,$\pm$\,0.12&14.5\,$\pm$\,0.5&13.3\,$\pm$\,0.8\\
     \noalign{\smallskip}
     &MSS&30.2\,$\pm$\,4.8&46.8\,$\pm$\,3.3&7410\,$\pm$\,370&7200\,$\pm$\,280&3.90\,$\pm$\,0.13&3.54\,$\pm$\,0.22&17.0\,$\pm$\,2.0&19.5\,$\pm$\,0.5\\
     &&12.3\,$\pm$\,1.5&62.4\,$\pm$\,1.2&7380\,$\pm$\,390&7080\,$\pm$\,260&4.0 &4.0&16.3\,$\pm$\,4.0&17.3\,$\pm$\,3.8\\
    \noalign{\smallskip}
    HD\,76310&HESP&18.9\,$\pm$\,2.3&10.1\,$\pm$\,1.8&7420\,$\pm$\,210&6220\,$\pm$\,180&4.0 & 4.0 &96\,$\pm$\,7&9\,$\pm$\,1\\
     &MRES&27.4\,$\pm$\,1.8&19.8\,$\pm$\,2.8&7030\,$\pm$\,170&6470\,$\pm$\,220&4.0 & 4.0 &103\,$\pm$\,8&20.3\,$\pm$\,3.4\\
%\noalign{\smallskip}
\hline
\noalign{\smallskip}
 HD\,73045* &  ESPaDOnS & 27.9 &- & 7570\,$\pm$\,200 & -& 4.05\,$\pm$\,0.2 &- &10\,$\pm$\,0.5 \\
    \noalign{\smallskip}
 HD\,73574 & HESP & 29.2\,$\pm$\,2.8 &- &7700 $\pm$ 160&-&4.12 $\pm$ 0.21&-&99 $\pm$ 5\\
    \noalign{\smallskip}
  HD\,73618* & ESPaDOnS & 15.1 &-& 8170 $\pm$ 200&-&4.00 $\pm$ 0.19 &-&  47 $ \pm$ 3  \\
  &  HESP & 39.2\,$\pm$\,0.5 &-& 7960 $\pm$ 180&-&3.76 $\pm$ 0.19 &-& 56 $ \pm$ 3  \\
%\noalign{\smallskip}
\hline
%\noalign{\smallskip}
 \end{tabular}\\
 \end{center}
  *\cite{2007A&A...476..911F}
\end{table*}

\section{Atmospheric Parameters} 
\label{para}
Much of the success of astronomy and astrophysics relies on the accurate knowledge about the basic parameters and structure of stars. Using high-resolution spectroscopic data, HD\,73045 has been extensively studied by \cite{2007A&A...476..911F, 2008A&A...483..891F}; hence, we refrained from re-deriving its atmospheric parameters. To determine the relevant atmospheric parameters, e.g., \teff, \logg, etc., for the remaining targets, we used two approaches, namely, photometric calibrations and high-resolution spectral data. We obtained the fundamental atmospheric parameters from photometry using the Geneva, $uvby\beta$, and 2MASS photometric systems \citep{2003yCat.2246....0C}, as well as via data obtained from the General Catalogue of Photometric Data{\footnote{ http://obswww.unige.ch/gcpd/indexform.html}} \citep{1997A&AS..124..349M}. The reddening parameters $E({\rm B-V})$ and $E({\rm B2-V1})$ were obtained using a 3-D dust-map \citep{bayestar,2019arXiv190502734G} and ratios of total-to-selective absorption computed by \cite{1976PASP...88..917C}, respectively. In Table\,\ref{table:basic}, we compiled the known and derived parameters of the programme stars based on multi-colour photometry. 
 
\begin{table*}
% \footnotesize
 \centering
\caption{Basic parameters of the studied stars extracted from various online databases. We list the atmospheric parameters ($T_{\rm eff}$, {\logg}, and \mh) computed from the three photometric systems (Geneva, uvby$\beta$ and 2MASS) as well as log$\left(L_\star/{\rm L_\odot}\right)$ as calculated from the {\it Gaia} parallaxes.}
\label{table:basic}
\fontsize{7.2}{9.0}\selectfont
\begin{tabular}{lccccccccccccc}
 \hline\hline
 \noalign{\smallskip}
 Star & V$\rm^a$ &  $\pi\rm^b$ & $E{\rm (B-V)}$ & $E{\rm (B2-V1)}$ & $T\rm_{eff}$ & [M/H] & $T_{\rm eff}$ & \logg & [M/H] & $T_{\rm eff}$ & log$\left[L_*/{\rm L_{\odot}}\right]$ & Sp. Type$\rm^c$ & Peculiarity\\
 &&&&&$\rm{}^{Geneva}$ &$\rm{}^{Geneva}$&${}^{uvby\beta}$&${}^{uvby\beta}$&${}^{uvby\beta}$& $\rm {}^{2MASS}$&&&\\
 &(mag)&(mas)&(mag)&(mag)&  ($\pm 70$\,K) & $\pm 0.08$& ($\pm 200$\,K)&($\pm 0.10$\,cgs)& $\pm 0.13$ &($\pm 190$\,K)& &&\\
 %&&&&&&$\pm 70$&$\pm 0.08$&$\pm 200$&$\pm 0.10$&$\pm 0.13$&$\pm 190$ &&&\\
 \noalign{\smallskip}
 \hline
 \noalign{\smallskip}
HD\,73045\, & 8.62   &  3.7587 & 0.035 & 0.0272 & 7440 & 0.44 & 7470 & 4.28 & 0.37 & 7390 &  1.29 $\pm$0.10  & A7  &  Am \\
 HD\,73574&7.75&  5.7064&0.018&0.0138&7870&0.16 & 7660&3.91&0.15&7730& 1.24 $\pm$ 0.17 &A3&-\\
 HD\,73618&7.30& 6.9834 &0.015&0.0118&7940&0.33& 8090&3.90&0.39&7910& 1.27 $\pm$ 0.13&A0&Am$\rm ^{d,e}$\\
 HD\,73619 & 7.52 &   5.4033 & 0.015 & 0.0120 & 7670 & 0.52 & 7890 & 4.17 & 0.47 & 7860 &  1.20 $\pm$ 0.06 & A0 & Am$\rm ^f$\\
 HD\,76310 & 8.52  &  5.6101 & 0.015 & 0.0117 & 7310 & 0.14 & 7470 & 4.09 & 0.25 & 7320 &  0.94 $\pm$ 0.11 & A2 & Am$\rm ^g$\\
 \noalign{\smallskip}
\hline
\noalign{\smallskip}
 \end{tabular}\\
$\rm^a$\cite{2000A&A...355L..27H}, $\rm ^b$\cite{2018yCat.1345....0G}, $\rm^c$\cite{1993yCat.3135....0C}, $\rm ^d$\cite{1960JO.....43..129B}, $\rm ^e$\cite{1966POHP....8...24R}, $\rm ^f$\cite{1956PASP...68..318B}, and $\rm ^g$\cite{1965PASP...77..184C}  
\end{table*}
 
\begin{figure}
\begin{center}
\includegraphics[width=\columnwidth]{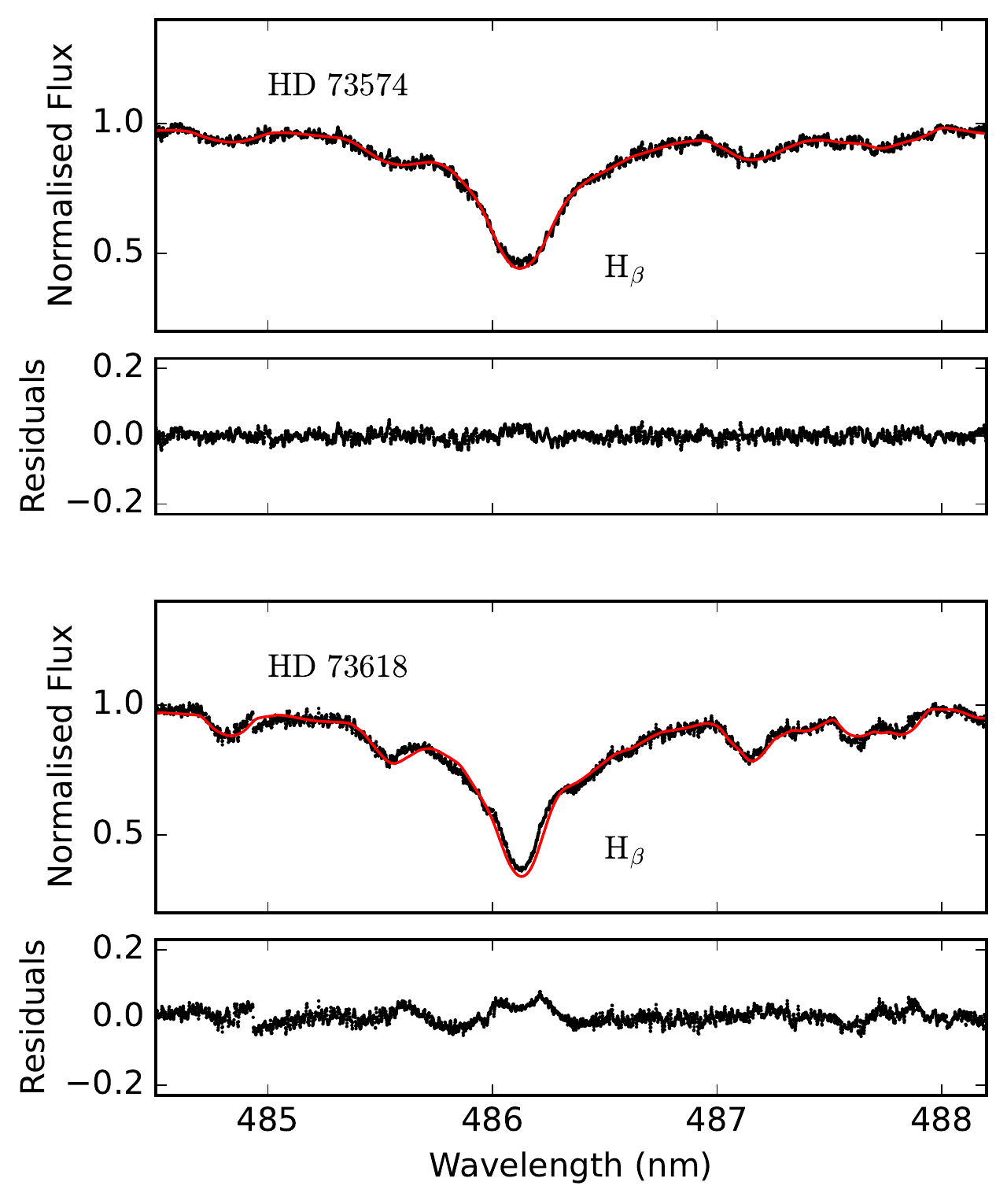}
\caption{H${_\beta}$ region of the spectra used for the determination of the respective stellar effective temperature. The synthetic spectrum (red) is over-plotted with the observed spectrum (black).}
\label{sphd73574-73618}
\end{center}
\end{figure}

The fundamental atmospheric parameters determined from photometry were used as initial input to facilitate further analyses. Spectroscopic parameters were derived by comparing observed spectra with synthetic ones computed for the adopted atmospheric parameters by using the $\chi^2$-minimisation method (see \citealt{2021MNRAS.504.5528T} for more detail). The used the radiative transfer code MOOG \citep{2012ascl.soft02009S}, together with the line-list version 5 of the Gaia-ESO Survey (GES) (wavelength range of 475\,--\,685\,nm) \citep{2015PhyS...90e4010H}, the solar abundances by \cite{2009ARA&A..47..481A}, and the ATLAS9{\footnote{ http://www.stsci.edu/hst/observatory/crds/castelli\_kurucz\_atlas.html}} atmospheric models of \cite{2003IAUS..210P.A20C} through an integrated software \textsc{iSpec} \citep{2014ASInC..11...85B, 2019arXiv190209558B}. 
The atmospheric parameters estimated from the best fits between the calculated and observed spectra are listed in Table\,\ref{table:hd76319-binary}. 
The synthetic spectra based on the values of the atmospheric parameters fit the observed spectra well, as shown in Fig.\,\ref{sphd73574-73618}.

\begin{figure}
\centering
 \includegraphics[width=\columnwidth]{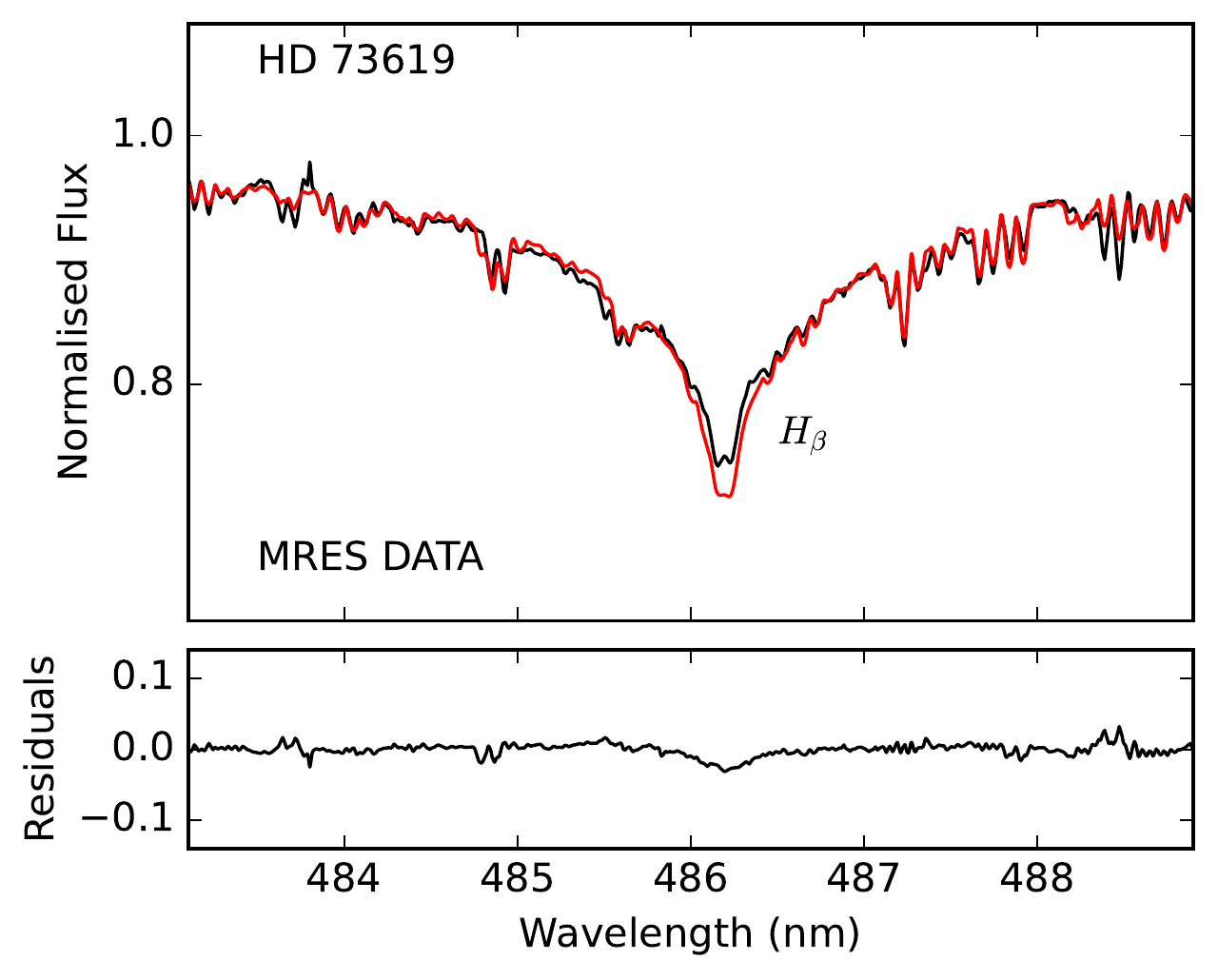}
\caption{H$_\beta$ section of the MRES spectrum (black) for HD\,73619 and the fit (red) using a two-component model and the 2-D version of \textsc{girfit}. 
}
\label{figure:76319-binary1}
\end{figure}

For HD\,73619, the HESP and MSS spectra clearly show the presence of two components (see Fig.\,\ref{figure:76319-binary1} and the LSD profiles of Fig.\,\ref{fig:polarim}). We derived the atmospheric parameters from HESP and MSS spectra using the two-dimensional version of the \textsc{girfit} code developed by \cite{2006A&A...451.1053F}. We reconstructed the composite spectra by searching for the best pair of Doppler shifts and synthetic spectra, one for each component, with solar-like composition in the spectral regions 439\,--\,450\,nm, 450\,--\,470\,nm, 470\,--\,500\,nm, and 510\,--\,550\,nm. Fig.\,\ref{figure:76319-binary1} shows a portion of the HESP spectrum (black) and the fit (red) using a two-component model. The radial velocities as well as the derived atmospheric parameters are listed in Table\,\ref{table:hd76319-binary}.

\section{Magnetic Field Measurement of \texorpdfstring{HD\,73619}{}}
\label{magnetic}
Magnetic field determinations allow us to distinguish between the two types of CP stars of interest, namely Ap and Am stars. Though spots are thought to occur in the presence of intense magnetic fields, \citet{2013MNRAS.431.2240B} attributed the photometric variability observed in some A-type stars as due to rotational modulation of stellar surfaces with inhomogeneities in cases where the magnetic field is either weak or absent. The magnetic field can be detected through the Zeeman effect where the left and right circular polarization spectra are shifted with respect to each other with the shift being proportional to the longitudinal magnetic field averaged over the stellar disk \citep[][and references therein]{2012MNRAS.424.2002J}. 

\begin{figure}
\centering
\includegraphics[width=\columnwidth]{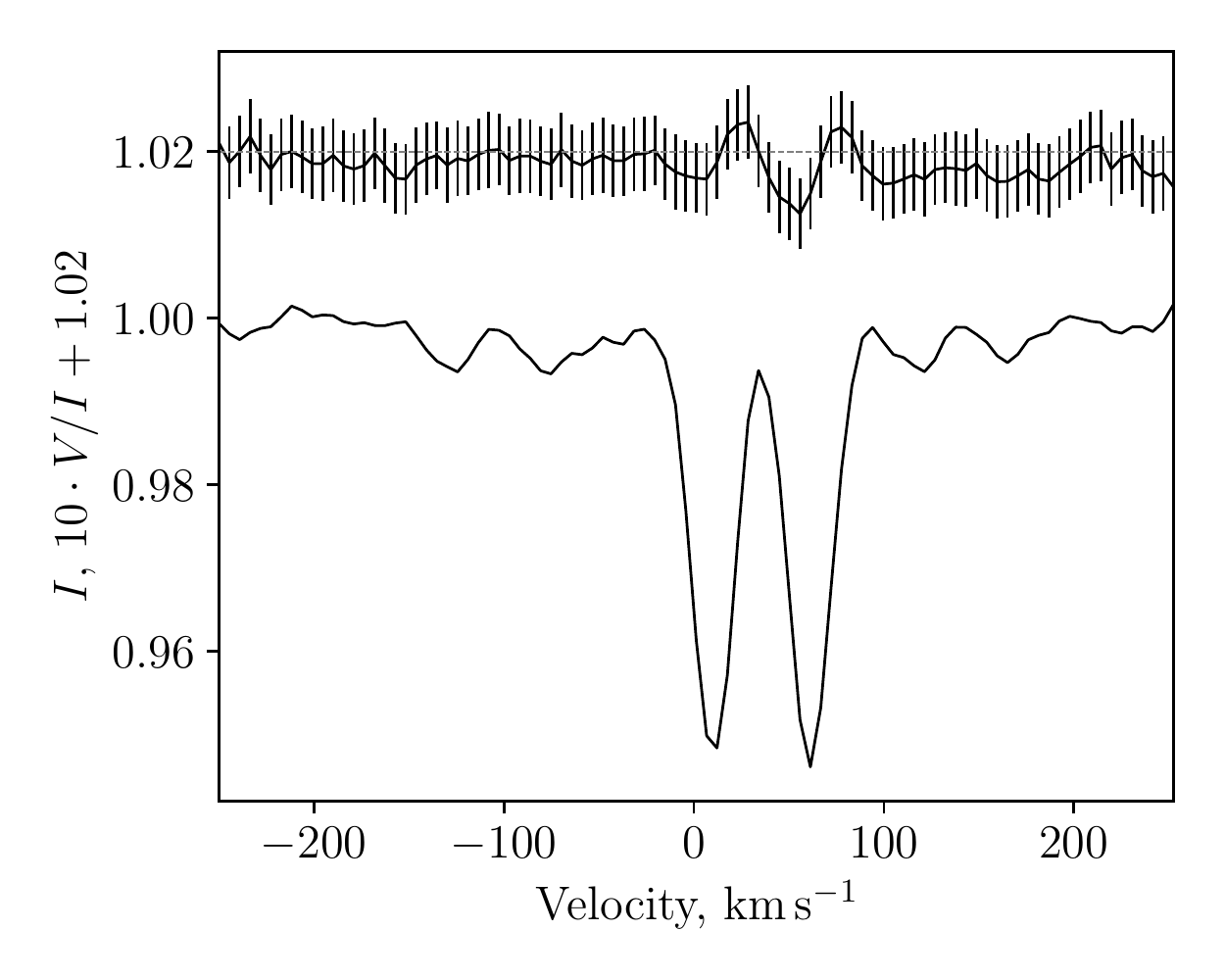}
\caption{The intensity and circularly polarized LSD profiles of HD\,73619.
}
\label{fig:polarim}
\end{figure}

We used three techniques to measure the magnetic fields: center of gravity, regression, and LSD. A mean value of 1.23 was adopted for the effective Land\'e factor that is slightly higher than the corresponding parameter in the original paper by \citet{2002A&A...389..191B}. An example of the intensity and circularly polarized LSD profiles of HD\,73619 is depicted in Fig.\,\ref{fig:polarim}. The magnetic field measurements are summarized in Table\,\ref{table:longfield}. 

\begin{table}
\caption{Summary of measurements of longitudinal magnetic field in HD\,73619 and the standard stars. The third, fourth, and fifth columns represent the magnetic field measured using classical center of gravity, regression and LSD profiles, respectively. The errors of the magnetic field are given by $\sigma$, $n$ is a number of measured lines.}
\label{table:longfield}
\begin{center}
\scalebox{0.78}{
\begin{tabular}{lcccc}
\hline
\hline\noalign{\smallskip}
Star Name &HJD (2\,450\,000+)& $B_{z}^\mathrm{g}\pm\sigma (n)$ & $B_{z}^\mathrm{r}\pm\sigma$ & $B_{z}^\mathrm{LSD}\pm\sigma$ \\
\noalign{\smallskip}
                  &  (day)& (G)                    &        (G)  &    (G)  \\   
\noalign{\smallskip}
\hline
\noalign{\smallskip}
HD\,73619        & 8578.300 & $125\pm50$ (215)   & $120 \pm 30$ & $70 \pm 50$ \\
\noalign{\smallskip}
                 & 8579.293 & $-150 \pm 50$ (82)  & $-115 \pm 30$ & $-30 \pm 30$   \\
                 &          & $-180 \pm 50$ (74)  & $-110 \pm 25$ &                \\
\noalign{\smallskip}
HD\,52711         & 8578.321 & $3 \pm 50$ (267)  & $10 \pm 20$ & $0 \pm 40$ \\
\noalign{\smallskip}
                 & 8579.307 & $170 \pm 50$ (244) & $150 \pm 40$ & $110 \pm 30$ \\
\noalign{\smallskip}                
53\,Cam          & 8578.330 & $-4960 \pm 150$ (204) & $-3730 \pm 80$ & $-2837 \pm 50$ \\
\noalign{\smallskip}
                 & 8579.314 & $-3640 \pm 150$ (204) & $-2680 \pm 90$ & $-2170 \pm 50$ \\
\hline
\end{tabular}
}
\end{center}
\end{table}

Besides the target star HD\,73619, during the same nights we also observed HD\,52711 and 53\,Cam as non-magnetic and magnetic standard stars, respectively. For each standard star, two spectra were obtained, one per night of observation. Within the observational uncertainties, the measured values of the magnetic field in 53\,Cam are consistent with the published results \citep[e.g.,][]{1998MNRAS.297..236H, 2004A&A...414..613K, 2010A&A...520A..79M}. Unlike the magnetic standard stars, a solar-type star such as HD\,52711 should not manifest a magnetic field within the typical $B_z$ measurement uncertainties. A spurious signal of circular polarisation in one spectrum of HD\,52711 obtained during the second night, is attributable to instrumental effects.

This signal, in combination with an incomplete separation of lines in the spectrum of HD\,73619, is the apparent cause for the measured negative longitudinal magnetic field of the star. To check the reliability of this result, we constructed a mask of not-blended lines belonging to the different components and measured the signal only within the mask. Taking into account this finding, we conclude that both components of the HD\,73619 system do not possess a magnetic field above 200\,G. Nevertheless, in light of multiple observations of the extremely weak magnetic fields seen in some Am stars, we propose additional high-resolution spectropolarimetric observations prior to any confirmation of magnetic field strengths at the sub-hundred Gauss level.

\section{Surface Rotation}
\label{discussion}
As a part of the Nainital-Cape survey, a total of 337 Ap and Am stars were monitored to search for photometric variability and most of them turned out to be non-variables. The most plausible reason is because the light variations (if any) are of low amplitude, and thus undetectable from the ground. Using the high-precision space-based {\it K2} data, we confirm the presence of sub-milli-magnitude amplitude light variability in HD\,73045, HD\,73574, HD\,73618, HD\,73619, and HD\,76310, which were previously classified as non-variable stars as part of the Nainital-Cape survey. In the following subsections, we discuss the possible sources of light variation. 

\subsection{Rotational Modulation}
\label{rotational_modulation}
Rotational modulation in the light curve of a periodic variable is an indication of the presence of star spots or co-rotating clouds. The periodogram associated with rotational modulation can be characterized by the presence of harmonics of the rotation period \citep[e.g.,][]{Santos2017}. However, rotational variables may not always exhibit detectable harmonics besides the fundamental period. Therefore, at low frequencies, it becomes increasingly difficult to distinguish between binarity and rotation. Nevertheless, the amplitude of light variations due to co-rotating star spots may change in time because of spot evolution, whereas the amplitude of light variations due to orbital motion is expected to remain unchanged during the time of observation. 
 
In order to understand the nature of low-frequency variations in A-type stars, \cite{2011MNRAS.415.1691B} assumed that the frequency of highest amplitude in the frequency range $0.1$\,d$^{-1}<f< 5.0$\,d$^{-1}$ is the rotation frequency. There are two ways of validating that the variability is due to rotational modulation. One way is to demonstrate that there is a relationship between the projected rotational velocities, \vsini, and the predicted equatorial rotational velocities, \vrot. Since sin\,$ i \leq 1$, the expectation is that in a plot of \vsini\ as a function of \vrot, all points will lie below or (less frequent) on the line \vsini\,$ = $\,\vrot, subject to measurement errors \citep{2017MNRAS.467.1830B}. Another method is to show that for stars in the main-sequence band the distribution of \vrot, derived from the rotational frequency $f_{\rm rot}$ and from an estimate of the stellar radius, matches the distribution of \vrot\ derived from spectroscopic measurements of \vsini\ for stars in the same stellar effective temperature range \citep{2013MNRAS.431.2240B}. 

\begin{figure}
\centering
\includegraphics[width=\columnwidth]{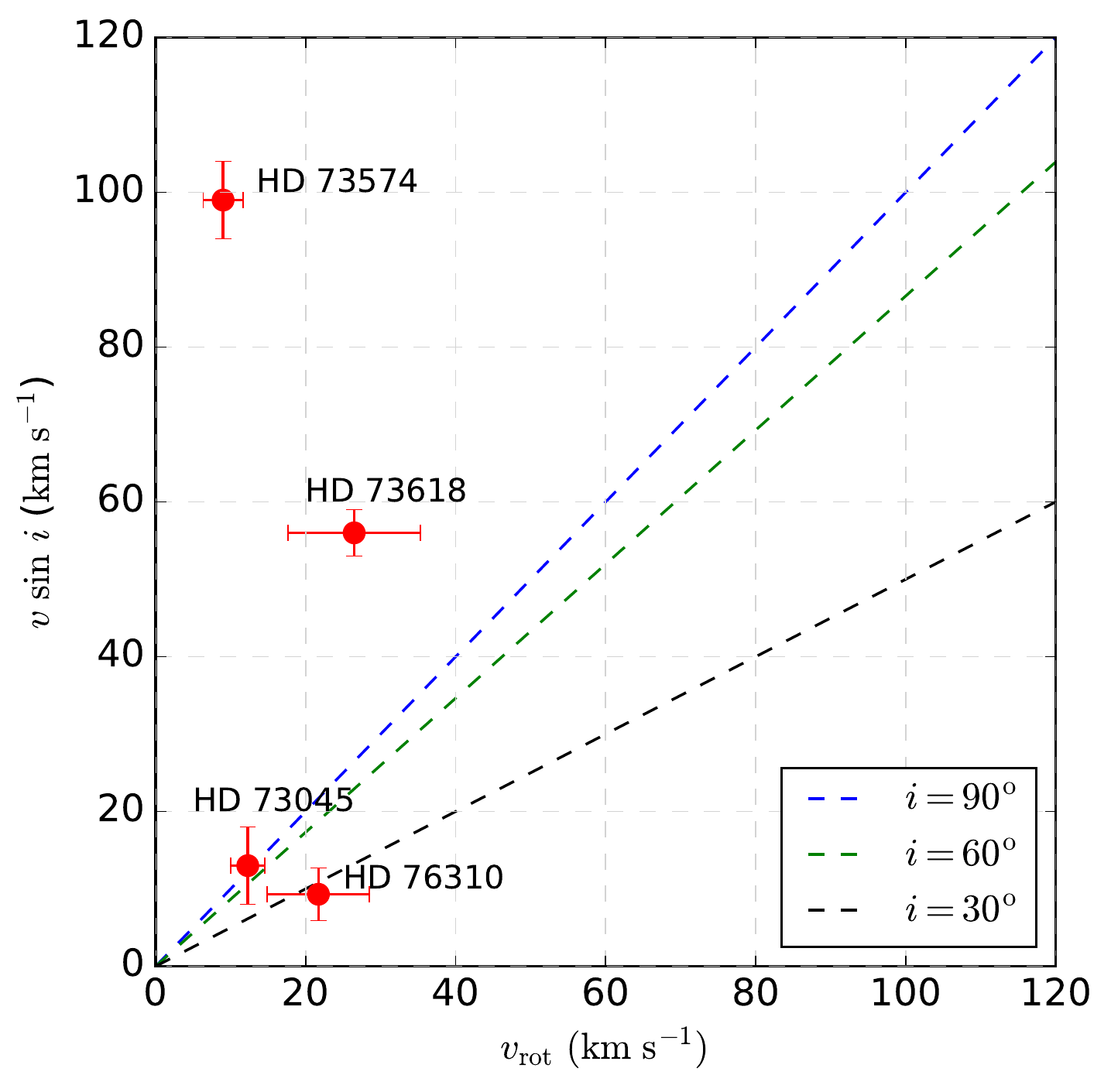}
\caption{The projected rotational velocity \vsini\ as a function of predicted equatorial rotational velocity \vrot\ for the target stars. The blue, green, and black dashed lines correspond to the inclination angle $i=90$\textdegree, 60\textdegree, and 30\textdegree, respectively.  Note that HD\,73619 is not included in this diagram because its photometric signal is orbital in nature.}
\label{rot}
\end{figure}

Fig.\,\ref{rot} shows the distribution of \vsini\ as a function of the \vrot.  
From this figure, it is evident that two stars, namely HD\,73045 and HD\,76310, are located below the line corresponding to the inclination angle $i=90$ \textdegree. Therefore, the observed variations are compatible with an interpretation of spot-like features. Based on the two observations available to us, we classify HD\,73045 and HD\,76310 as rotational variables. For the remaining two stars, HD\,73574 and HD\,73618, positioned above the diagonal line, the source of variability is unknown, thus demanding further investigations. We did not include HD\,73619 in Fig.\,\ref{rot} because its dominant signal is clearly of orbital origin.    

\subsection{Surface Mapping}
\label{map}
In order to confirm the presence and locations of spots on stellar surfaces, the phased light curves were transformed into stellar images using the inversion technique developed by \cite{2008AN....329..364S}. The surface mapping was performed for HD\,73045 and HD\,76310, classified as rotational variables in Section\,\ref{rotational_modulation}. The data for the other stars contain many artefacts that are most likely spacecraft-related (the regular thruster firings to maintain pointing), hence the poor quality data prevent us to construct high-precision spot maps.

The time-series data were divided into different observational data sets, each covering one rotational period of the star.  Hence, we analysed 43 and 8 data sets for HD\,76310 and HD\,73045, respectively. The stellar surface was divided into a grid of 6\textdegree\,$\times$\,6\textdegree\ pixels of unit area and the values of the filling factor were determined for each grid pixel. The temperature inhomogeneity maps were constructed for inclination angle $i$= 45\textdegree. 

\begin{figure*}
\begin{center}
\includegraphics[width=\textwidth]{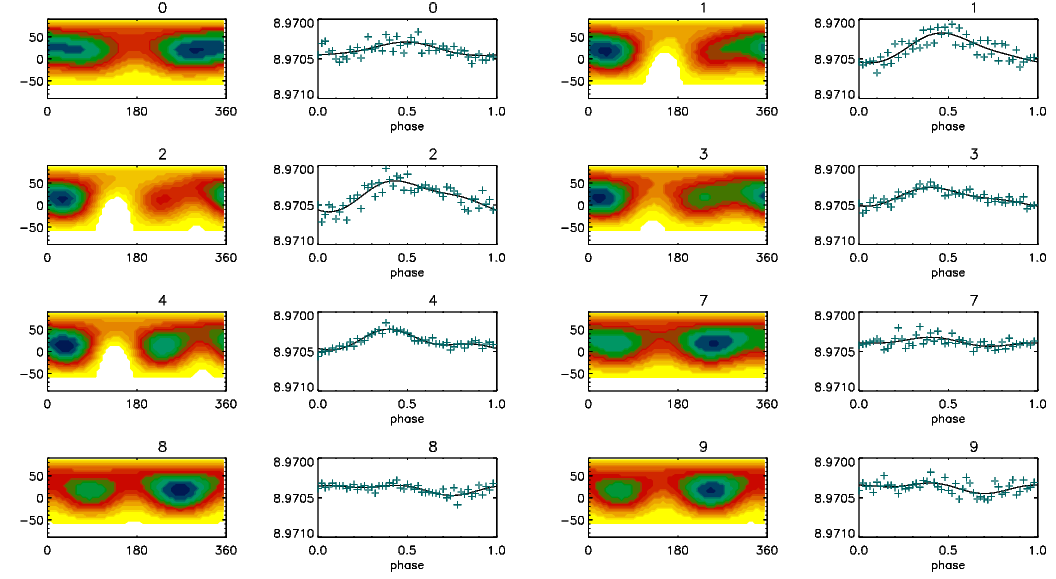}
\caption{Temperature inhomogeneity maps of HD\,73045 constructed for eight data sets. All maps use the same scale. Darker areas imply a higher spot filling factor (i.e., darkest regions correspond to a filling factor of about 0.994). The abscissa denotes the longitude in degrees and the ordinate denotes the latitude in degrees. The phase diagrams show both the observed and the model reconstructed light curves.}
\label{spot73045}
\end{center}
\end{figure*}

Fig.\,\ref{spot73045} shows the surface temperature inhomogeneity maps of HD\,73045. There are obvious concentrations of spots at two longitudes registered as two independent spotted regions. The positions of the spots on the surface of HD\,73045 vary rapidly on time-scales of one rotational period of about 13 days. The maps clearly reveal that spots are continuously changing their location in longitudinal direction, a clear indication of rotation. For HD\,76310 (Fig.\,\ref{spot76310}), there is typically only one spotted region at longitudes corresponding to phases 0.5\,--\,0.7, but for the map numbers from 10 to 15, 29, 32, and 35, two spot-like structures are visible. Furthermore, the light curves become more flattened. The spot shape in most of the maps is elongated; one can also assume that for maps from 10\, to \,15, a single spotted region is split into two regions. This result supports our previous detailed analysis of spot activity for several Am stars \citep{2018ARep...62..814S} exhibiting complex behaviours depending on the positions of spots on the stellar surfaces. 

\subsection{Wavelet Analysis}
\label{wave}
Akin to sunspots, the star spots are well-established tracers of stellar rotation, but their dynamic behaviour may also be used to analyze other relevant phenomena, such as stability, stellar magnetic activity, and cycles \citep{2010Sci...329.1032G,2014A&A...562A.124M}.  Similar to  sunspots, the size and location of star spots also change; sometimes they suddenly appear or vanish. One might expect the same behaviour for spots on A-type stars, leading to variations in the amplitude of the frequencies. In order to investigate the change in frequency, we constructed time–frequency diagrams for the {\it K2} data sets of each star based on the wavelet technique that allows a better interpretation of the physical features (such as spots) prior to their consideration for period refinement \citep{10.1175/1520-0477(1998)079-61, 2010A&A...511A..46M}.

\begin{figure*}
\centering
\includegraphics[width=0.85\textwidth]{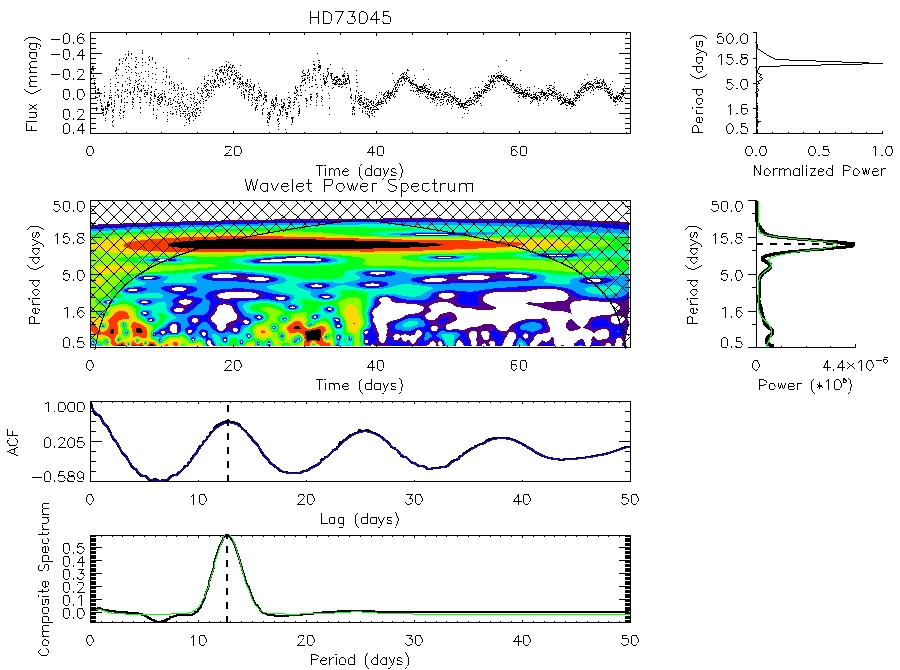}
\caption{Wavelet map of HD\,73045 using the C05 data. The top left panel shows the light curve while on the top right is the associated power density spectrum as a function of period between 0.5 and 50\,d. The left panel of the second row depicts the wavelet power spectrum (WPS) computed using a Morlet wavelet between 0.5 and 50\,d on a logarithmic scale and the associated global wavelet power spectrum (GWPS) is in right panel of the second row. The black and blue colours correspond to high and low power, respectively. The black-crossed area is the cone of influence corresponding to the unreliable results.
The panel on the third row shows the autocorrelation function (ACF) of the full light curve plotted between 0 and 50\,d. Finally, the composite spectrum \citep{2016MNRAS.456..119C, 2017A&A...605A.111C} is shown in the bottom panel.
The black dashed lines mark the respective rotation period estimates.}
\label{wavel1} 
\end{figure*}

For a given signal, this technique permits the analysis of frequency (or period) variations with time (non-stationary signals).  The Morlet wavelet, which is interpreted as the convolution of a sinusoidal and a Gaussian function, was used as the reference wavelet \citep{GOUPILLAUD198485, 1989wtfm.conf..286H}.  For a given frequency, we calculated the correlation between the mother wavelet and the data by sliding the wavelet along the time axis of the light curves, resulting in a wavelet power spectrum (WPS). The WPS was then projected on the period axis to obtain the global wavelets power spectrum (GWPS). The time--frequency plots of the studied stars are shown in Fig.\,\ref{wavel1} and Figs.\,\ref{wavel45c05}-- \ref{wavel10c18}. The black and blue colours in the maps indicate regions of high and low power, respectively.  

In all these cases, the variable (non-stationary) signal is noticeable. The variable signals in HD\,73045 and HD\,76310 could be due to the drifting of spot-like features or spot growth and decay. For HD\,73574, HD\,73618 and HD\,73619, variable signals are detected but do not seem to be of rotational origin (spots). To confirm the nature of the signal in HD\,73574, a light curve of longer timescale is required while that in HD\,73618 seems to originate from the beating of close frequencies. The rotation periods ($P_{\rm GWPS}$) extracted from the GWPS are 12.5\,$\pm$\,1.05\,d and 4.8\,$\pm$\,0.45\,d for HD\,73045 and HD\,76310, respectively. The detected periods $P_{\rm GWPS}$ for HD\,73574, HD\,73618, and HD\,73619 are 14.0\,$\pm$\,1.05\,d, 3.8\,$\pm$\,0.44\,d, and 12.5\,/12.7\,$\pm$\,1.04\,d, respectively. The error in the periods are derived from the error in the frequencies, which are equivalent to the HWHM of the peaks in the GWPS.

\subsection{Autocorrelation Function Analysis}
\label{decay}
The surface mapping and the time-frequency analysis imprint the existence and evolution of spots on the surface of rotational variables. We autocorrelated the time-series data at a given time lag to reveal the spots possibly co-rotating with the stars \citep{2013MNRAS.432.1203M, 2014ApJS..211...24M,2014A&A...572A..34G, Santos_2019}. The positive and normalised resultant autocorrelation functions (ACFs) have local maxima and minima that show the presence of spots on these stars; they mimic the displacement equation for an under-damped simple harmonic oscillator (uSHO) \citep{2017MNRAS.472.1618G}. In each ACF, the first local maximum represents the dominant period ($P\rm_{ACF}$). The $P\rm_{ACF}$ values for HD\,73045, HD\,73574, HD\,73618, and HD\,76310 are 12.8\,d, 13.5\,d, 4.1\,d, and 4.7\,d, respectively for the {\it K2} C05 data. These values are in good agreement with the periods derived from the amplitude spectra, wavelet maps and composite spectra (e.g., Fig.\,\ref{wavel1}, and Table\,\ref{table:acfs}).
 
 \begin{table*}
\centering
\caption{The approximate spot radius ($R\rm_{spot}$) as derived from the amplitude, period ($P\rm_{ACF}$), and variation time-scale ($\tau_{\rm VT}$) as estimated from the autocorrelation functions and the rotation periods ($P\rm_{GWPS}$).  The composite spectrum power estimate ($P\rm_{CS}$) is obtained from the respective global wavelet power spectrum and composite spectrum. }
\label{table:acfs} 
%\fontsize{7.5}{9.0}\selectfont
\begin{tabular}{lcccccc}
\hline
\hline
Star & Campaign & $R\rm_{spot}$ & $P\rm_{ACF}$  & $\rm \tau_{VT}$ & $P\rm_{GWPS}$ & $P\rm_{CS}$ \\ %&$\Delta\Omega$\\\\
Name & & ($R\rm_E$) & ($\rm d$) & (d) & (d) & (d) \\ %&(rad\,d$^{-1}$)\\
\hline\\
HD\,73045& C05 & 3.44 $\pm$	0.18 & 12.8 & 18.67 $\pm$ 0.38 & 12.5\,$\pm$\,1.04 & 12.7\,$\pm$\,0.61 \\ %&0.476 $\pm$ 0.008\\
& C18 & 2.02 $\pm$ 0.17 & 12.8 & 16.07 $\pm$ 0.24 & 12.5\,$\pm$\,1.05 & 12.8\,$\pm$\,0.57 \\
%\noalign{\smallskip}
\hline
%\noalign{\smallskip}
%HD\,73574 & C05 & 7.72 $\pm$ 0.50 & 13.5 & 20.14 $\pm$ 0.48 & 14.0 $\pm$ 1.05 & 13.8 $\pm$ 0.75 \\
HD\,73574 & C05 &  & 13.5 &  & 14.0 $\pm$ 1.05 & 13.8 $\pm$ 0.75 \\
%\noalign{\smallskip}
\hline
HD\,73618 & C05 &  & 4.1 & & 3.8 $\pm$ 0.44 & 3.9 $\pm$ 0.21 \\
& C16 &  & 4.3 &   & 4.1 $\pm$ 0.44 & 4.3 $\pm$ 0.23 \\
& C18 &   & 3.9 &   & 3.8 $\pm$ 0.44 & 3.8 $\pm$ 0.20 \\
%\noalign{\smallskip}
\hline
%\noalign{\smallskip}
HD\,73619 & C05 &  & 12.9 &  & 12.7 $\pm$ 1.04 & 12.8 $\pm$ 0.50 \\
& C18 &  & 12.8 &  & 12.5 $\pm$ 1.04 & 12.7 $\pm$ 0.57 \\
\hline
%\noalign{\smallskip}
HD\,76310 & C05 & 5.04 $\pm$ 0.27 & 4.7 & 19.31 $\pm$ 0.13 & 4.6\,$\pm$\,0.45 & 4.7\,$\pm$\,0.21 \\ %&1.278 $\pm$ 0.004\\
& C16 & 4.49 $\pm$ 0.26  & 4.9 & 18.55 $\pm$ 0.27 & 4.8\,$\pm$\,0.45 & 4.9\,$\pm$\,0.23 \\
& C18 & 7.44 $\pm$ 0.30  & 4.8 & 21.68 $\pm$ 0.13 & 4.8\,$\pm$\,0.45 & 4.8\,$\pm$\,0.22 \\
\hline
\end{tabular}
% \end{minipage}
\end{table*}
 
The values of the spot variation time-scale ($\tau_{\rm VT}$; related to the time of visibility of the spot on the stellar disk) obtained from fitting the ACF (C05) with an uSHO model are 18.67\,$\pm$\,0.38\,d for HD\,73045 and 19.31\,$\pm$\,0.13\,d for HD\,76310. We note, however, that the dominant frequencies in the target stars, HD\,73045 and HD\,76310, are sustained for a duration longer than the $\tau_{\rm VT}$ as estimated from the ACF analysis. \citet{Santos2021} demonstrated that the $e$-folding time of the uSHO systematically underestimates the characteristic spot time-scale. Moreover, the length of the observations drastically limits the retrieved timescale \citep{Santos2021}. As a result, the length of the {\it K2} time-series might be insufficient to constrain spot lifetimes through the ACF. Nevertheless, we list the values for $\tau_{\rm VT}$, which should be considered as lower limits, and we advise caution while interpreting the retrieved $\tau_{\rm VT}$. The additional values of $\tau_{\rm VT}$ obtained using the remainder of the campaigns are listed in Table\,\ref{table:acfs}. 
 
Assuming that the amplitude of the rotational frequency is produced by a black circular spot, the scenario is identical to the determination of exoplanet sizes from a transit. Using {\it K2} C05 data for  HD\,73045 and HD\,76310, respectively, the minimum spot radius ($R\rm_{spot}$) --- as derived from the photometric amplitude following a procedure described by \cite{2020MNRAS.492.3143T} --- are 3.44\,$\pm$\,0.18\,$R_{\rm E}$ and 5.04\,$\pm$\,0.27\,$R_{\rm E}$, respectively (where $R_{\rm E}$ denotes Earth's radius). The rest of the $R\rm_{spot}$ values obtained through other campaigns, when available, are given in Table\,\ref{table:acfs}. The $\tau_{\rm VT}$ of the spotted regions in these stars is of the order of a few weeks. Although, as discussed above, $\tau_{\rm VT}$ might not be a good constraint for the spot lifetimes due to the length of the observations, these results might point towards weak or non-existing magnetic fields in the program stars. The origin and nature of spots on such stars is a topic that will require detailed future modelling in consideration of that, e.g., a star like HD\,73045 known to be non-magnetic \citep{2007A&A...476..911F} shows unequivocal evidence of spots.
 
The composite spectrum (CS) combines the time-frequency analysis and the ACF, while being the product of the normalized GWPS and the normalized ACF \citep{2016MNRAS.456..119C, 2017A&A...605A.111C}. The period estimate, $P_\text{CS}$, including the respective uncertainty, corresponds to the central period and HWHM of the Gaussian function that fits the highest peak in the CS. The results from the wavelet maps, and the ACF and CS analyses are listed in Table\,\ref{table:acfs}. 

\section{An observational H-R diagram}
\label{evolve}
In order to infer the evolutionary status, the location of the target stars within the H-R diagram needs to be precisely determined.  Therefore, accurate values for the effective temperature \teff\ and the luminosity $L_\star$ need to be obtained; note that a typical error of 150\,K in \teff\
(as given by many traditional methods) entails an error of about 0.2\,mag in the bolometric magnitude \citep{2015MNRAS.454L..86N}. Fortunately, high-resolution spectroscopy allows to deduce accurate values of \teff\ and \luminosity\ as this method takes into account line blanketing caused by the chemical peculiarities. 

The values of \luminosity\ of the programme stars have been calculated using the standard relation as discussed in \citet{10.1093/mnras/sty1511}. The bolometric corrections for the Sloan Digital Sky Survey (SDSS) ugriz photometric system were obtained from \citet{2004A&A...422..205G}. For the derivation of the absolute magnitude, the {\it Gaia EDR3} parallaxes \citep{2021A&A...649A...1G} were used. Combining all these values yielded adequate \luminosity\ values, which are presented in Table\,\ref{table:basic}. 

\begin{figure} 
\centering
\includegraphics[width=\columnwidth]{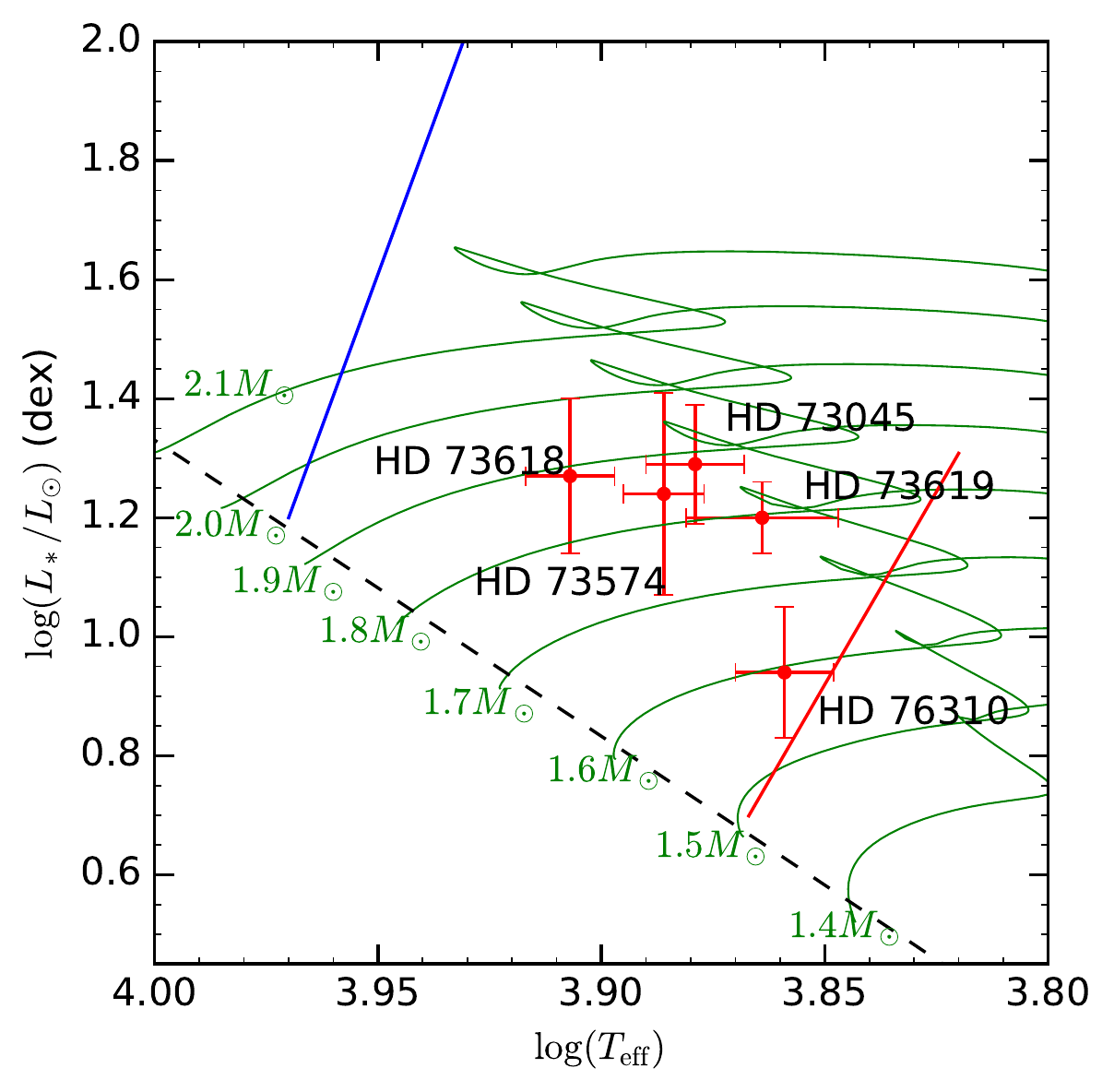}
\caption{Location of the target stars in the H-R diagram. The solid green lines represent the theoretical evolutionary tracks for the masses from 1.4\,$M_{\odot}$ to 2.1\,$M_{\odot}$ taken from  \citet{2019MNRAS.485.2380M}. The blue and red edges of the observational $\delta$\,Scuti instability strip are over-plotted with blue and red lines, respectively. The black dashed line represents the ZAMS. }
\label{HR}
\end{figure}

Based on the derived \teff\ and \luminosity\ values, we placed the programme stars in the H-R diagram and their location is shown in Fig.\,\ref{HR}. On inspection of their position, we conclude that all stars moved away from the ZAMS and are heading towards the Terminal Age Main Sequence (TAMS). Considering the error boxes, all targets lie within the $\delta$\,Scuti observational instability strip as determined by \cite{2019MNRAS.485.2380M}.

\section{Comments on Individual Stars}
\label{com_rot}
Capitalizing on the unprecedented quality of photometric data obtained from space missions, searches for photometric variability followed by determinations of the various stellar physical properties using high-resolution spectroscopy help us to ascertain the various stellar properties and activities such as chemical peculiarities, spots, pulsation, and rotation. The following subsections provide a brief summary and discussion about individual stars along with information on their basic physical properties.

\subsection{\texorpdfstring{HD\,73045/KW\,538 (Praesepe)}{}} 
HD\,73045 is classified as an Am star by \cite{1956PASP...68..318B} and \cite{1960JO.....43..129B} with a spectral type A7.2 \citep{2007AJ....134.2340K} and classified as an SB1 \citep{2000A&A...354..881D, 2007MNRAS.380.1064C}. This star was extensively studied by \cite{2007A&A...476..911F, 2008A&A...483..891F} who derived the basic physical parameters, chemical composition.  They reported that the star is of non-magnetic nature.

HD\,73045 was photometrically analysed using the HI-1A photometer of the Heliospheric Imager \citep{2009SoPh..254..387Eyles} aboard the STEREO spacecraft \citep{2008SSRv..136...17Driesman} --- a study done by \cite{2013MNRAS.429..119P} who reported a variability period of 1.25\,d. However, this variation was a consequence of the pixel response function (PRF) of HD\,73045 blending with that of its visual neighbor BD+19\textdegree2046 (P.~E.~Williams, 2019, private communication). The blended PRFs vary in shape and periodically bleed outside of the circular collection aperture as the sources track across the instrument's field-of-view. The sources were found to shift from one $y$-pixel to the next with a period of 1.25\,d, coinciding with the period reported by \cite{2013MNRAS.429..119P}. Our analysis did not find any evidence for a periodicity of 1.25\,d in the long cadence as well as the short cadence {\it K2} data, hence contradicting the findings of the variability of 1.25\,d reported by \cite{2013MNRAS.429..119P}.

Applying the analyses described in Section\,\ref{space} to the LC {\it K2} data, the photometric variation exhibits a dominant component close to 12.65\,d (12.68\,d for C05 and 12.64\,d for C18). Both data sets also show similar secondary signals at 6.42\,d and 6.51\,d, respectively.  The WPS and GWPS (Fig.\,\ref{wavel1}) also shows an excess of power at around the half of the dominant period. These secondary signals are close to harmonics of the dominant components. However, they do not correspond exactly to harmonics as can be seen in the irregular characteristic of the light curve, highlighted by the models (e.g., Fig.\,\ref{K2_HD73045_C18}, top-left, red), whereas a harmonic signal would exhibit a repetitive period-to-period structure. This observed feature is also visible in other stars such as HD\,174356 \citep{2018AandA...616A..77Bowman} and, with more complexity, in Atlas in the Pleiades \citep{2017MNRAS.471.2882W}. Based on the surface map, time-frequency and the ACF analyses, the origin of the periodic variability in HD\,73045 could be attributable to star spots situated at different latitudes on a star exhibiting differential rotation. Particularly, the rotational modulation in Fig.~\ref{K2_HD73045_C18} shows a double dip feature which indicates the presence of at least two dominant spots.

Analysing the C18 SC data (Fig.\,\ref{K2_HD73045_C18_SC}, top), the two main components described above are observed. The main goal in studying these data, however, was to confirm short-term variability (i.e., 16, 28.8, and 36\,d$^{-1}$) as previously suspected by \cite{2015IAUS..307..218J}. However, as revealed in the Lomb-Scargle spectral frequency distribution (Fig.\,\ref{K2_HD73045_C18_SC}, bottom), the range of interest (as shown in the inset) exhibits no clear signals above the noise level that would meet the FAP threshold criterion. In addition, the time-resolved radial velocity analysis did not reveal any short-term variability. Therefore, within the given photometric and spectroscopic detection limits, we conclude that HD\,73045 is pulsationally stable. The RV values derived from our observations agrees
well with the previous measurements by \cite{2007MNRAS.380.1064C}. Indeed, the radial velocities, after being averaged over each series, are in good agreement with theoretical predictions based on the published stellar orbital parameters. 

\subsection{\texorpdfstring{HD\,73574/KW 203 (Praesepe)}{}}

HD\,73574 is classified as A5V and A5III by \cite{1956PASP...68..318B} and \cite{1966POHP....8...24R}, respectively. \cite{2008A&A...483..891F} determined the values for $T\rm_{eff}$, \logg\, and $\rm[F_e/H]$ as 7662\,K, 4.00 (cgs) and 0.10, respectively. This star was identified as a moderate rotator with \vsini\ of 120\,\kms\ \citep{1967MNRAS.137..303M}. The {\it K2} C05 data set exhibits two periods, namely, 10.8\,d and 14.27\,d (see Fig.\,\ref{K2_HD73574_C05}). From the time-frequency as well as the ACF and CS analyses of the {\it K2} C05 data set, it is found that HD\,73574 shows a periodicity of about 13.8\,d (Fig.\,\ref{wavel74}).

Using the spectroscopic observations, we derived values of $T\rm_{eff}$, \logg, [M/H], $\rm \xi$,  $\rm \nu$\,sin\,$i$ and RV as $7700\pm160$\,K, $4.12\pm0.21$\,(cgs), $0.07\pm0.11$, $2.12\pm0.17$\,\kms, $99\pm5$\,\kms, and $29.2\pm2.8$\,\kms, respectively. 

\subsection{\texorpdfstring{HD\,73618/KW\,224 (Praesepe)}{}}
HD\,73618 is classified as A0 spectral type \citep{1993yCat.3135....0C}. It is also referred to as an Am star \citep{1956PASP...68..318B,1960JO.....43..129B,1966POHP....8...24R}.  The \vsini\ value for this star was reported as 51\,\kms\ \citep{1999ApJ...521..682A} and 44\,\kms\ \citep{2000A&A...354..881D}.

\cite{1993AJ....106..637M} identified this star as the primary component of a SB1 binary system in which the flux of the secondary star does not significantly influence the combined total flux spectrum.  There are various studies about the atmospheric parameters.  Based on those determinations, $T_{\rm eff}$, {\logg}, and $\rm [F_e/H]$ are given as 8060\,K, 3.87\,(cgs), and 0.46 by \citet{1997A&A...323..901}, 8100\,K, 4.00\,(cgs), and 0.50 by \cite{1998A&A...338.1073B}, and 8170\,K, 4.00\,(cgs), and 0.46 by \cite{2007A&A...476..911F}, respectively. \cite{2007A&A...476..911F} did not find any detectable magnetic field.

The {\it K2} C05 data set (Fig.\,\ref{K2_HD73618_C05}) shows a dominant peak at 3.78\,d.  In the C16 data set (Fig.\,\ref{K2_HD73618_C16}), the three peaks are given as 3.83\,d, 4.87\,d, and 2.49\,d, whereas in the C18 data set (Fig.\,\ref{K2_HD73618_C18}), they are given as 3.76\,d, 4.81\,d, and 2.37\,d. These main features are accompanied by lower amplitude, seemingly unrelated, components, producing  a rather complex time-series. The time-frequency, ACF and CS analyses suggest a periodic variability of about 4\,d with its origin still undetermined. 

An attempt was made to analyse the C16 and C18 time-series together, in a $\sim$208-day dataset that includes a non-centered gap extending across 36\,per\,cent of the temporal baseline. While issues arise when applying Fourier methods to gapped data \citep{2016AnGeo..34..437M}, the Lomb-Scargle algorithm did produce a periodogram with a frequency resolution of 0.05\,d$^{-1}$. Applying oversampling to the Lomb-Scargle process introduced sidelobes that completely corrupting the periodogram.

Based on our spectroscopic analysis, the derived parameters are given as $T\rm_{eff}=7960\pm180\,K$, \logg\,$\rm=3.76\pm0.19$\,(cgs), $\rm [M/H]=0.34\pm0.11$, $\rm \xi=2.77\pm0.16$\,\kms, \vsini\,$=56\pm3$\,\kms, and RV\,$=39.5\pm0.5$\,\kms. Based on age and metallicity, \cite{2007A&A...463..789A} classified HD 73618 as a blue straggler; hence, it can be considered an ideal object for testing current models of collisionally formed blue stragglers. The wide and detailed knowledge available on this star and its environment should allow us to test the reliability of current models and provide important constraints for future model development.

\subsection{\texorpdfstring{HD\,73619/KW\,229 (Praesepe)}{}}

HD\,73619 is a member of the Praesepe open cluster \citep{1931ApJ....74..201S}. \cite{1956PASP...68..318B} identified this star as a classical Am star, with a spectral type given as A4 and F0\,III based on Ca\,\textsc{ii}\,K and metallic lines, respectively. \cite{1999ApJ...521..682A} and \cite{2000A&A...354..881D} identified the projected stellar rotational velocity as 20 and 11.2\,\kms, respectively. 

The light curve of HD\,73619 shown in the upper-left panel of Fig.\,\ref{K2_HD73619_C05} is similar to that of a heartbeat star \citep{2012ApJ...753...86T, 2017MNRAS.472.1538F,2018MNRAS.473.5165H, 2019MNRAS.485.5498W, 2020MNRAS.499.2817V}. Analysis of the frequency spectrum provided orbital periodicities of 12.97 and 12.91\,d from the C05 and C18 data, respectively. These values were confirmed by using time-domain autocorrelation and box fitting \citep{2002A&A...391..369K} methods. Although HD\,73619 was observed during Campaign 16 of the {\it K2} mission, its image was blended with HD\,73598; thus, these C16 observations have been excluded from this study.

A thorough search for any pulsational variability induced by tidal interaction requires an accurate model to be constructed and subtracted from the light curve, and the residuals analysed. While conservative sinusoidal models were produced from a limited number of components  (Table\,\ref{table:KepSineComps}), many more components would be required to reconstruct accurately the light curve. None of these components relate to pulsational signals. Construction of a detailed model was also attempted using the PHOEBE package \citep{2005ApJ...628..426P}, but the procedure failed to converge. However, two other methods were performed to probe the existence of pulsational phenomena.

\begin{figure*}
\centering
 \includegraphics[width=0.7\textwidth]{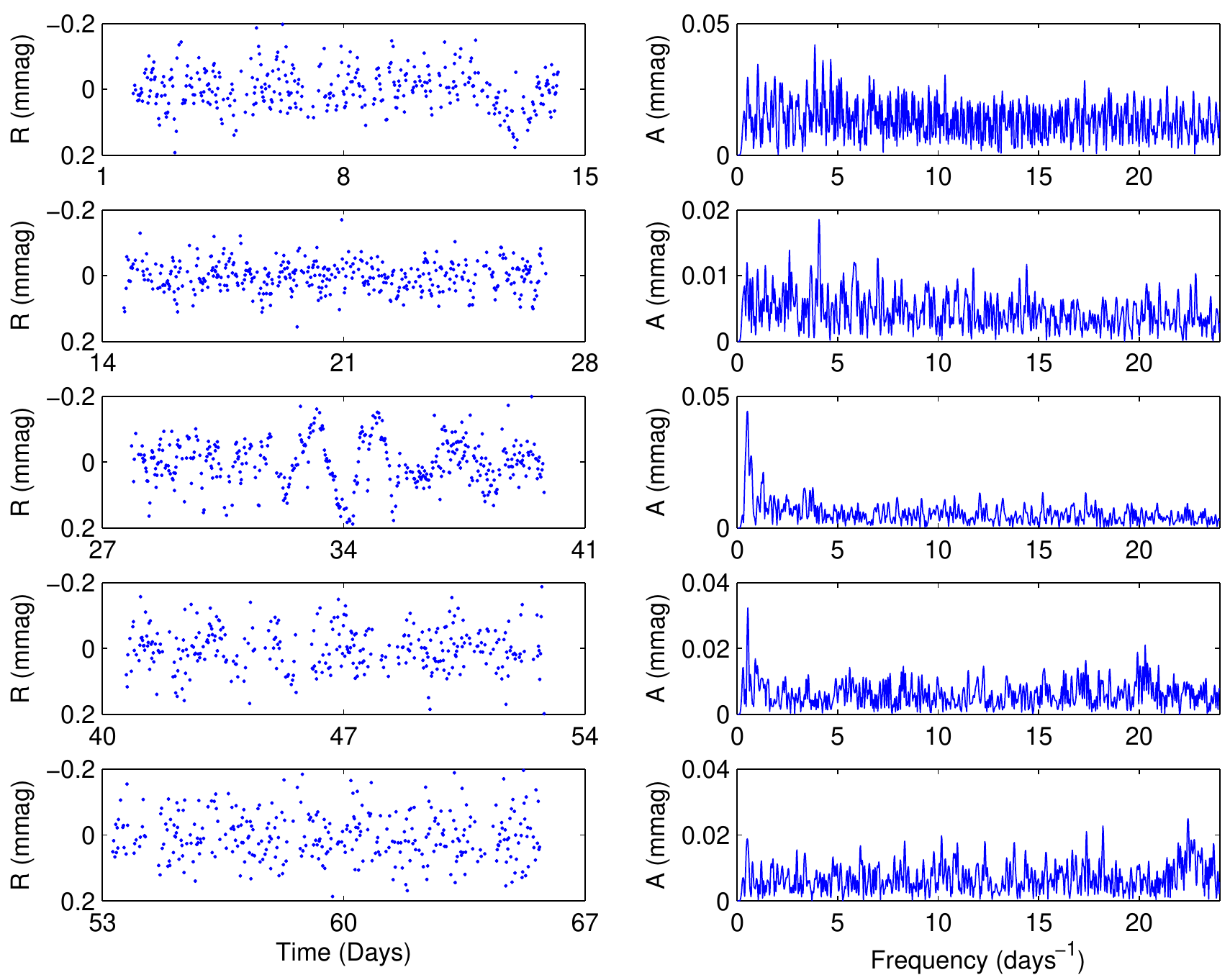}
\includegraphics[width=0.7\textwidth]{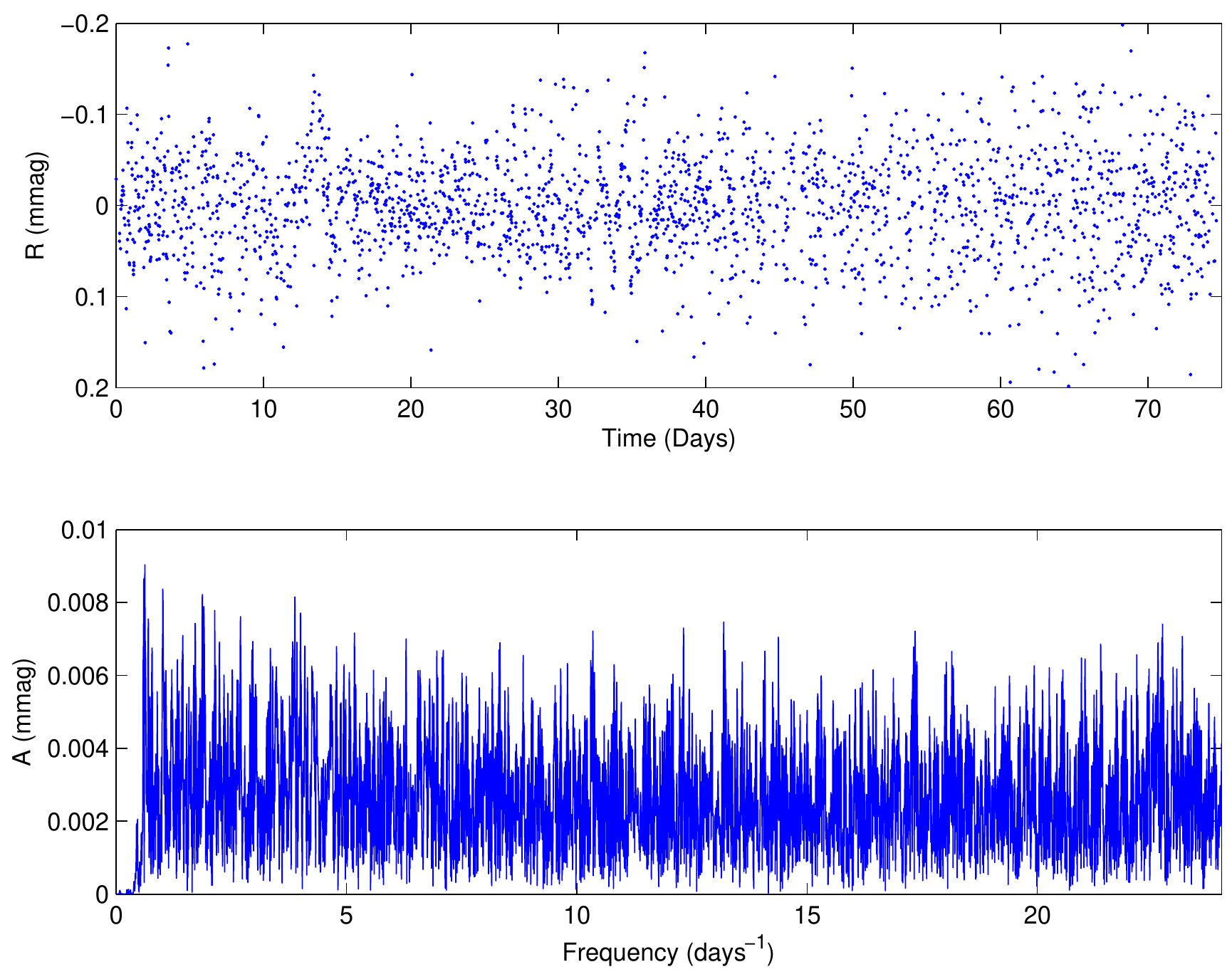}
\caption{The top five rows illustrate the residual light curve (left) and amplitude spectrum (right) of HD\,73619 after removing the heartbeat signal from five segments of the C05 time-series. The $\sim$0.5\,d$^{-1}$ spectral peak in segments 3 and 4 are related to an instrumental artifact. The middle panel shows the complete time-series after being de-trended by the k2sc algorithm with its respective amplitude spectrum in the bottom panel. Neither analysis indicates any evidence for the existence of pulsation phenomena.}
  \label{pulsation2_c05}
\end{figure*}

One method involved separating the heartbeat light curve into separate intra-period segments. Each segment was fit with an eighth-order polynomial, which was subsequently subtracted from the data, resulting in respective sets of residuals (Fig.\,\ref{pulsation2_c05}, top left). These residual data-sets were analysed with the LS algorithm to produce a frequency spectrum (Fig.\,\ref{pulsation2_c05}, top right). While an instrumental artifact appeared in the third and fourth segments of the C05 data, neither evidence of pulsation was observed nor any such evidence was seen in the C18 data (Fig.\,\ref{pulsation2_c18}). The other method involved using the K2SC algorithm previously applied to clean SAP light curves. This process also produced a de-trended residual data-set with the dominant heartbeat signal removed (Fig.\,\ref{pulsation2_c05}, middle). Again, the data-sets were analysed using the LS algorithm with no evidence of pulsation signals neither in the C05 (Fig.\,\ref{pulsation2_c05}, bottom) nor the C18 (Fig.\,\ref{pulsation2_c18}, middle
 and bottom) found.  In conclusion, within the limits of our analysis, HD\,73619 shows no evidence of pulsational variability.

Based on the parameters given in Table\,\ref{table:hd76319-binary}, we ratify that the two components of the heartbeat system have similar characteristics. 
From Fig.\,\ref{orb_hd73619}, our revised radial velocities support the orbital solution determined by \citet{2000A&A...354..881D}. Moreover, \citet{2000A&A...354..881D} reported an orbital period of 12.91124\,$\pm$\,0.00004\,d. This implies that the dominant 12.91\,d signal detected in both {\it K2} data sets is orbital in nature. The spectropolarimetric analysis indicates that, within the observational uncertainty of 200\,G, magnetic fields in HD\,73619 are either weak or absent. 

\subsection{\texorpdfstring{HD\,76310}{}}

\cite{Cowley_1965} classified HD\,76310 as an Am star.  The projected rotational velocity \vsini\ as a function of the predicted equatorial rotational velocity \vrot\ for the target stars. The blue, green, and black dashed lines correspond to the inclination angle $i=90$\textdegree, 60\textdegree, and 30\textdegree, respectively.  Note that HD\,73619 is not included in this diagram because its photometric signal is orbital in nature.

All three {\it K2} data sets tend to be dominated by two components at $\sim$4.7\,d and $\sim$5.4\,d, with variation around those values across the campaigns. These close frequencies result in well-pronounced beating effects in C05 (Fig.\,\ref{K2_HD76310_C05}) and C16 (Fig.\,\ref{K2_HD76310_C16}), but less interaction between the components during C18 (Fig.\,\ref{K2_HD76310_C18}) where the shape of the time-series varies across the periods. The level of these interactions are defined by the ratio of their amplitudes. Also note that overtones of the $\sim$4.8\,d feature seem to exist, although they do not always clear the FAP threshold and, in some cases, are obscured by the background noise. Finally, it is noteworthy that the modeling of the C05 time-series failed to produce a good fit to the data when the single component relating to the dominant 4.622 d peak was included, and 
two other components at around $\pm$0.3\,d either side of the peak were required to better fit the data.  From the time-series analysis of the {\it K2} data, the surface map, the wavelet, the ACF, and the composite spectrum, 
a dominant period close to 4.7\,d was obtained.  Given that the photometric amplitude of HD\,73610 changes significantly within only a few dozen variability cycles, we may not rule out the possibility of having unresolved multiperiodic pulsational signals, for instance of the $\gamma$ Doradus type.  An attempt to analyse the combined time-series of C16 and C18, corresponding to that for HD\,73618,  was performed with a similar outcome.

\begin{figure}
\centering
\includegraphics[width=\columnwidth]{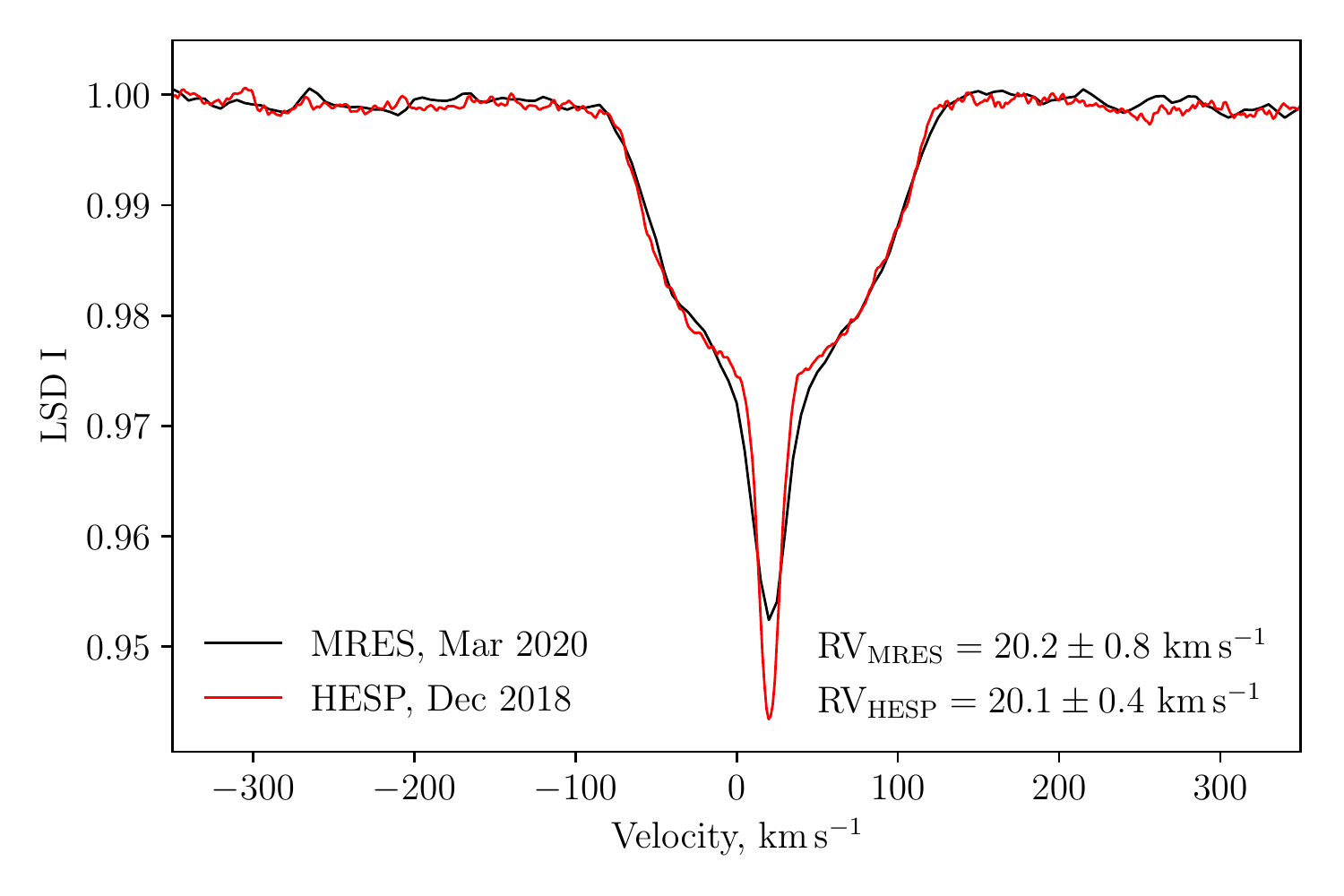}
\caption{The LSD profiles for HD\,76310 based on HESP (red) and MRES (black) data.}
\label{fig:lsd}
\end{figure}

On investigating the \vsini\, vs. \vrot\ diagram in Fig.\,\ref{rot}, HD\,76310 was found to be positioned below the line of sin\,$i=1$. Based on the surface maps, wavelet, and ACF, we thus categorise this star as a rotational variable. The LSD profiles shown in Fig.\,\ref{fig:lsd} indicate a complex structure. The RV values obtained at different epochs (December 2018 and April 2020) are the same within the observational uncertainty (i.e., $20.13\pm0.40$\,\kms\ and $20.22\pm0.83$\,\kms, respectively). The similarity between the shapes of the LSD profile of HD\,76310 and the CCF profile of KIC\,11572666 \citep{2018A&A...610A..17L} implies that the system is an SB2 system.
We fitted the MRES spectrum of HD\,76310 with a composite model using the 2D version of \textsc{girfit} in spectral regions around H$_\gamma$ (430\,--\,440\,nm), H$_\beta$ (470\,--\,500\,nm), H$_\alpha$ (630\,--\,680\,nm), and the Mg\,\textsc{i} triplet (510\,--\,520\,nm). We treated the component $T_{\rm eff}$, \vsini, RV and the luminosity ratio as free parameters while keeping \logg\ fixed at 4.0\,(cgs).

\begin{figure}
\centering
\includegraphics[width=\columnwidth]{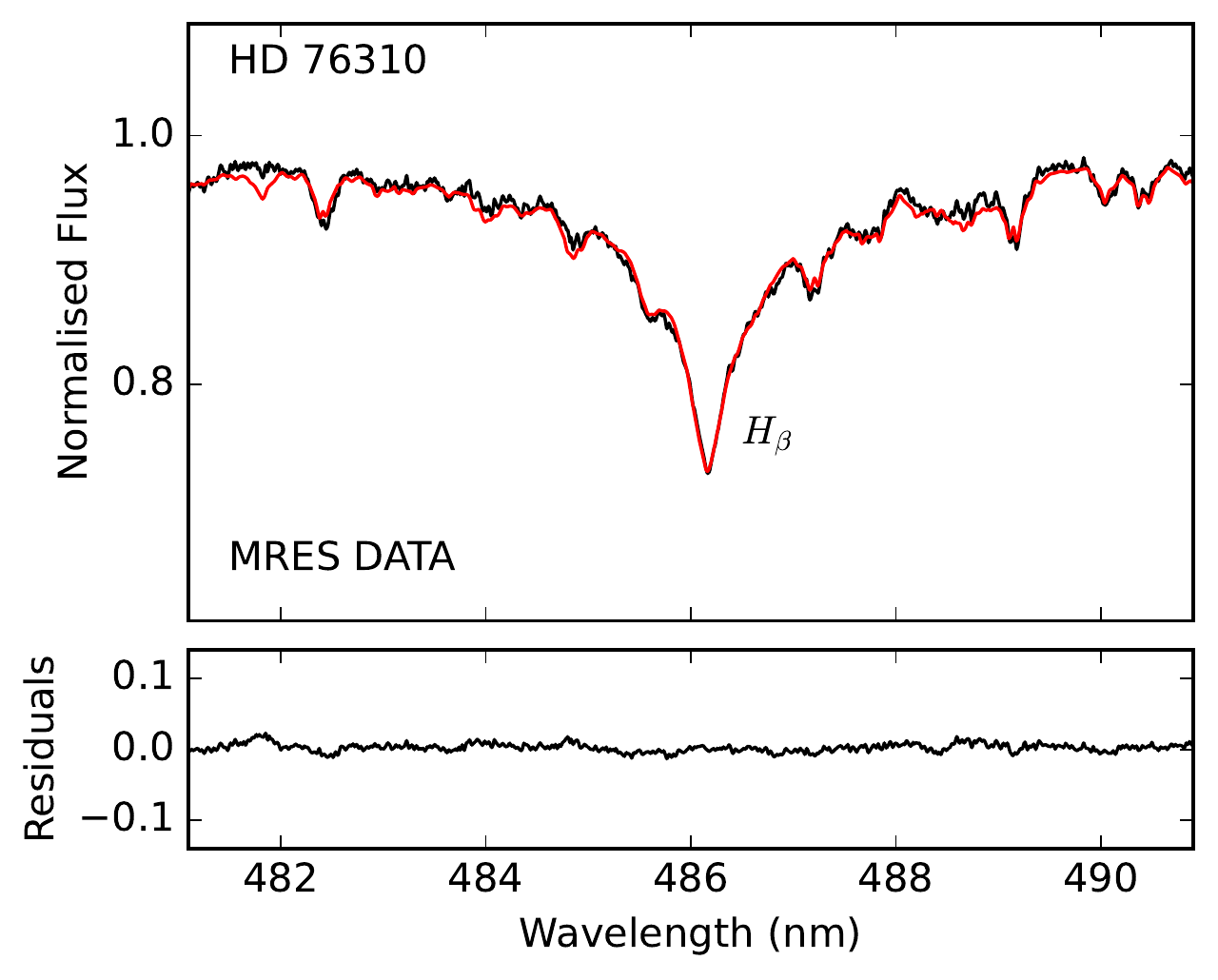}
\caption{The H$_\beta$ MRES spectrum for HD\,76310 (black) and the composite model spectrum (red) obtained using the 2D version of \textsc{girfit}.}
\label{fig:h_beta}
\end{figure}

The composite spectrum was fitted well (e.g., see H$_\beta$ line in Fig.\,\ref{fig:h_beta}); thus, we adopted the mean values of the stellar parameters from these spectral regions. The obtained values are: effective temperature of $T_{\rm eff1}=7030\pm170$\,K and $T_{\rm eff2}=6470\pm220$\,K, \vsini\ of  $97.1\pm0.4$\,\kms\ and  $9.3\pm0.7$\,\kms, RV of $27.4\pm1.8$\,\kms\ and $19.8\pm2.8$\,\kms, respectively, with a light ratio $l_1$ of $0.887\pm0.023$ (see Table\,\ref{table:hd76319-binary}). 

Based on the rotational velocities and periods of HD\,76310, the source of the stellar photometric variability, as indicated, could be the sharp-lined and cooler component, thus favouring the stellar binarity hypothesis. It is also noteworthy that HD\,76310 is included in the Washington Double Star Catalogue\,\citep{2001AJ....122.3466M} as a system consisting of two equally bright stars separated by $0.1$\,arcsec. Nevertheless, there is still the possibility that HD\,76310 is a single star surrounded by a cloud shell \citep{1986PASP...98..867S, 2003AJ....125.2196F}. Future observations would be crucial for exploring the variability and possible binarity nature of this object \citep{Welsh_2013}.

\section{Conclusions and Future Prospects}
\label{conclusion}
	 
Based on the time-series photometry obtained from the ground, the {\it K2} mission and from contemporaneous high-resolution spectroscopy and spectropolarimetry, we arrived at the following conclusions:

\begin{itemize}

\item 

All targets in the study sample, previously part of the Nainital-Cape survey and classified as constant stars, are now (potentially) identified as variables. Our analysis suggests that HD\,73045 and HD\,76310 are rotational variables. The LSD profiles of HD\,76310 indicates that this star is part of binary system or surrounded by a cloud shell. The light variation recorded for HD\,73619 is consistent with orbital signals and likely to be the first chemically peculiar Am star in a heartbeat system without tidally induced pulsations. As we could not ascertain the origin of the variability of HD\,73574 and HD\,73618, further time-domain studies are needed to apprehend the nature of these objects.

\item 

Within the detection limits of the {\it K2} photometry and radial velocity measurements, we conclude that HD\,73045 is pulsationally stable contrary to the previous prediction by \cite{2015IAUS..307..218J}.  Similarly, the analysis of the {\it K2} data revealed the absence of a periodicity of 1.25\,d previously reported by \cite{2013MNRAS.429..119P}; it is confirmed as previously erroneous.

\item 

High-resolution spectroscopy measurements obtained with two spectrographs at different epochs revealed that the HD\,73619 system is comprised of two stars with similar properties. The individual radial velocities of HD\,73619 at the different epochs are in good agreement with the orbital solution by \citet{2000A&A...354..881D}. The spectropolarimetric data analysis indicates that HD\,73619 may possess a weak magnetic field, hence additional high-resolution spectropolarimetric observations would be essential for confirmation.

 \item

The atmospheric parameters of the target stars have been derived using high-resolution spectroscopy. Considering observational uncertainties, it is found that all the studied stars are evolved away from the ZAMS and located within the $\delta$\,Scuti instability strip. 

 \end{itemize}

The potential presence of a weak magnetic field in HD\,73619, and the surface spots seen on HD\,73045 and HD\,76310, constitute further evidence that magnetic fields cannot be disregarded when exploring the origin of the peculiar chemical abundances in Am stars.   The magnetic fields leading to the formation of the surface spots  might not be stable, suggesting that they may be dominated by the dynamo effect. It is unclear if (or how) magnetic fields would affect the diffusion model for the formation of the peculiar chemical abundances in these stars, or whether alternative formation mechanism(s) would be needed to explain these phenomena.

The origin of weak magnetic fields in metallic line (Am) stars and even in normal $\delta$ Sct stars (hotter than Sun) are fossil or related to a dynamo effect is still largely an unsolved problem, although copious valuable contributions for A-type stars have previously been made \citep[e.g.,][]{2004MNRAS.348..702M}. A detailed study of the various underlying processes, particularly the diverse theoretical aspects require more attention. Future studies are relevant to the bigger picture of stellar and galactic evolution, in consideration of the significance of massive stars, notably for the production and proliferation of heavy elements as well as the overall galactic chemistry \citep[e.g.,][and references therein]{2021A&ARv..29....5M}.

Our overall long-term goal of this work is to continue the study of stellar structure and atmospheres of CP stars as well as of magnetic fields, inhomogeneities (such as spots), and tidal interaction. In the framework of our ongoing research, our future plan is to extend the detailed analysis of heretofore unstudied CP stars with {\it TESS} data in combination with future high-resolution spectroscopic and spectropolarimetric observations.

\section*{Acknowledgments}
The work presented here is supported by the Belgo-Indian Network for Astronomy and astrophysics (BINA), sanctioned by Department of Science and Technology (DST, Govt. of India; DST/INT/Belg/P-09/2017) and the Belgian Federal Science Policy Office (BELSPO, Govt. of Belgium; BL/33/IN12). 
%SJ is thankful to R. Sagar, D. W. Kurtz, P. Martinez, S. Seetha and B. N. Ashoka for initiating the field of asteroseismology research at ARIES. 
OT and EJ thank the International Science Programme (ISP) of Uppsala University and African Astronomical Society (AfAS)  for financial support. DLH acknowledges financial support from the Science and Technology Facilities Council (STFC) via grant ST/M000877/1. AG, DM, and SJ are  grateful for the support received from  the Indo-Thailand Programme of co-operation in Science and Technology through the Indo-Thai joint project DST/INT/Thai/P-16/2019. 
FKA and SC thanks the Polish National Center for Science (NCN) for grant 2015/18/A/ST9/00578. RAG acknowledges support from the CNES PLATO grant. SM acknowledges support from the Spanish Ministry with the Ramony Cajal fellowship number RYC-2015-17697. ARGS acknowledges support from NASA under Grant No. NNX17AF27G and from STFC consolidated grant ST/T000252/1. IS thanks the Government of the Russian Federation and the Ministry of Higher Education and Science of the Russian Federation, grant no. 075-15-2020-780 (N13.1902.21.0039). This paper includes data collected by the {\it K2} mission available at Mikulski Archive for Space Telescopes (MAST), SIMBAD and NASA’s ADS. The authors thank the anonymous reviewer for the insightful comments and suggestions which led to the improvement of the manuscript.

\section*{DATA AVAILABILITY}

The data underlying this article will be shared on reasonable request to the corresponding author.

 \bibliographystyle{mnras}
 \bibliography{ref}
 
 \appendix
 
\section{Author Affiliations}
$^{1}$Aryabhatta Research Institute of Observational Sciences, Manora Peak, Nainital-263002, India\\
$^{2}$Department of Physics, Mbarara University of Science and Technology, P.O. Box 1410, Mbarara, Uganda \\
$^{3}$National Astronomical Research Institute of Thailand, Chiangmai 50180, Thailand\\
$^{4}$Special Astrophysical Observatory, Russian Academy of Sciences, Nizhnii Arkhyz 369167, Russia\\
$^{5}$Affiliate Faculty, Physics and Astronomy Department, George Mason University, Fairfax, VA 22030, USA\\
$^{6}$Royal Observatory of Belgium, Ringlaan 3, B-1180 Brussel, Belgium\\
$^{7}$Jeremiah Horrocks Institute, University of Central Lancashire, Preston PR1 2HE, UK\\
$^{8}$CEA, Universit\'e Paris-Saclay, F-91191 Gif-sur-Yvette, France
$^{9}$AIM, CEA, CNRS, Universit\'e Paris-Saclay, Universit\'e Paris Diderot, Sorbonne Paris Cit\'e, F-91191 Gif-sur-Yvette, France\\
$^{10}$Universidad de La Laguna (ULL), Departamento de Astrof\'isica, E-38206 La Laguna, Tenerife, Spain\\
$^{11}$Instituto de Astrof\'isica de Canarias (IAC), E-38205 La Laguna, Tenerife, Spain\\
$^{12}$Department of Physics, University of Warwick, Coventry, CV4 7AL, UK\\
$^{13}$Space Science Institute, 4765 Walnut Street, Suite B, Boulder, CO 80301, USA \\
$^{14}$Indian Institute of Astrophysics, Koramangala, Bangalore-560034, India\\ 
$^{15}$Department of Physics, University of Texas at Arlington, Arlington, TX 76019, USA\\
$^{16}$Department of Physics and Astronomy, National Institute of Technology, Rourkela - 769008, Odisha, India\\
$^{17}$Copernicus Astronomical Center, Bartycka 18, PL-00-716 Warsaw, Poland\\
$^{18}$\c{C}anakkale Onsekiz Mart University, Faculty of Sciences and Arts, Physics Department, 17100 Canakkale, Turkey\\
$^{19}$ Centre for Space Research, North-West University, South Africa\\
$^{20}$Stellar Astrophysics Centre, Department of Physics and Astronomy, Aarhus University, Ny Munkegade 120, Denmark\\
$^{21}$Institute of Astronomy, Russian Academy of Sciences, Moscow-119017, Russia\\
$^{22}$Institut d'Astronomie et d'Astrophysique, Université Libre de Bruxelles, Boulevard du Triomphe, 1050-Bruxelles, Belgium\\
$^{23}$Kourovka Astronomical Observatory, Ural Federal University, Yekaterinburg, Russia\\
$^{24}$M. V. Lomonosov Moscow State University, P. K. Shternberg State Astronomical Institute, Moscow, Russia\\
$^{25}$SoS in Physics and Astrophysics, Pt. R. S. University, Raipur-492010, India\\
$^{26}$Institute of Informatics and Communication, University of Delhi South Campus, New Delhi, India\\
$^{27}$Department of Physics, Kyambogo University, P. O. Box 1, Kyambogo, Kampala, Uganda\\
$^{28}$ Indian Institute of Science (IISc), Bengaluru, India \\

\section{K2 Photometric Data Analysis}
\label{k2_appendix}

We analysed the K2 data discussed in Section\,\ref{space}. From these data sets, the flux (in e$^{-}$\,s$^{-1}$), $F_{i}$, over a set of $N$ sample times, $t_{i}$, with indices $i = \{0, 1, 2, ..., N-1\}$, was analysed to search for photometric variability. The analysis processes, as described below, were applied to the long cadence data of all stars. The analysis of the short cadence data acquired for HD\,73045 was performed in a similar manner.

To begin, each time step, $t_{i}$, was defined in a way that $t_{0}$ = 0; this was done by subtracting the original Barycentric Julian Date (BJD) times, $\tau_{i}$, by that of the first point, $\tau_{0}$, using

\begin{equation}
t_{i} = \tau_{i} - \tau_{0},
\label{eq:timeshift}
\end{equation}

\noindent where $\tau_{0}$ = 2\,457\,139.63 days, 2\,458\,095.49 days, and 2\,458\,251.57 days for C05, C16, and C18, respectively. 

The
%PDC\_SAP
flux data were re-calibrated as a magnitude deviation, $\Delta \it{m}_{i}$, of each flux, $F_{i}$, from its mean, $\langle F \rangle$. For all $N$ data points in a data set, at a corresponding time, $\it{t}_{i}$, the $i$-th magnitude deviation data point in the series is defined by

\begin{equation}
\Delta m_{i} = -2500\log_{10}\left(\frac{F_{i}}{\langle F \rangle}\right),
\label{eq:mmagdev}
\end{equation}

\noindent where $\Delta \it{m}_{i}$ is calculated in mmag (hence the incorporated factor of 1000). To prepare these data for the signal analysis process, variable artifacts, such as those possibly left over from the PDC process were removed by subtracting an eighth order polynomial, although it is noted that such artifact removal may also cancel any intrinsic variability component of a corresponding frequency. Subsequently, the signal search was focused on periods  shorter than 20 days. The median value of the data was also subtracted. As an example, the analysed photometric light curve of HD\,73045  obtained during C18 campaign is shown in Fig.\,\ref{K2_HD73045_C18} (top-left, blue points).

%%%%
%Position of the figure 3
%%%%

To analyse a light curve in the frequency domain, the Lomb-Scargle (LS) algorithm \citep{1976Ap&SS..39..447L, 1989BAAS...21.1069S} was applied to the time-series data, noting that the original SAP data are irregularly sampled. The Nyquist frequency $(2dt)^{-1}$ for LC (29.4\,min cadence) and SC (58.8\,s cadence) sampling are $\sim$24.5\,d$^{-1}$ and $\sim$735\,d$^{-1}$, respectively. Each campaign occupies its own temporal range $T$, giving in a respective frequency resolution, $\Delta \it{f}$. For example, C18 has a temporal window of $T = 50.814$\,d, providing a frequency resolution of $\Delta \it{f}$ = 0.0197\,d$^{-1}$ that allows an  estimate for the peak frequencies and corresponding amplitudes. To improve signal peak definition and to better estimate the peak frequencies and amplitudes, oversampling was applied to provide a frequency domain sampled every 0.001\,d$^{-1}$. Although the half-width at half-maximum (HWHM) of a peak is often used to estimate the accuracy of the peak frequency, \cite{2018ApJS..236...16V} suggests avoiding this approach through computing the false-alarm probability (FAP) for each peak. This criterion estimates the probability that a peak was produced by noise as opposed to an inherent signal. The lower the value of the FAP, the more likely that a given peak is real. Furthermore, \cite{2008MNRAS.388.1693F} noted the subjectivity of the noise level determination in calculating the signal-to-noise ratio, which is often used to select relevant peaks from the background noise spectrum \citep{1993A&A...271..482B}. Therefore, while the amplitude signal-to-noise value was calculated for candidate signal peaks, through the use of a smoothing function across the frequency spectrum to calculate the noise spectrum, the FAP was used as the selection criterion to determine a signal's validity. For this study, the independent frequency method \citep{2018ApJS..236...16V} was applied to calculate the FAPs of respective spectral amplitudes across the frequency domain. After a provisional analysis, a threshold FAP of $10^{-8}$ was adopted across all data sets to provide a consistent, conservative criterion for identifying true signals amongst the noise, which is noted by \cite{2018AandA...616A..77Bowman} to be particularly significant at low frequencies. An example periodogram (logarithm of the period versus amplitude) is shown in Fig.\,\ref{K2_HD73045_C18} (bottom, blue-line) for the C18 data set for HD\,73045. The frequencies, periods, and amplitudes for peaks with FAP values of $10^{-8}$ and below, for all stars during their respective campaigns as a result of performing the Lomb-Scargle analysis are listed in Table\,\ref{table:KepLSresults}.

Next, the time-series data were analysed to determine the frequency, $f$, amplitude, $A$, and the phase at $t_{0} = 0$, $\phi(t_{0})$, from which a model of the light curve was constructed. Using the results of the Lomb-Scargle analysis as initial starting points, sinusoidal components incorporating these parameters were determined by sequentially fitting a series of components in the form of a sinusoid plus an offset, $Z$, to the data:

\begin{equation}
\Delta m_{i} = Z + A \sin(2\pi f t_{i} + \phi),
\label{eq:whitenfit}
\end{equation}

\noindent where $\phi$ is the phase derived from the fit and is described below.

Using the methods of \citet{1999DSSN...13...28M},  the corresponding errors, $\sigma (f)$, $\sigma (A)$, and $\sigma (\phi)$, were also derived using:

\begin{equation}
\sigma (f) = \sqrt{\frac{6}{\mathcal{N}}} \frac{1}{\pi \mathcal{T}} \frac{\sigma{(\Delta m)}}{A},
\label{eq:errfreq}
\end{equation}

\begin{equation}
\sigma (A) = \sqrt{\frac{2}{\mathcal{N}}} \sigma{(\Delta m)},
\label{eq:erramp}
\end{equation}

\begin{equation}
\sigma (\phi) = \sqrt{\frac{2}{\mathcal{N}}} \frac{\sigma{(\Delta m)}}{A}.
\label{eq:errphs}
\end{equation}

\noindent Here, $\mathcal{N}$ is the number of terms in a regularly sampled time series with an interval $\Delta t$, $\mathcal{T} = \mathcal{N}\Delta t$ is the time range of the data set, $A$ is the amplitude of the sinusoidal fit, and $\sigma{(\Delta m)}$ is the root mean square of the flux deviation values.

To determine these sinusoidal parameters and their uncertainties, the data were resampled at the regular 29.4-min cadence using cubic spline interpolation.  Next, the time was adjusted such that the zero point was placed at the center of the data set by subtracting the average of the resampled time array. Using the Lomb-Scargle results as a starting point, single sinusoids were sequentially fit using  Eqn.\,\ref{eq:whitenfit} and subtracted from the time-series, until all components (i.e., those with Lomb-Scargle FAP values below the threshold) were accounted for. The fitting process allowed the amplitude to vary within 20\,per\,cent about its LS value, while the frequency was allowed to vary within 0.01 d$^{-1}$ about its input value. From the resulting phase value, $\phi$, which is determined at this central point, the phase of the first point of the fit, $\phi(t_{0})$, was calculated.

For any component with a variation period $P$, the phase of each data point $\phi_{i}$, at time $t_{i}$ relative to the first point in the data series at time $t_{0}$, can be calculated using

\begin{equation}
\phi_{i} = 2 \pi \left(\frac{t_{i} - t_{0}}{P} - \left\lfloor \frac{t_{i} - t_{0}}{P} \right\rfloor \right),
\label{eq:phasefold}
\end{equation}

\noindent which simplifies when $t_{0} = 0$. From this, phase-folded plots of the time-series were produced corresponding to the dominant period of each star for each campaign. The C18 HD\,73045 data are folded to the dominant period of 12.64\,d in Fig.\,\ref{K2_HD73045_C18} (top-right) along with a sinusoidal fit to this component (red-line) with the parameters given in Table\,\ref{table:KepSineComps}.

\begin{table}
\centering
\caption{Flux variation signal properties (ordered by amplitude) derived from applying the Lomb-Scargle analysis to the all flux-deviation data sets. For HD\,73045, C18, refer to the blue-line peaks of Fig.\,\ref{K2_HD73045_C18} (bottom).}
 \smallskip
\label{table:KepLSresults}
\fontsize{7.1}{9.0}\selectfont
\begin{tabular}{lccccc}  % l = left, c = centered
\hline
\hline
\noalign{\smallskip}
Star Name & Campaign & Frequency & Period &          Amplitude      & log$_{10}$(FAP)  \\
                   &                  &     (d$^{-1}$)      &  (d) &           (mmag)        &    \\
\hline
\noalign{\smallskip}
 HD\,73045 & C05 & 0.078 & 12.82  & 0.112 & -302 \\
                   &         & 0.155 & 6.452  & 0.039 &  -33   \\  
                  &         & 0.054 & 18.52  & 0.0309 &  -18   \\  
                   &         & 0.138 & 7.246  & 0.029 &  -17   \\  
                   &         & 0.171 & 5.848  & 0.022 &  -8    \\  
\noalign{\smallskip}
                   & C18 & 0.079 & 12.66  & 0.039 &  -95   \\
                   &         & 0.155 & 6.452  & 0.022 &  -29   \\  
\hline    
\noalign{\smallskip}
 HD\,76310 & C05 & 0.217 & 4.608    & 0.498 & -361  \\
                   &         & 0.186 & 5.376    & 0.189 &  -49   \\  
                   &         & 0.421 & 2.375    & 0.113 &  -15   \\  
                   &         & 0.049 & 20.41    & 0.105 &  -13   \\  
                   &         & 0.159 & 6.289    & 0.095 &  -10   \\  
\noalign{\smallskip}
                   & C16 & 0.209 & 4.785    & 0.398 &  -331  \\
                   &         & 0.187 & 5.348    & 0.333 &  -231   \\  
                   &         & 0.076 & 13.16    & 0.104 &  -19    \\  
                   &         & 0.430 & 2.326    & 0.089 &  -13    \\  
                   &         & 0.969 & 1.031    & 0.081 &  -11    \\  
                   &         & 0.058 & 17.24    & 0.072 &  -8    \\  
\noalign{\smallskip}
                   & C18 & 0.206 & 4.854  & 1.094 &  -420  \\
                   &         & 0.177 & 5.650  & 0.275 &  -24  \\  
                   &         & 0.430 & 2.326  & 0.188 &  -9   \\  
\hline    
\noalign{\smallskip}
 HD\,73574 & C05 & 0.070 & 14.29    & 0.8139 &  -76    \\
                   &         & 0.091 & 10.99    & 0.3827 &  -14   \\  
\noalign{\smallskip}
                   & C18 & 0.091 & 10.99    & 0.0513 &    -9   \\
\hline  
\noalign{\smallskip}
 HD\,73618 & C05 & 0.264 & 3.788    & 0.415 & -274  \\
                   &         & 0.205  & 4.878    & 0.228 & -81 \\  
                   &         & 0.410 & 2.439    & 0.152 & -34 \\  
                   &         & 0.244 & 4.098    & 0.104 &   -14 \\  
                   &         & 0.431 & 2.320    & 0.102 &   -13 \\  
                   &         & 0.184 & 5.435    & 0.086 &   -9 \\  
\noalign{\smallskip}
                   & C16 & 0.261 & 3.831    & 0.320 &  -254 \\
                   &         & 0.205 & 4.878    & 0.287 &  -203 \\  
                   &         & 0.401 & 2.494    & 0.137 &  -44  \\  
                   &         & 0.225 & 4.444    & 0.135 &  -43  \\  
                   &         & 0.418 & 2.392    & 0.104 &  -24  \\  
                   &         & 0.242 & 4.132    & 0.094 &  -19  \\  
                   &         & 0.290 & 3.448    & 0.074 &  -10  \\  
                   &         & 0.448 & 2.232    & 0.069 &  -9   \\  
\noalign{\smallskip}
                   &   C18   & 0.266 & 3.759    & 0.595 &  -256  \\
                   &         & 0.213 & 4.695    & 0.249 &  -42   \\  
                   &         & 0.196 & 5.102    & 0.194 &  -24   \\  
                   &         & 0.417 & 2.398    & 0.190 &  -23   \\  
                   &         & 0.238 & 4.202    & 0.125 &  -8    \\  
                   &         & 0.295 & 3.390    & 0.123 &  -8    \\  
\hline 
\noalign{\smallskip}
 HD\,73619 & C05 & 0.077 & 12.987    & 0.388 &  -289   \\
                   &         & 0.155  & 6.452    & 0.210 & -82 \\  
                   &         & 0.233 & 4.292    & 0.160 & -46 \\  
                   &         & 0.058 & 17.241    & 0.108 & -19 \\  
                   &         & 0.042 & 20.810    & 0.098 & -16 \\  
                   &         & 0.310 & 3.226    & 0.095 &  -14 \\  
                   &         & 0.096 & 10.417    & 0.082 & -10 \\  
\noalign{\smallskip}
                   & C18 & 0.077 & 12.987    & 0.324 &  -217   \\
                   &         & 0.156  & 6.410    & 0.163 & -52 \\  
                   &         & 0.235 & 4.256    & 0.103 & -19 \\  
                   &         & 0.112 & 8.929    & 0.075 &   -9 \\  
\noalign{\smallskip}
\hline
%\tableline
\end{tabular}
\end{table}

\section{Supplementary Materials}
\label{appendixC}

\begin{table*}
\begin{center}
\caption{Ground-based CCD photometric observations of HD\,73045 obtained from various optical telescopes located at different sites.}
\smallskip
\label{table:groundlog}
% {\small
\fontsize{7.2}{9.0}\selectfont
\begin{tabular}{lcccccccr}  % l = left, c = centered
\hline
\noalign{\smallskip}
S. No. & Observatory & Telescope  & Filters & Exposure Time   & Total Durations& Date  & HJD  \\
&  & Diameter (m)  &  & (s)   & (h)&dd-mm-yyyy & (d)  \\
\noalign{\smallskip}
\hline
% 1. & ARIES & 104 & B &  3  & 2.68& 07-01-2013 & 2456300  \\
% 2. &       &   & B & 10  & 8.78 &  08-01-2013 & 2456301  \\
1. & ARIES & 1.04   & V &  2  & 1.47 & 16-04-2014 & 2456764 \\
% 4. &       &  & V &  3  & 0.65 & 18-04-2014 & 2456766   \\
2. &       &  & V &  3  & 0.92 &  20-04-2014 & 2456768 \\
3. &       &  & B &  5  & 2.53 & 21-04-2014 & 2456769\\

4.& Devasthal & 1.30  & B          &  5 & 1.80 &  02-11-2014 & 2456964  \\ %02Nov14
5.&       &  & V          &  4 & 1.47 &  11-11-2014 & 2456973  \\ %11Nov14
6.&       &  & B          &  4 & 0.17 & 30-11-2014 & 2456992   \\ %30nov14
% 10.&       &  & V          &  3 & 0.06 & 30-11-2014 & 2456992   \\ %30nov14
% 11.&       &  & V          &  3 & 5.54 & 12-01-2015 &  2457034 \\ %12jan15
% 12.&       &  & V          &  3 & 3.39 & 13-01-2015 &  2457035  \\%13jan15
% 13.&       &  & V          &  7 & 1.68 &  29-01-2015 &  2457051  \\%29jan15
% 14.&       &  & V          & 15 & 6.17 & 29-12-2015 &  2457386  \\ %29dec15
7.&       &  & V          & 15 & 0.85 & 30-12-2015 &  2457387  \\ %30dec15
8.&       &  & V          & 10 & 0.57 & 27-01-2016 & 2457415  \\ %27jan16
9.&       &  & B          & 15 & 2.12 & 28-01-2016 & 2457416  \\ %28jan16
10.&       &  & V          &  3 & 2.02 & 19-12-2016 & 2457742  \\ %19dec16
% 19.&       &  & V          & 10 & 0.95 & 22-04-2017 & 2457866  \\ %22Apr17
% 20.&       &  & V          &  3 & 0.77 & 20-01-2018 & 2458139  \\ 
\hline
% 21. &  	& 40 &  V & 5  & 0.95 & 07-11-2015 & 2457334  \\
11. & MASTER-II   	& 0.40 & B, V& 5  & 2.92, 2.92 &  09-11-2015 & 2457336  \\
12. & -URAL	&  & B & 5 & 1.35 &  22-01-2016 & 2457410  \\
13. &  			&  & B, V & 5 & 0.13,0.13 & 13-02-2016 & 2457432*  \\
% 25. &  			&  &    V & 5  & 6.95 & 14-02-2016  & 2457433  \\
% 26. &  			&  &  V & 5 & 5.59 & 15-02-2016 & 2457434  \\
% 27. &  			&  & B & 5 & 0.92,0.92 & 16-02-2016 & 2457435  \\
% 28. &  			&  &  V   & 5 & 5.74     & 19-02-2016 & 2457438  \\
14. &  			&  & B & 5 & 4.00 & 11-03-2016 & 2457459  \\
15. &  			&  &  V & 5 & 3.97 & 14-03-2016 & 2457463  \\
16. &  			&  & B, V & 5 & 3.96, 3.96 & 15-03-2016 & 2457478  \\
% 32. &  			&  & V & 5 & 2.87  & 30-03-2016 & 2457478 \\
\hline
17. & PROMPT-8 & 0.60  & B & 4 & 1.26 & 12-05-2014 & 2456790 \\
18. &      &  & B & 4 & 0.27 & 14-05-2014 & 2456792 \\ 
19. &      &  & B & 4 & 1.65 & 16-05-2014 & 2456794 \\ 
% 36. &      &  & B & 4 & 1.16 & 17-05-2014 & 2456795 \\ 
% 37. &      &  & B & 4 & 0.37 & 18-05-2014 & 2456796 \\ 
20. &      &  & B & 15 & 1.10 & 11-01-2015 & 2457033 \\ 
21. &      &  & B & 20 & 1.21 & 09-02-2015 & 2457062 \\ 
\hline
\end{tabular}
% }
\end{center}
\end{table*}

\begin{figure}
\centering
 \includegraphics[width=0.93\columnwidth]{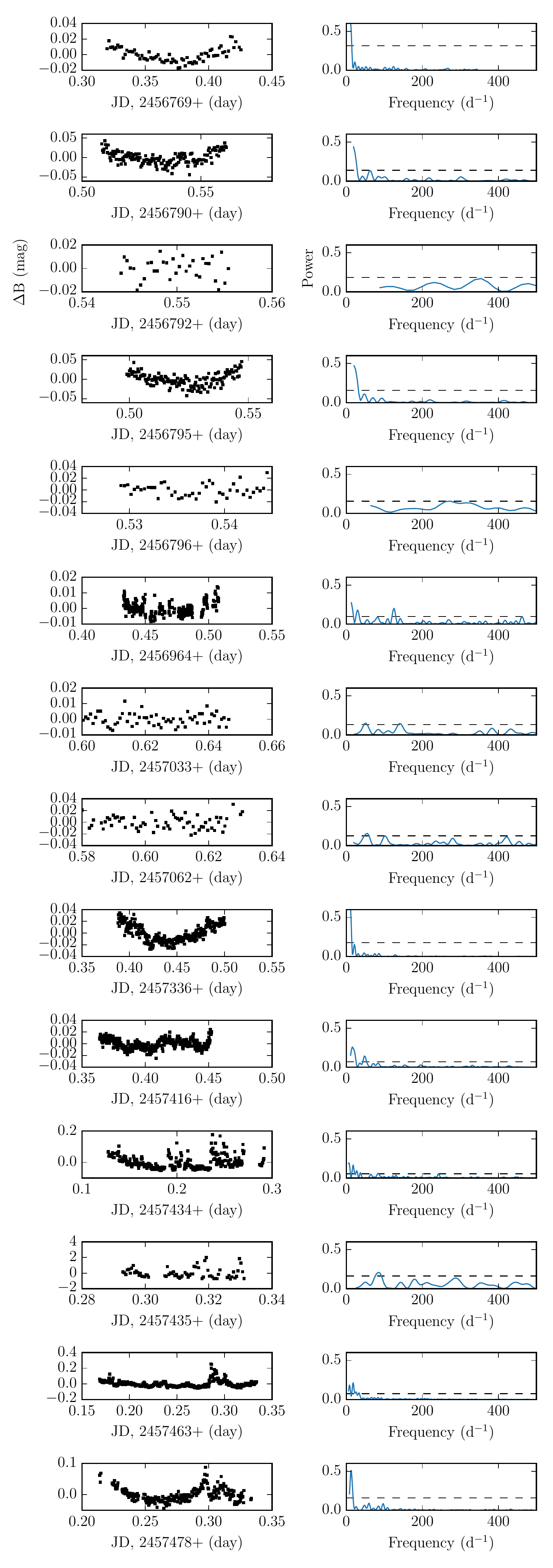}
\caption{Light curves for HD\,73045 from ground-based observations with B-filter.}
\label{lc:B}
\end{figure}

\begin{figure}
\centering
 \includegraphics[width=\columnwidth]{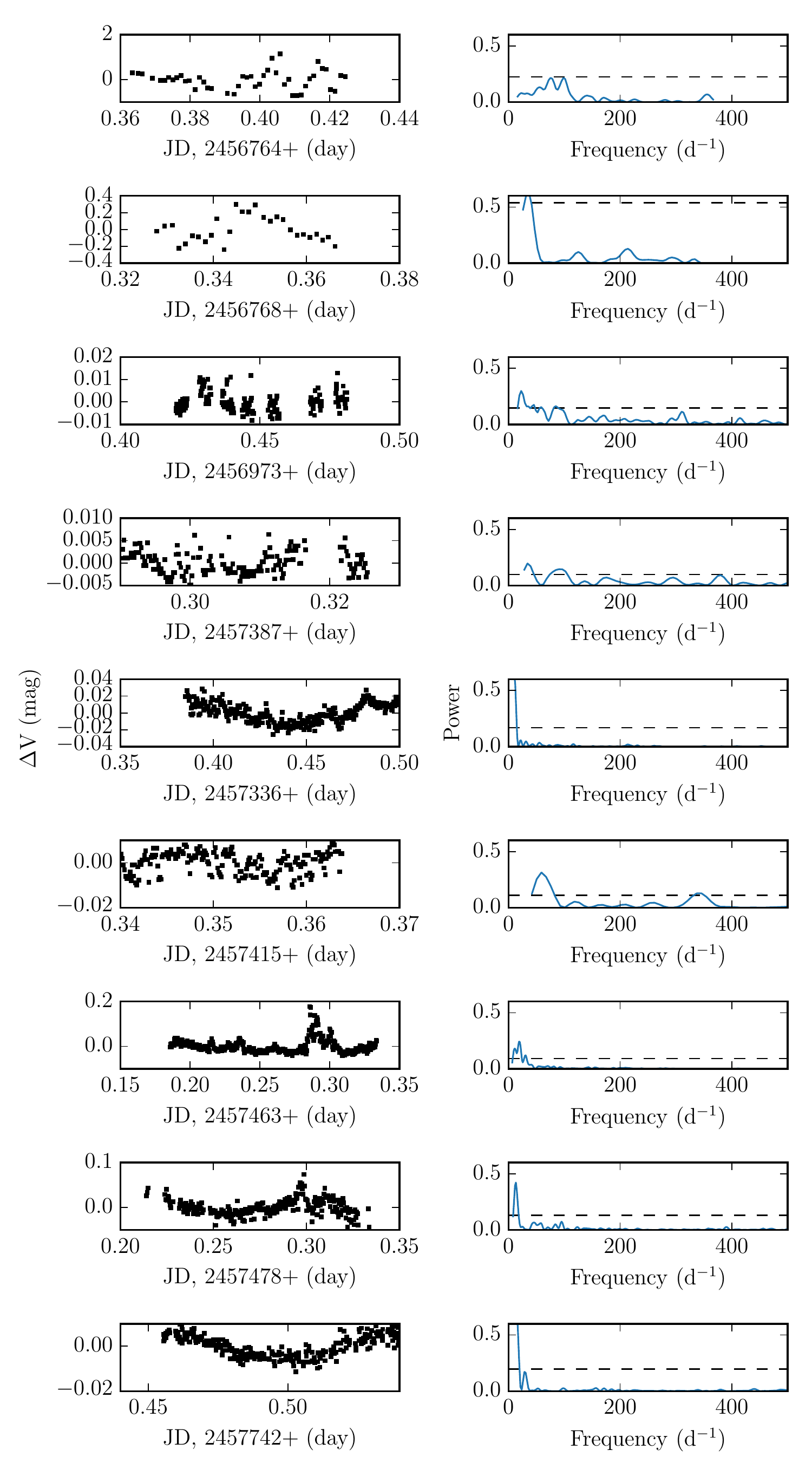}
\caption{Light curves for HD\,73045 from ground-observations with V-filter.}
\label{lc:V}
\end{figure}

\begin{figure}
\centering
 \includegraphics[width=\columnwidth]{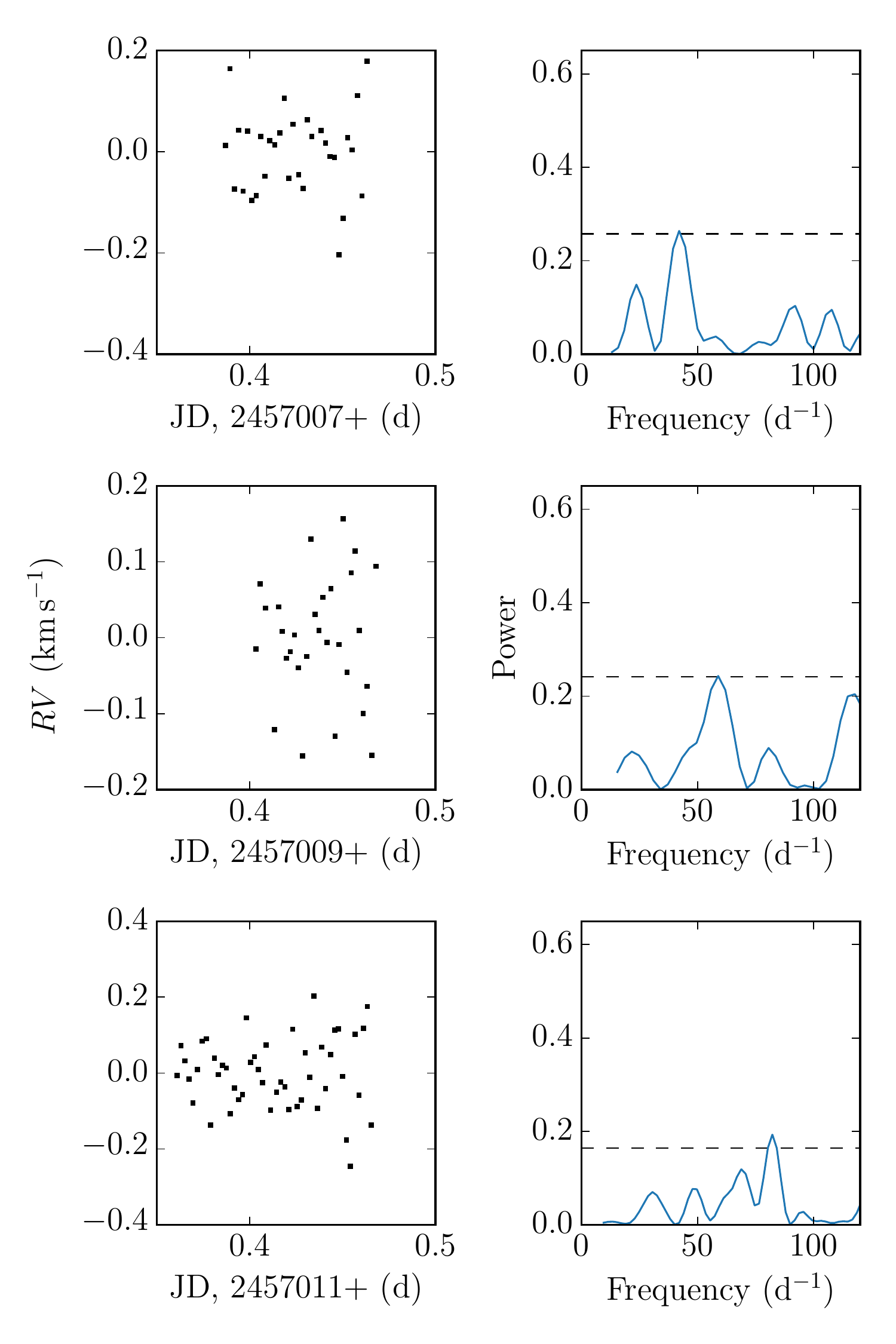}
\caption{RV curve ({\it panels on the left}) and power spectrum ({\it panels on the right}) of the radial velocity measurements for HD\,73045 from MRES data. The black horizontal dashed line in the power spectrum represents SNR of 4. }
\label{rv:fig_mres}
\end{figure} 

\begin{figure}
\centering
 \includegraphics[width=\columnwidth]{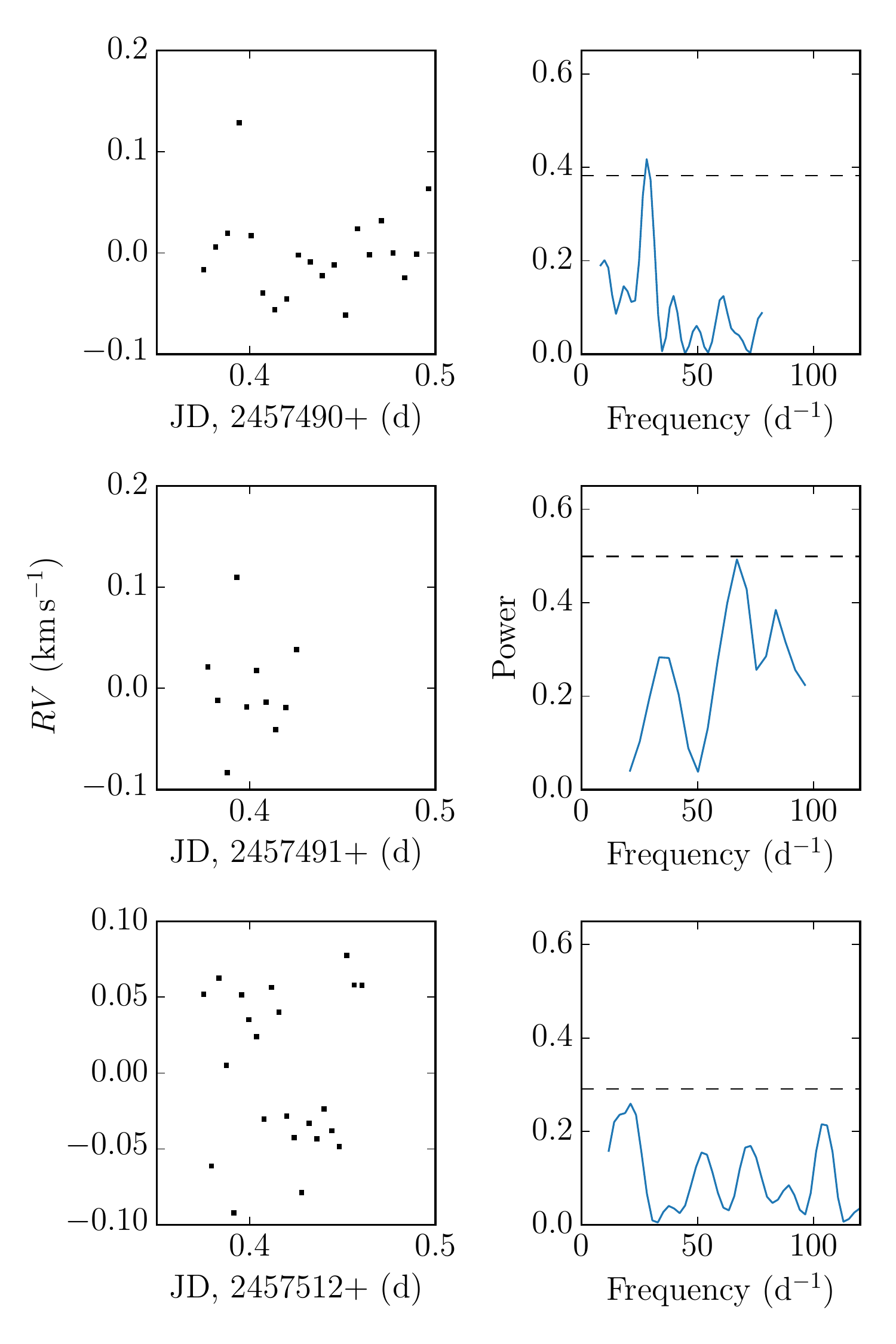}
\caption{
RV curve ({\it panels on the left}) and power spectrum ({\it panels on the right}) of the radial velocity measurements for HD\,73045 from HERMES data. The black horizontal dashed line in the power spectrum represents SNR of 4. }
\label{rv:fig_hermes}
\end{figure}

\begin{figure}
	\begin{center}
		\includegraphics[width=\columnwidth]{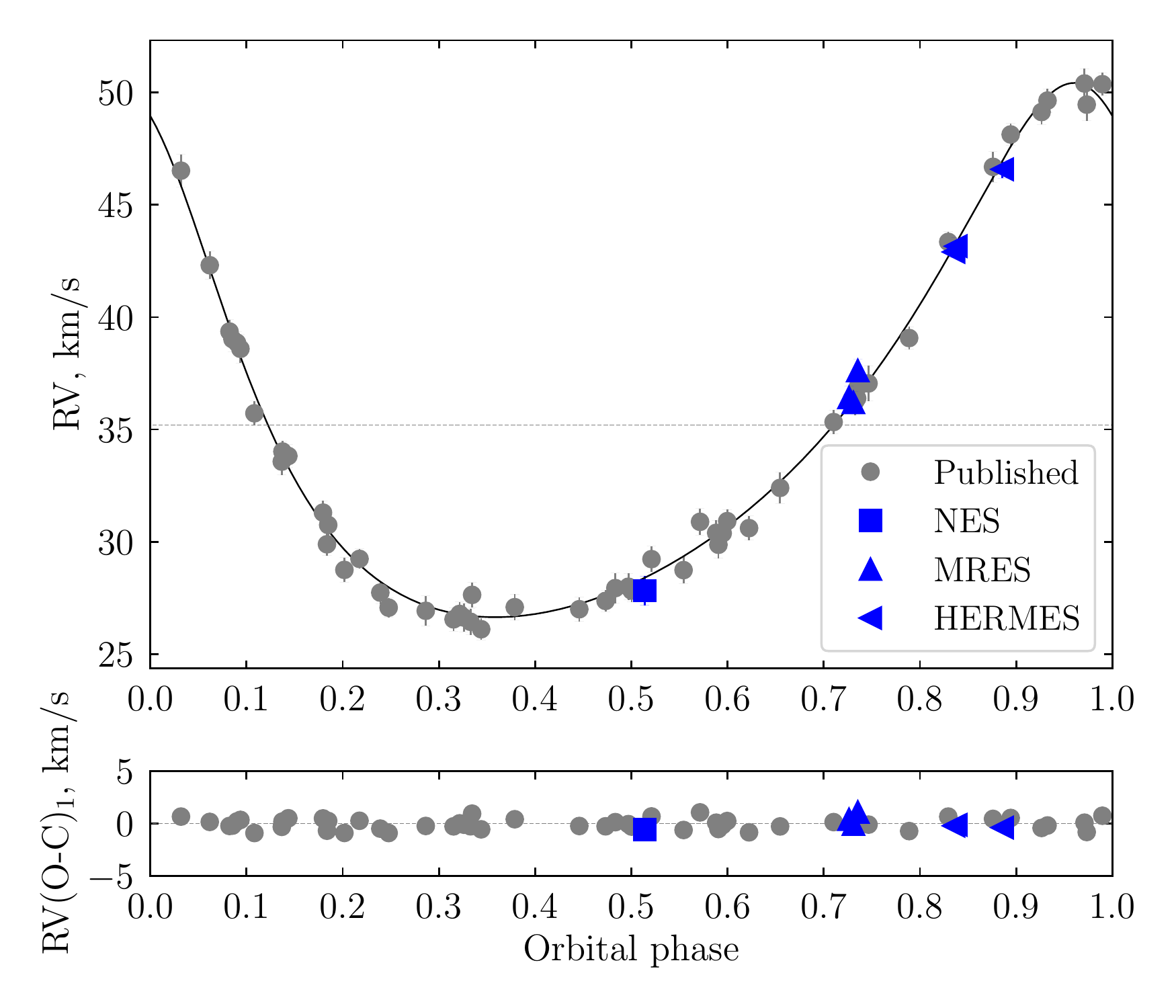}%&
		\caption{RV of HD\,73045, as measured by us (blue), compared to the orbital solution by \citet{2007MNRAS.380.1064C} (grey).}
		\label{hd73045orb}
	\end{center}
\end{figure}

\begin{figure*}
\centering
\includegraphics[width=0.95\textwidth]{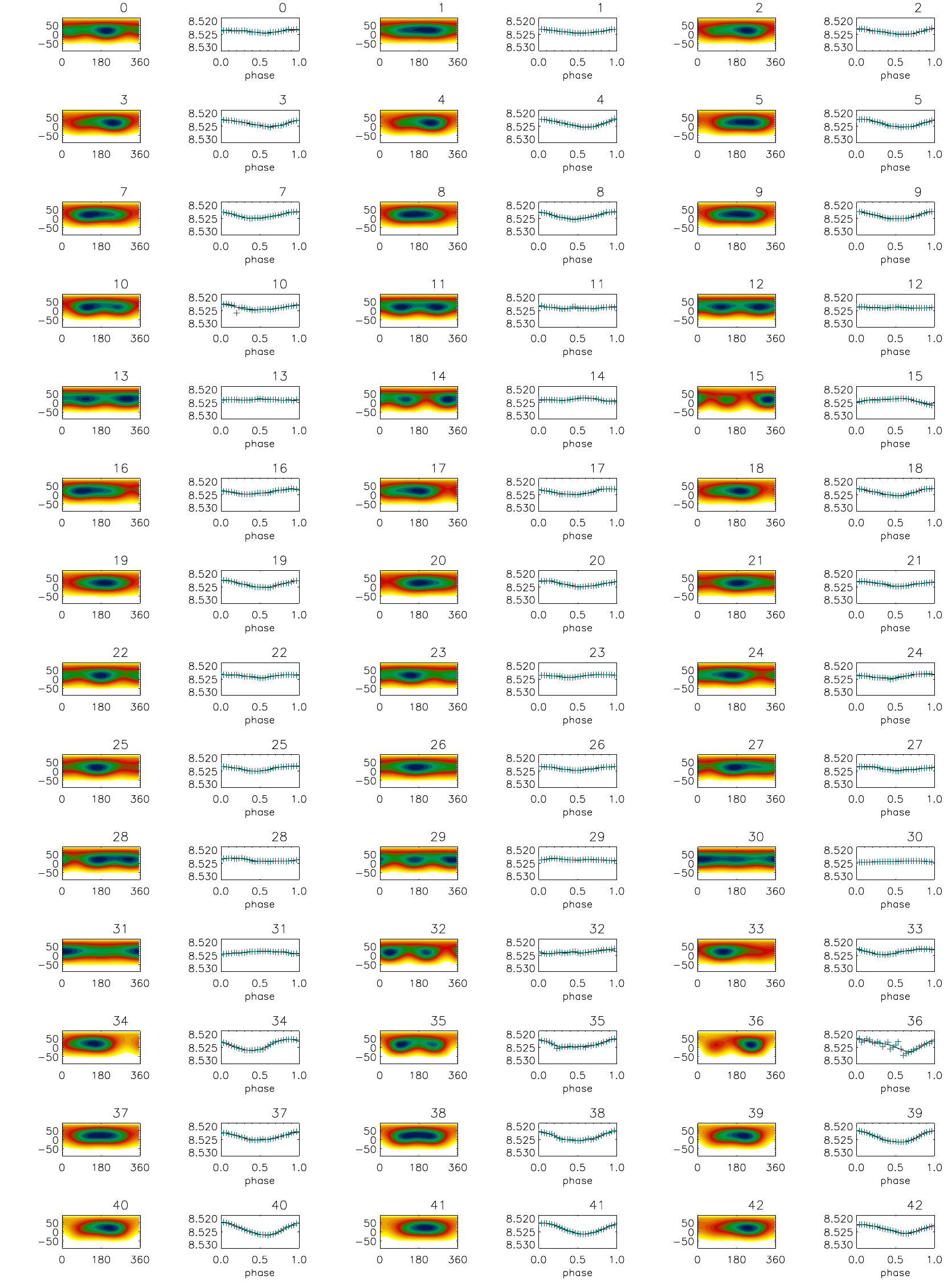}
\caption{As Fig.\,\ref{spot73045}, but spot maps for HD\,76310.}
\label{spot76310}
\end{figure*}

%\twocolumn

%\twocolumn
\begin{figure}
\begin{center}
%\begin{flushright}
\includegraphics[width=0.7\textwidth]{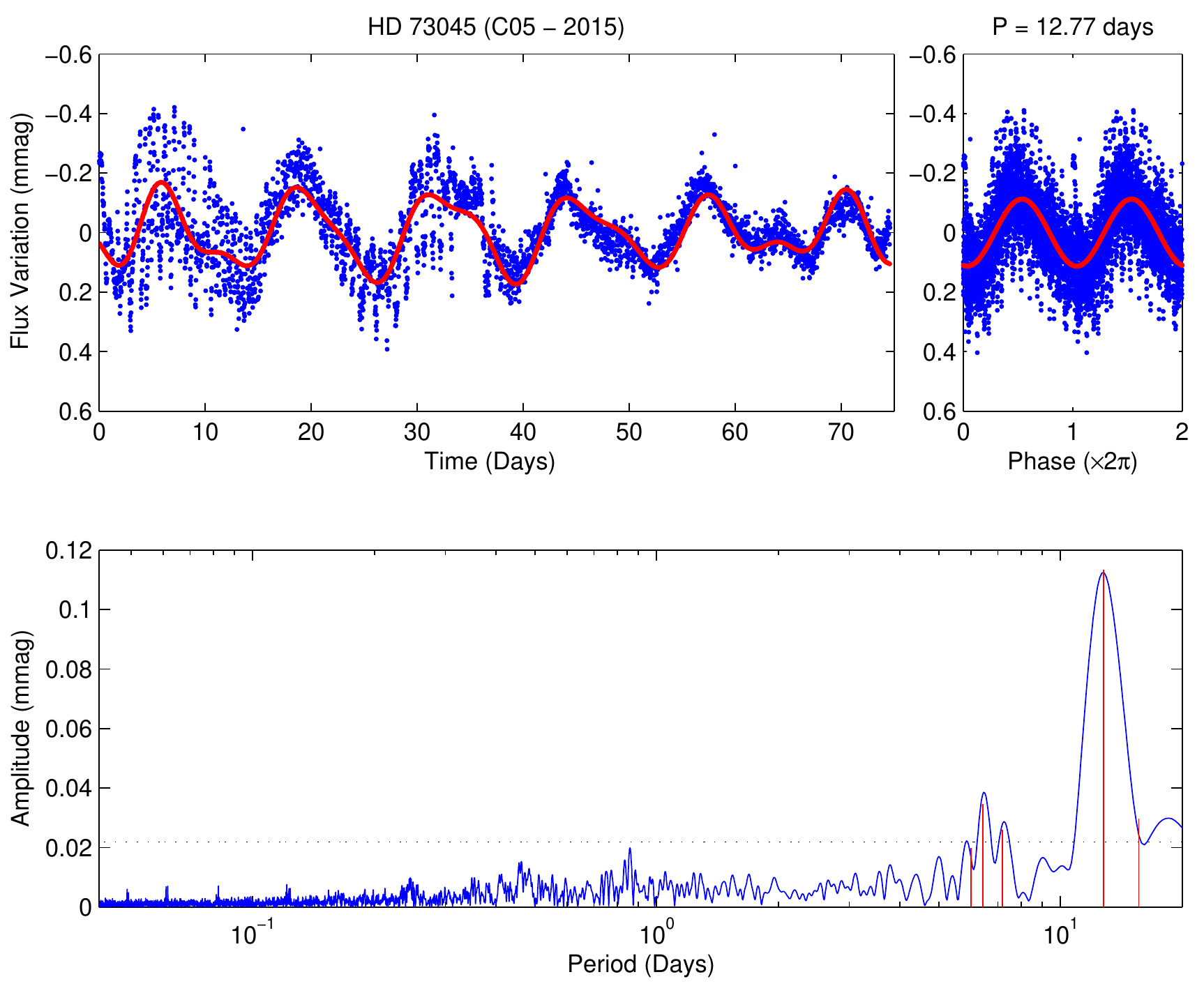}
%\vspace{-9.0cm}
%\end{flushright}
%\begin{center}
\caption{As Fig.\,\ref{K2_HD73045_C18}, but for the LC {\it K2} Campaign 5 observations of HD\,73045.}
\label{K2_HD73045_C05}
\end{center}
\end{figure} 

\begin{figure*}
\begin{center}
%\begin{flushright}
\includegraphics[width=0.7\textwidth]{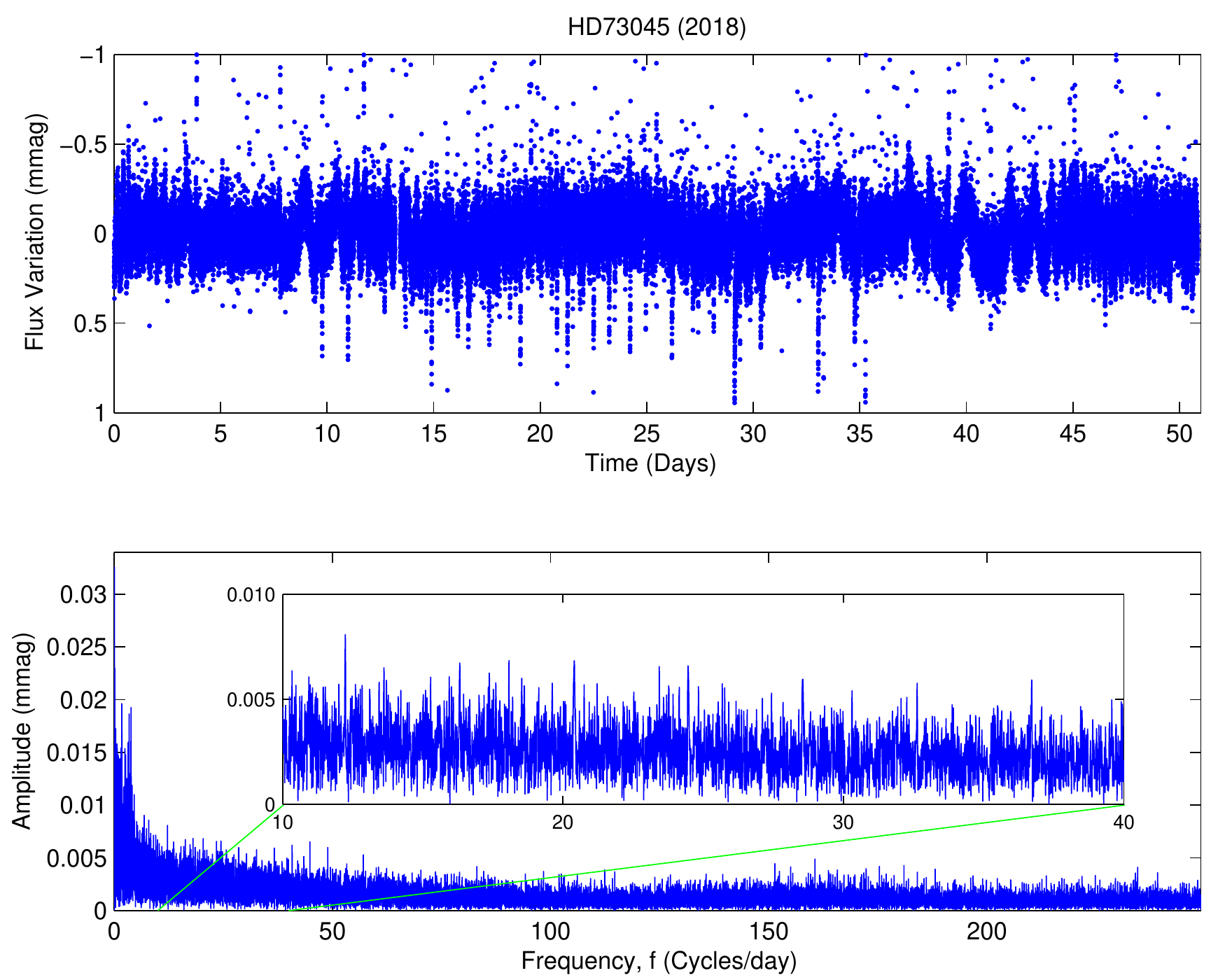}
%\vspace{-9.0cm}
%\end{flushright}
%\begin{center}
\caption{Top: Time series photometric variability (in mmag) of HD\,73045 from the SC 2018 {\it K2} Campaign 18. The PDC\_SAP data includes intrinsic signals as well as systemic and noise artifacts. Bottom: Frequency distribution of spectral amplitudes, in mmag, derived from applying the Lomb-Scargle algorithm to the time series. The frequency range from 10\,--\,40\,d$^{-1}$ is expanded into the inset to provide more detail of that range. It can be seen there are no signals related to those from the earlier ground-based observations and that this range is dominated by noise.}
\label{K2_HD73045_C18_SC}
\end{center}
\end{figure*}

\begin{figure*}
\begin{center}
%\begin{flushright}
\includegraphics[width=0.7\textwidth]{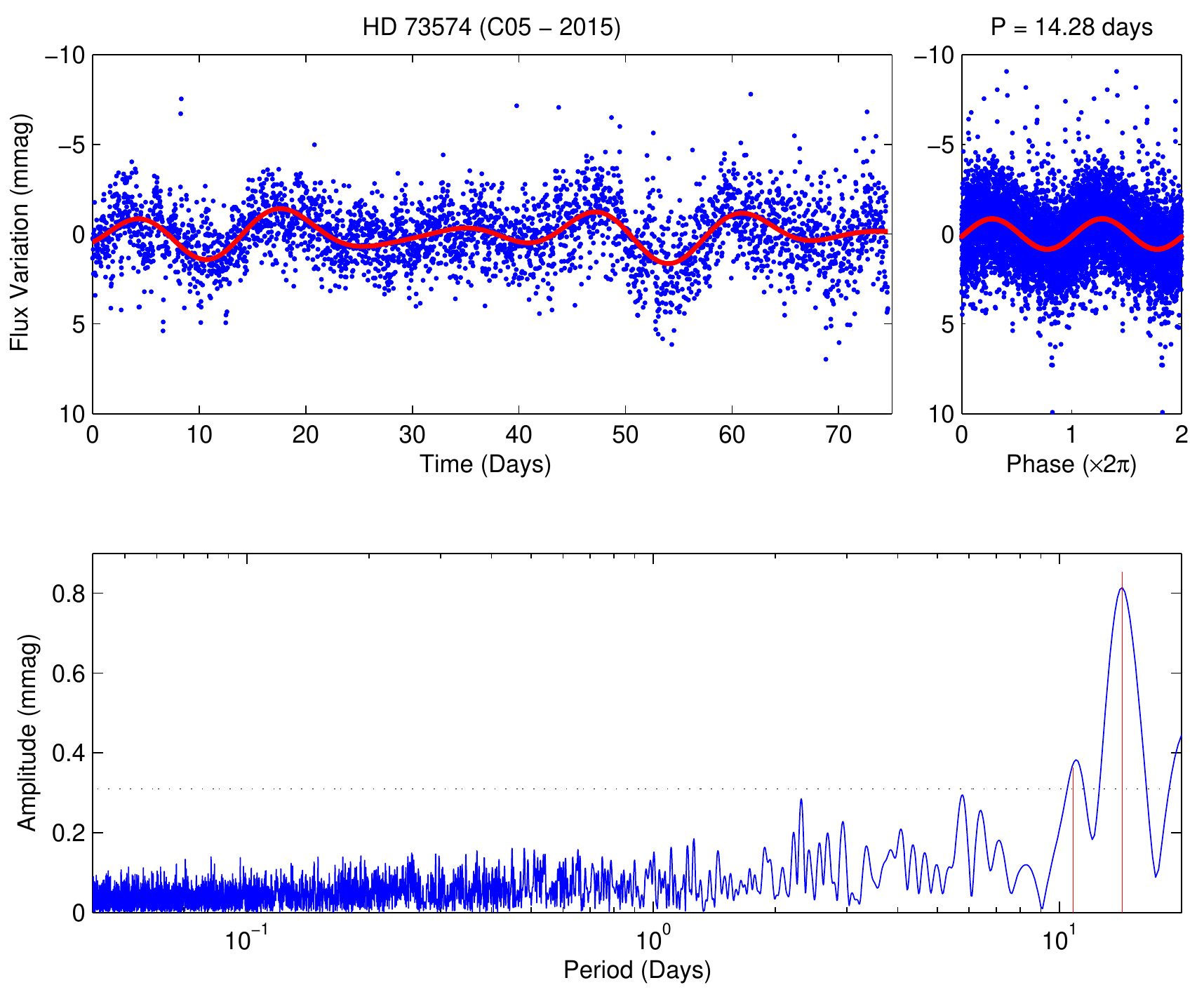}
%\vspace{-9.0cm}
%\end{flushright}
%\begin{center}
\caption{As Fig.\,\ref{K2_HD73045_C18}, but for the LC {\it K2} Campaign 5 observations of HD\,73754.}
\label{K2_HD73574_C05}
\end{center}
\end{figure*}

\begin{figure*}
\begin{center}
%\begin{flushright}
\includegraphics[width=0.7\textwidth]{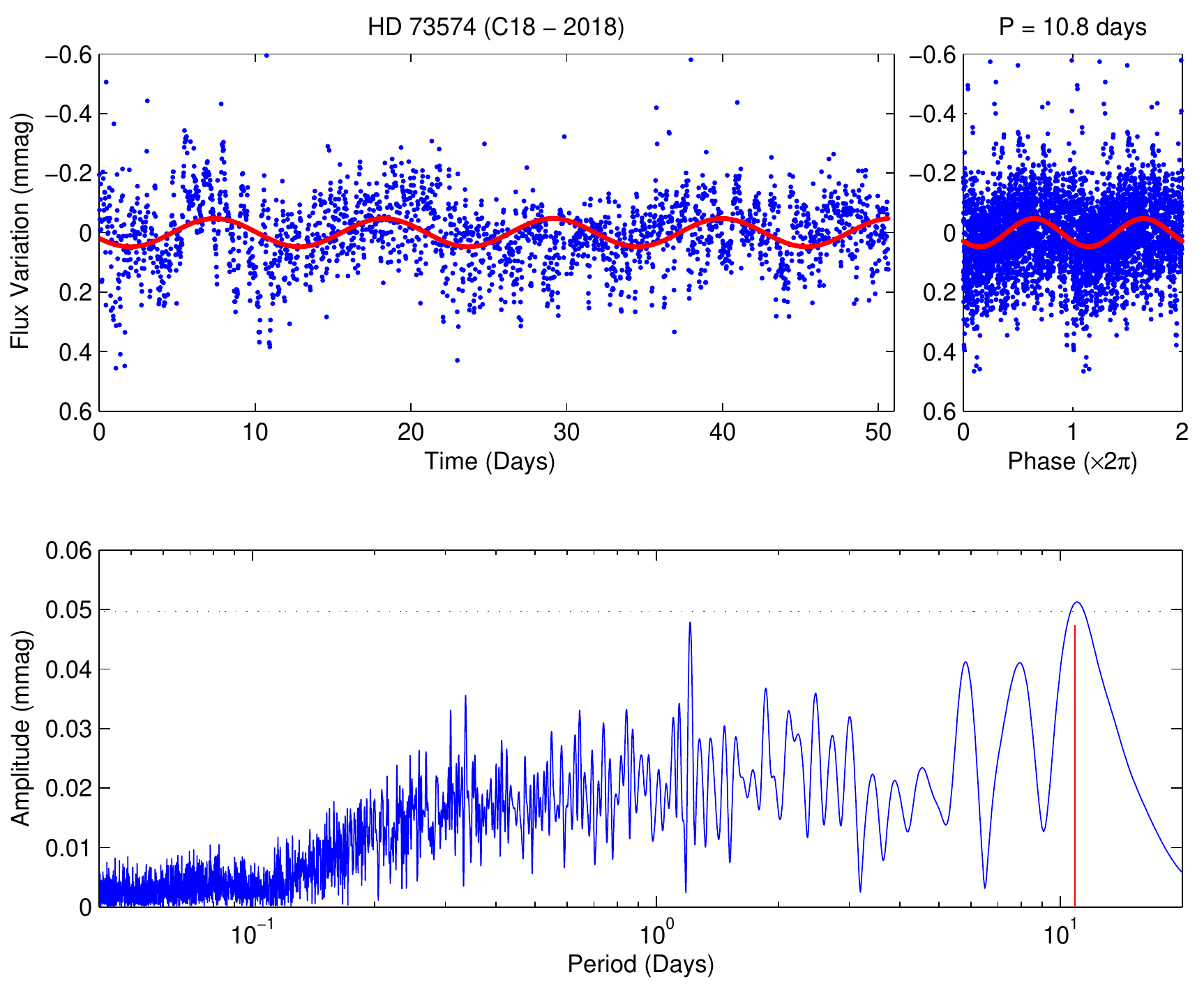}
%\vspace{-9.0cm}
%\end{flushright}
%\begin{center}
\caption{As Fig.\,\ref{K2_HD73045_C18}, but for the LC {\it K2} Campaign 18 observations of HD\,73754.}
\label{K2_HD73574_C18}
\end{center}
\end{figure*}

%\clearpage

\begin{figure*}
\begin{center}
%\begin{flushright}
\includegraphics[width=0.7\textwidth]{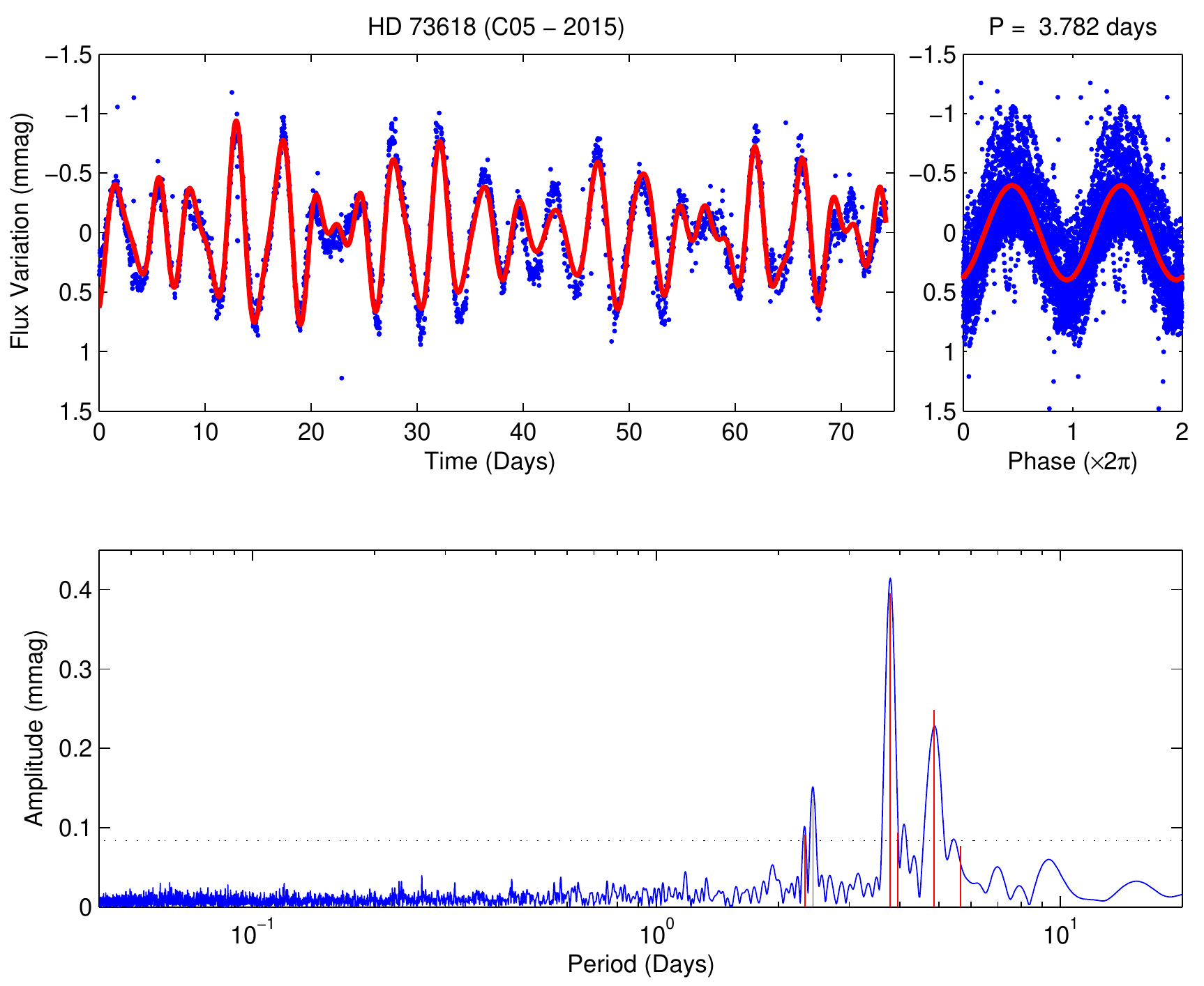}
%\vspace{-9.0cm}
%\end{flushright}
%\begin{center}
\caption{As Fig.\,\ref{K2_HD73045_C18}, but for the LC {\it K2} Campaign 5 observations of HD\,73618.}
\label{K2_HD73618_C05}
\end{center}
\end{figure*}

\begin{figure*}
\begin{center}
%\begin{flushright}
\includegraphics[width=0.7\textwidth]{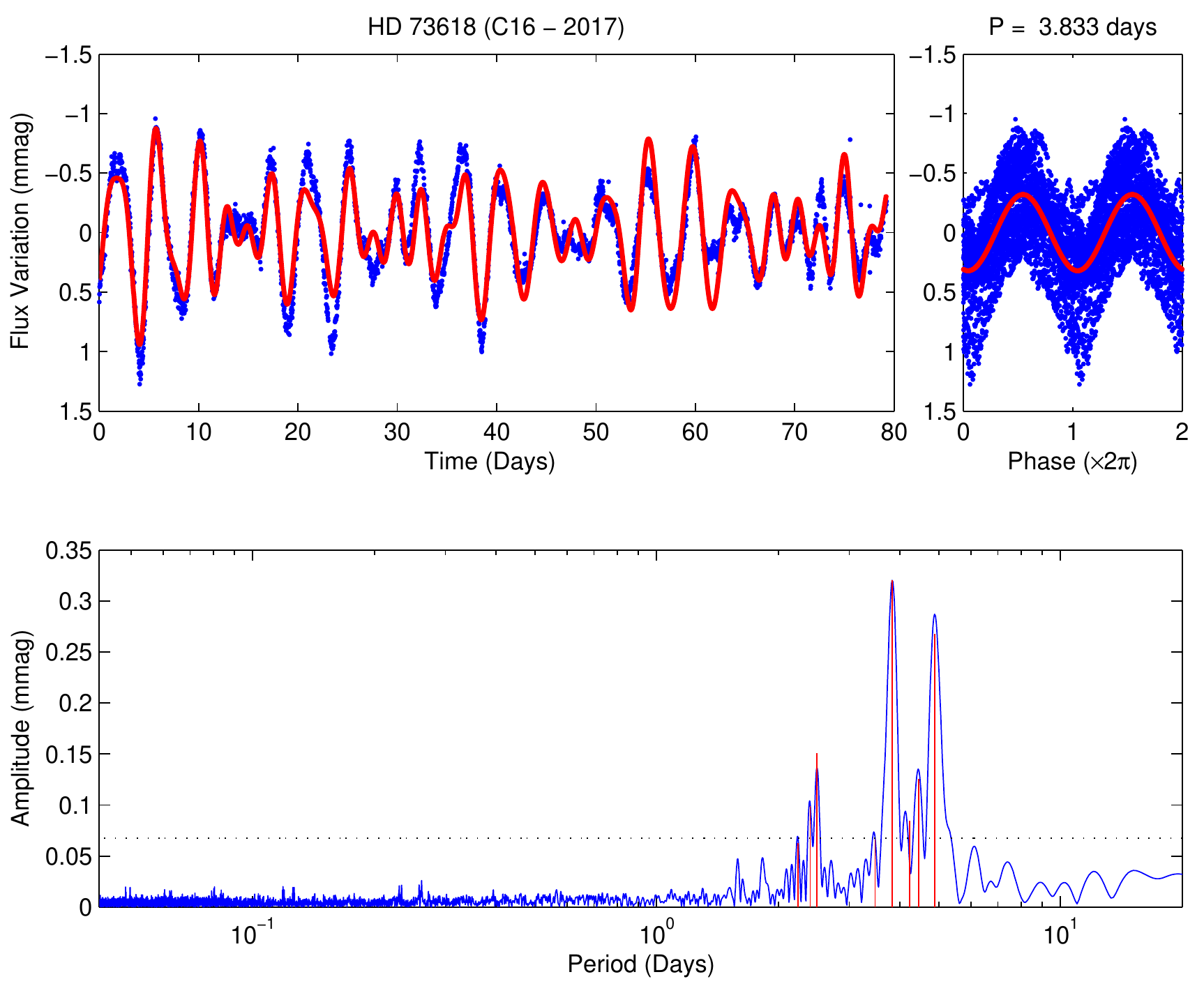}
%\vspace{-9.0cm}
%\end{flushright}
%\begin{center}
\caption{As Fig.\,\ref{K2_HD73045_C18}, but for the LC {\it K2} Campaign 16 observations of HD\,73618.}
\label{K2_HD73618_C16}
\end{center}
\end{figure*}

\begin{figure}
\begin{center}
%\begin{flushright}
\includegraphics[width=0.7\textwidth]{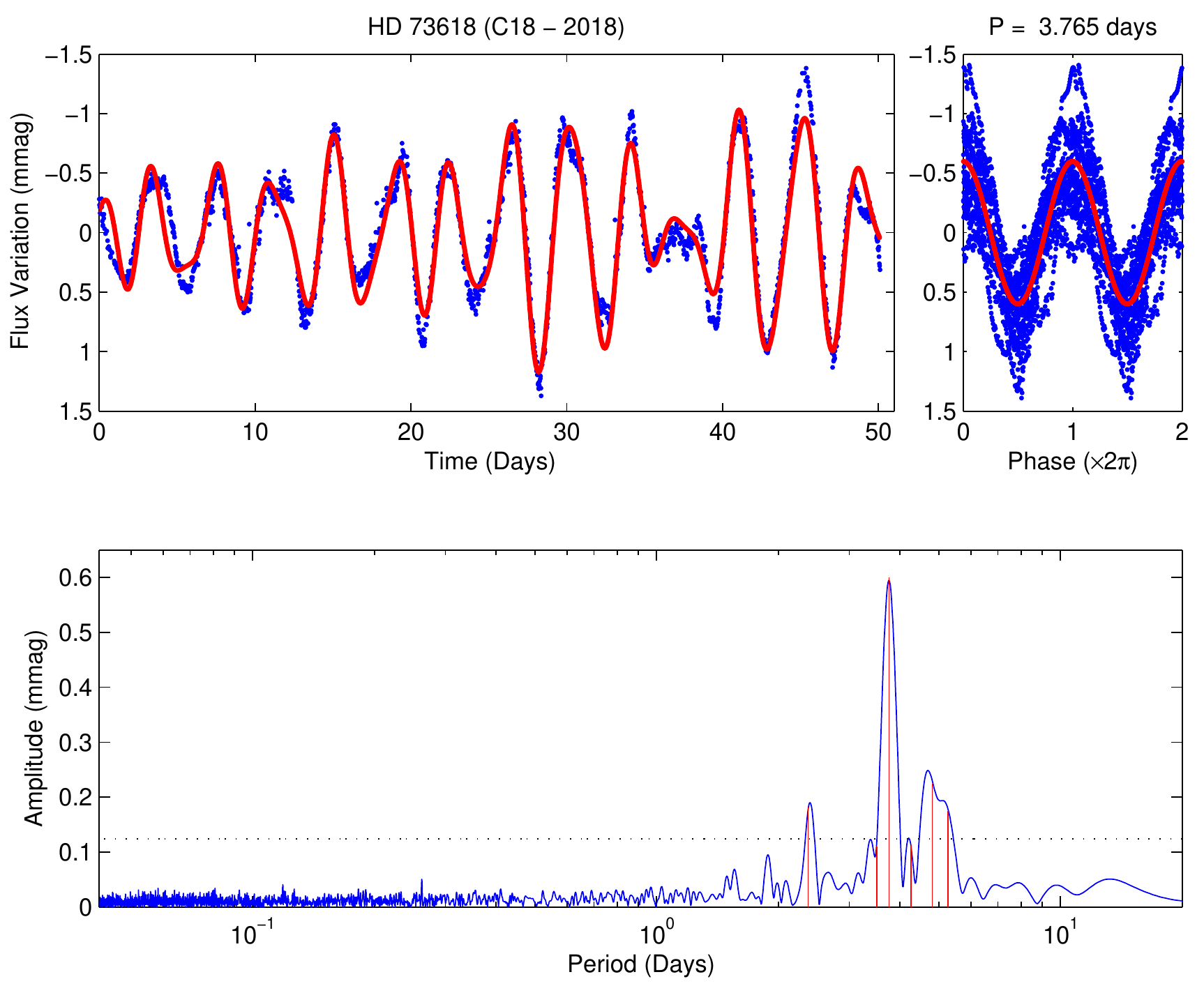}
%\vspace{-9.0cm}
%\end{flushright}
%\begin{center}
\caption{Time series (top) and respective Lomb-Scargle periodogram (bottom) for the LC {\it K2} Campaign 18 observations of HD\,73618. %No discernable variational signals are observed.
}
\label{K2_HD73618_C18}
\end{center}
\end{figure}
% \clearpage

\begin{figure*}
\begin{center}
%\begin{flushright}
\includegraphics[width=0.7\textwidth]{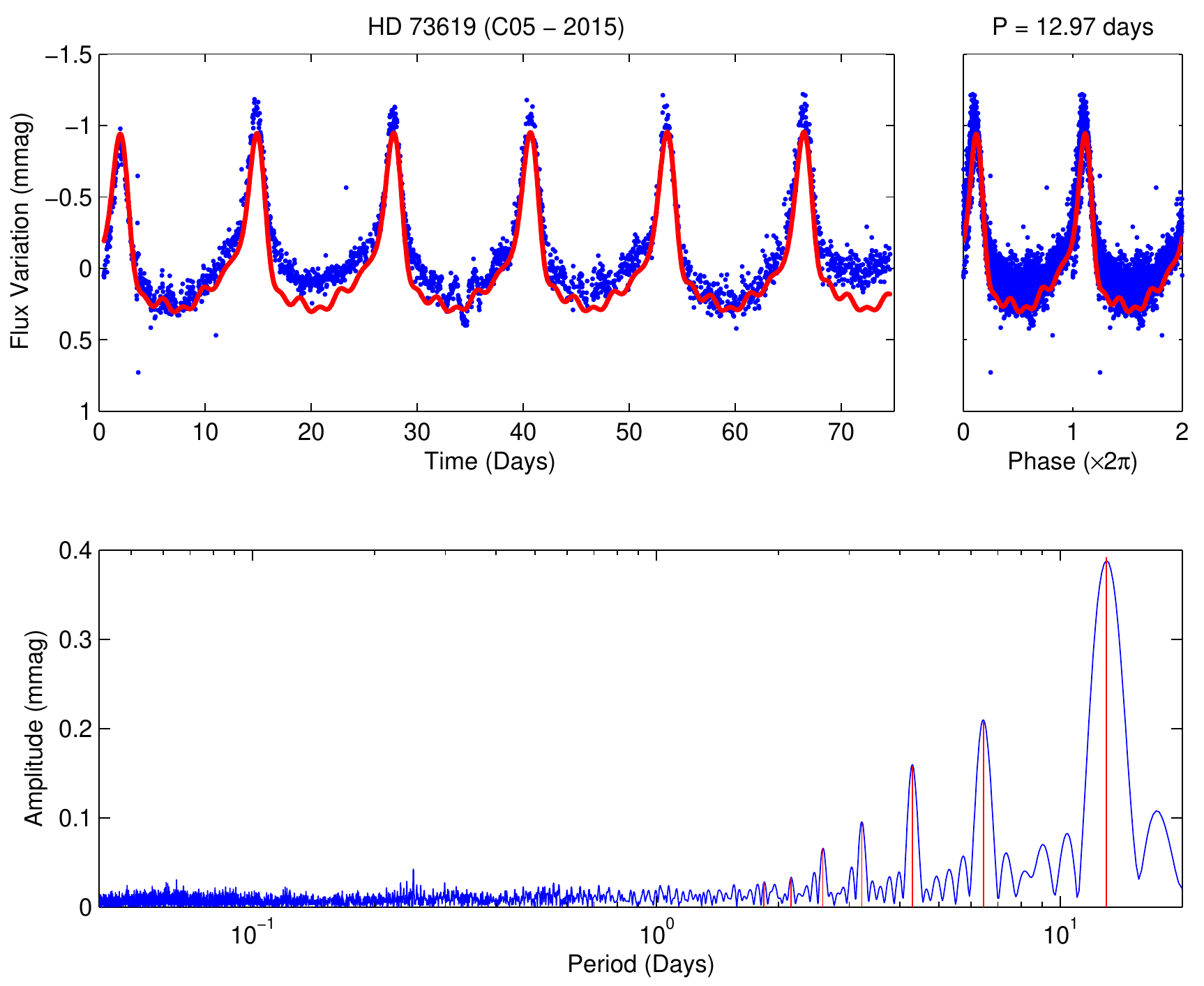}
%\vspace{-9.0cm}
%\end{flushright}
%\begin{center}
\caption{As Fig.\,\ref{K2_HD73045_C18}, but for the LC {\it K2} Campaign 5 observations of HD\,73619. However, in contrast to the remainder of this paper, the phase diagram is traced by two periods of the light curve model, and not by the dominant component. Additionally, the FAP threshold criterion has been removed, with the model being constructed from the fundamental component and its overtones.}
\label{K2_HD73619_C05}
\end{center}
\end{figure*}

\begin{figure*}
\begin{center}
%\begin{flushright}
\includegraphics[width=0.7\textwidth]{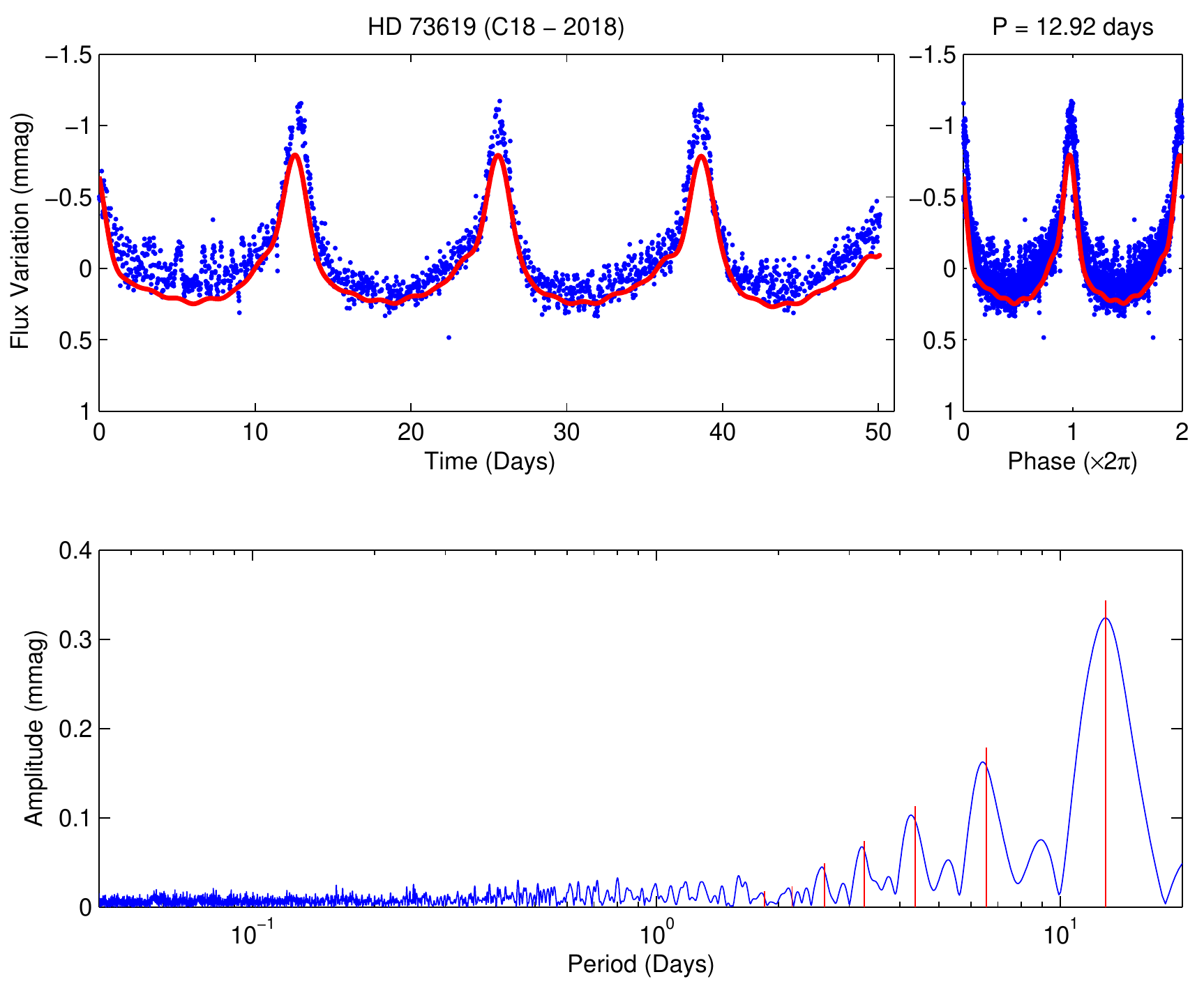}
%\vspace{-9.0cm}
%\end{flushright}
%\begin{center}
\caption{Time series (top) and respective Lomb-Scargle periodogram (bottom) for the LC {\it K2} Campaign 18 observations of HD\,73619. %No discernable variational signals are observed.
}
\label{K2_HD73619_C18}
\end{center}
\end{figure*}

\begin{figure*}
\begin{center}
%\begin{flushright}
\includegraphics[width=0.7\textwidth]{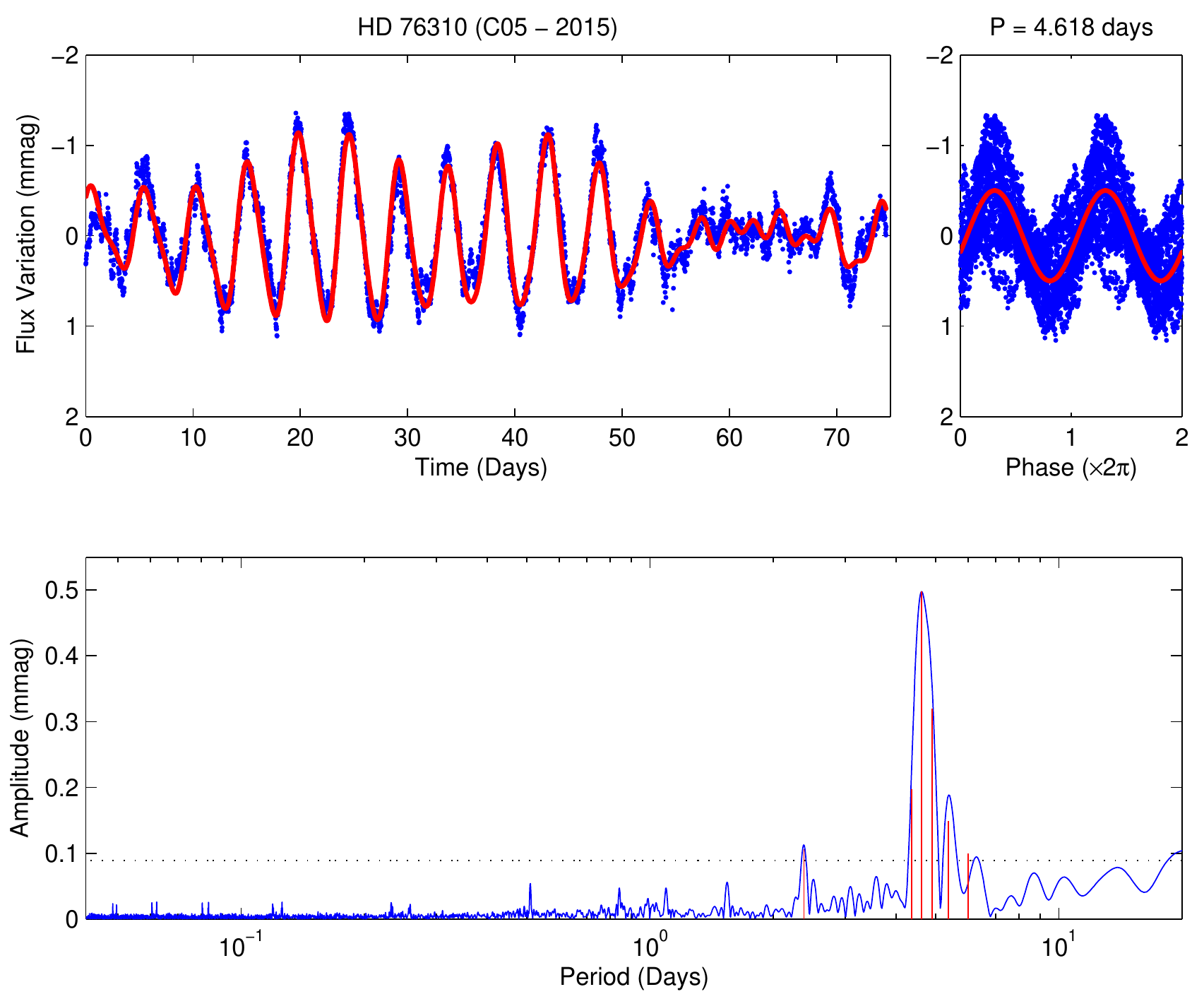}
%\vspace{-9.0cm}
%\end{flushright}
%\begin{center}
\caption{As Fig.\,\ref{K2_HD73045_C18}, but for the LC {\it K2} Campaign 5 observations of HD\,76310.}
\label{K2_HD76310_C05}
\end{center}
\end{figure*}

\begin{figure*}
\begin{center}
%\begin{flushright}
\includegraphics[width=0.7\textwidth]{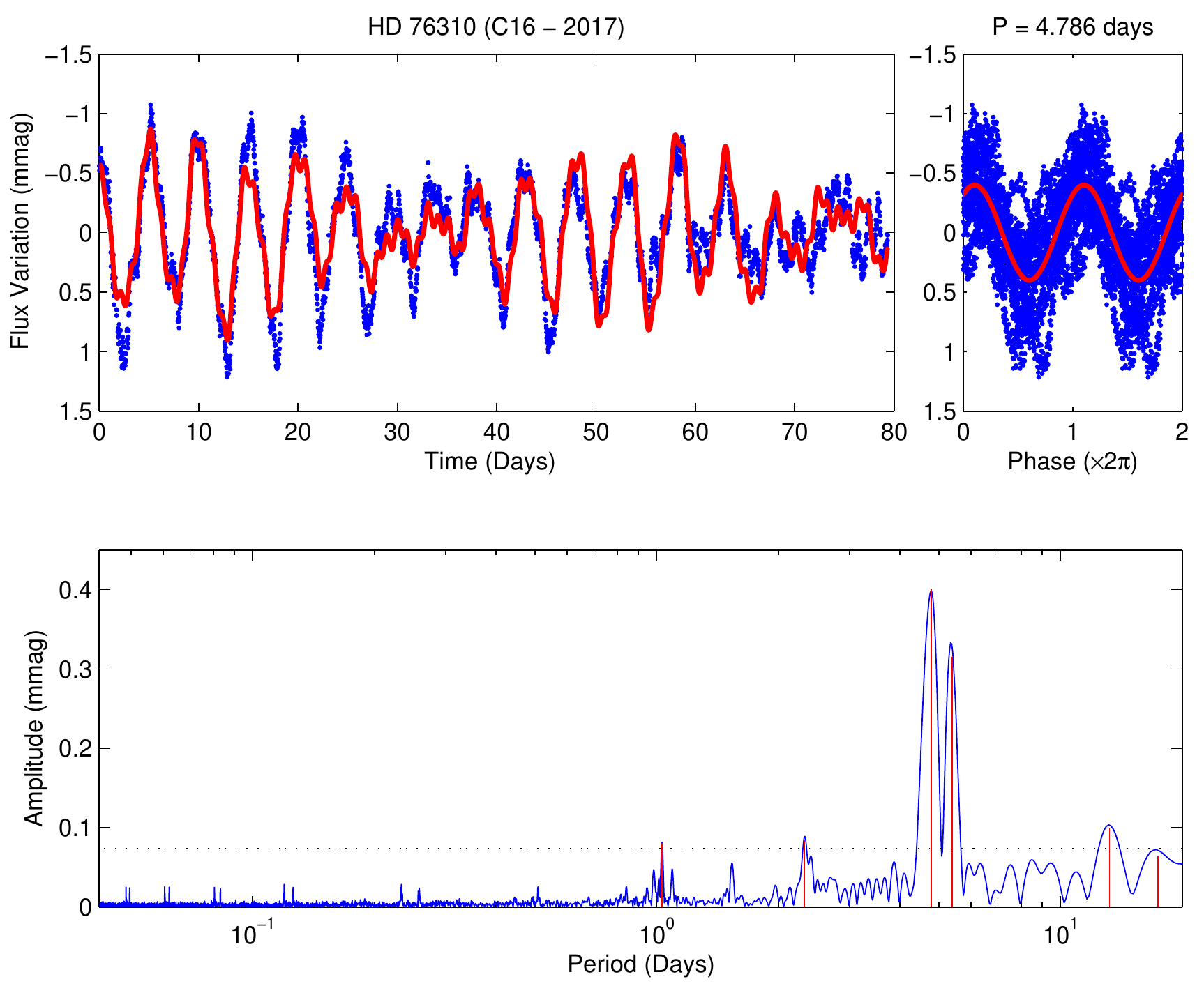}
%\vspace{-9.0cm}
%\end{flushright}
%\begin{center}
\caption{As Fig.\,\ref{K2_HD73045_C18}, but for the LC {\it K2} Campaign 16 observations of HD\,76310.}
\label{K2_HD76310_C16}
\end{center}
\end{figure*}

% \clearpage

\begin{figure*}
\begin{center}
%\begin{flushright}
\includegraphics[width=0.7\textwidth]{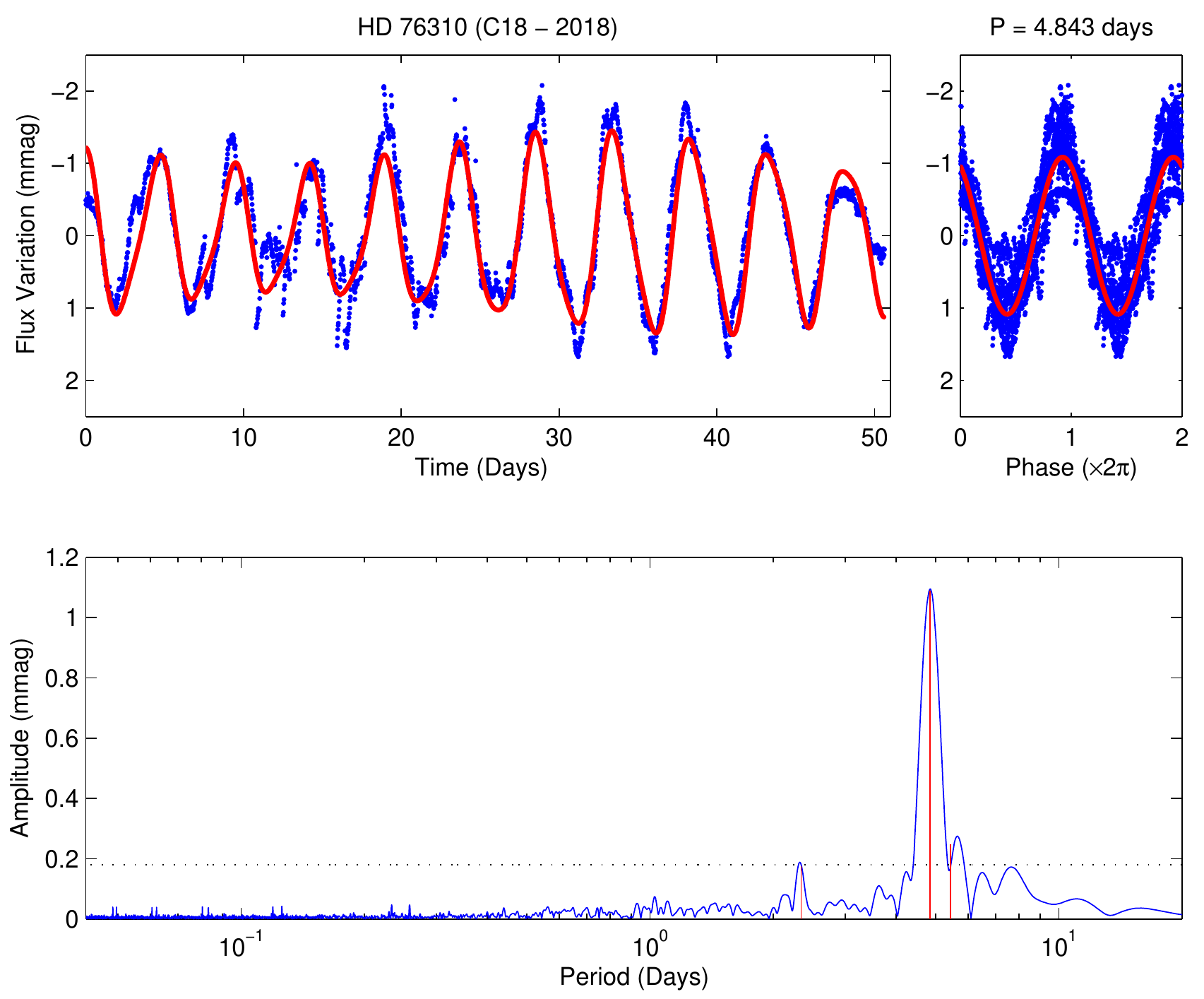}
%\vspace{-9.0cm}
%\end{flushright}
%\begin{center}
\caption{As Fig.\,\ref{K2_HD73045_C18}, but for the LC {\it K2} Campaign 18 observations of HD\,76310.}
\label{K2_HD76310_C18}
\end{center}
\end{figure*}

%\onecolumn
\begin{figure*}
\centering
 \includegraphics[width=0.85\textwidth]{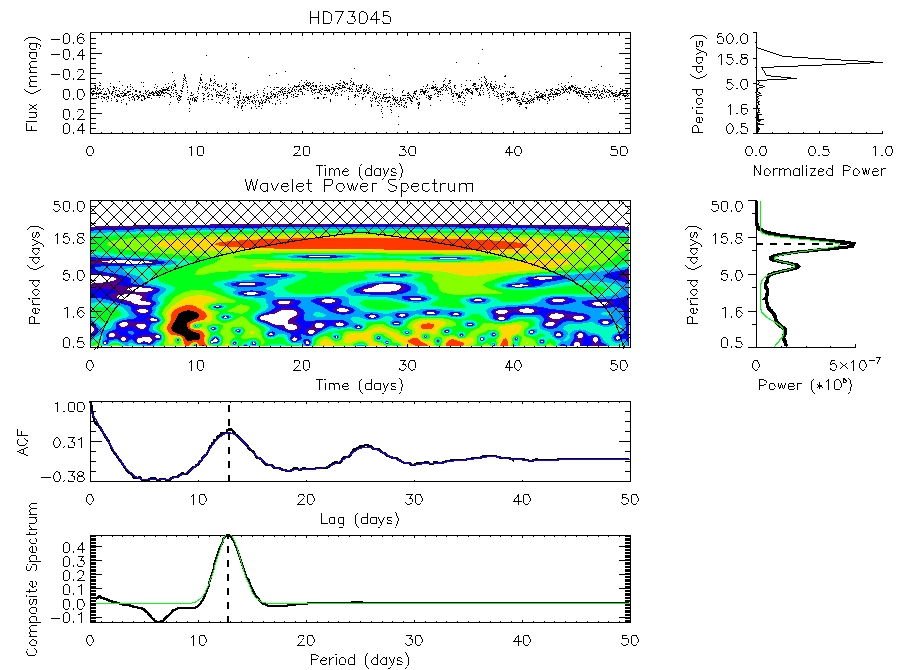}
  \caption{The wavelet map of HD\,73045 based on {\it K2} C18 data; see Fig.\,\ref{wavel1} for description details.}
  \label{wavel45c05}
\end{figure*}

\begin{figure*}
\centering
 \includegraphics[width=0.85\textwidth]{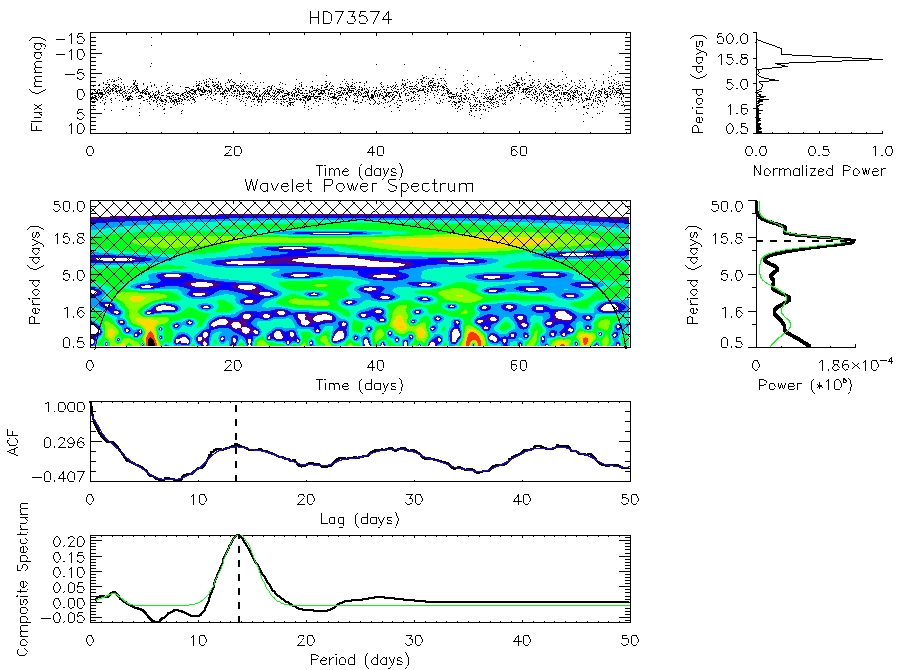}
  \caption{The wavelet map of HD\,73574 based on C05 data. The description follows that for HD\,73045, see Fig.\,\ref{wavel1}.}
  \label{wavel74}
\end{figure*}

\begin{figure*}
\centering
 \includegraphics[width=0.85\textwidth]{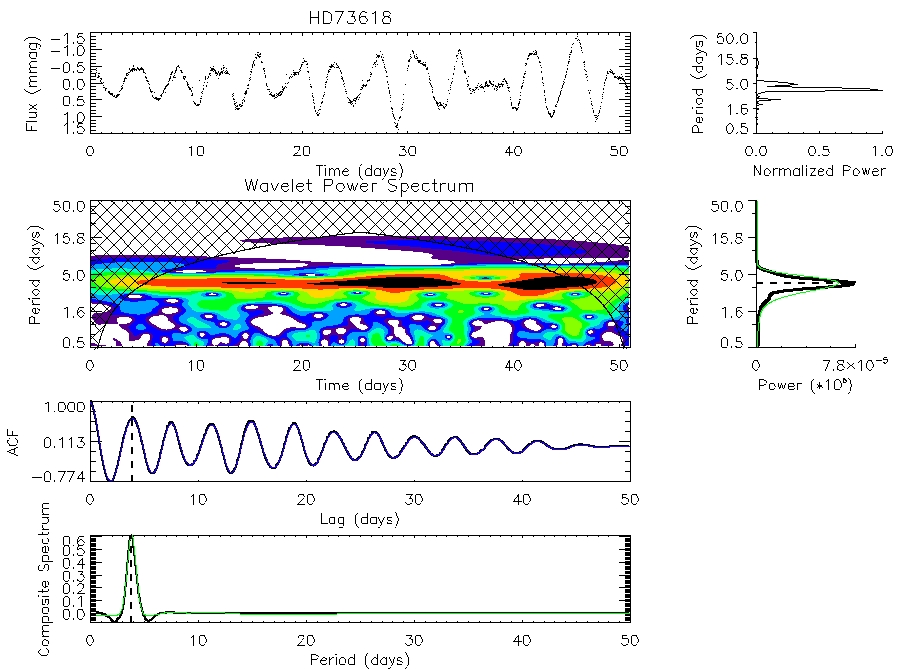}
  \caption{The wavelet map of HD\,73618 based on C18 data. The description follows that for HD\,73045, see Fig.\,\ref{wavel1}.}
  \label{wavel18}
\end{figure*}

\begin{figure*}
\centering
 \includegraphics[width=0.85\textwidth]{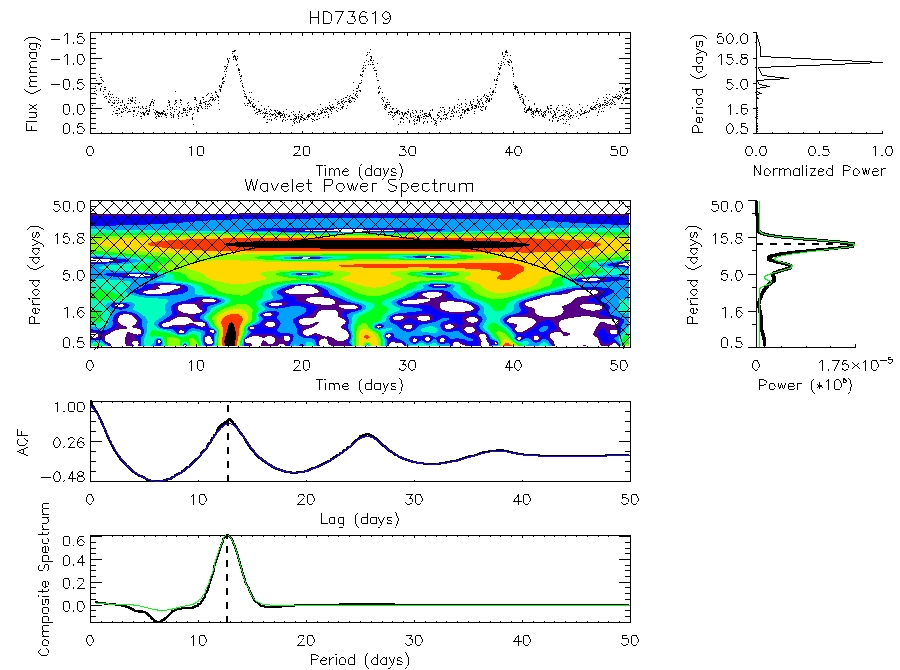}
  \caption{The wavelet map of HD\,73619 based on C18 data. The description follows that for HD\,73045, see Fig.\,\ref{wavel1}.}
  \label{wavel19}
\end{figure*}

\begin{figure*}
%\centering
 \includegraphics[width=0.85\textwidth]{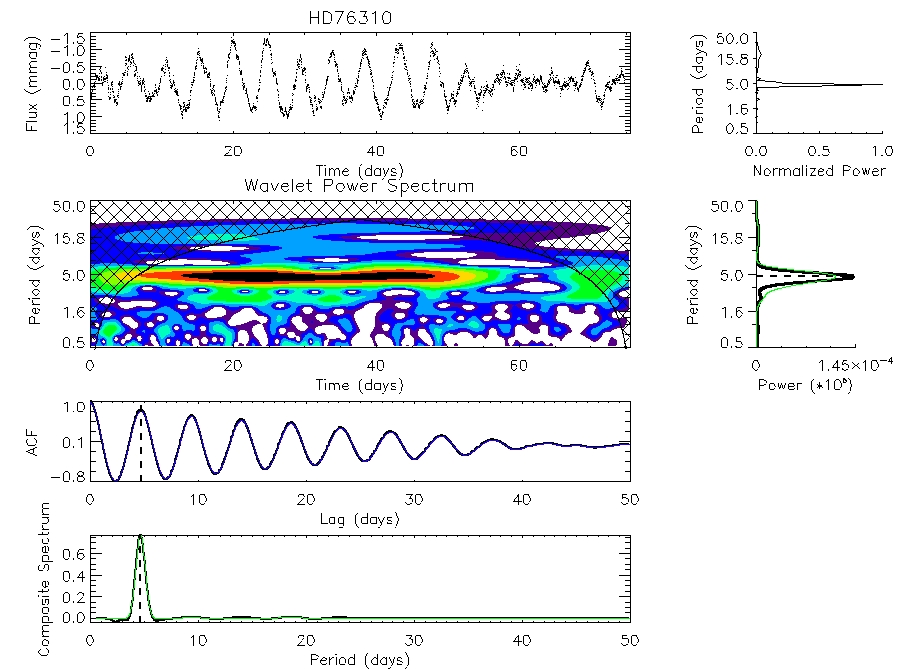}
  \caption{The wavelet map of HD\,73610 based on C05 data. The description follows that for HD\,73045, see Fig.\,\ref{wavel1}.}
  \label{wavel2}
\end{figure*}

\begin{figure*}
\centering
 \includegraphics[width=0.85\textwidth]{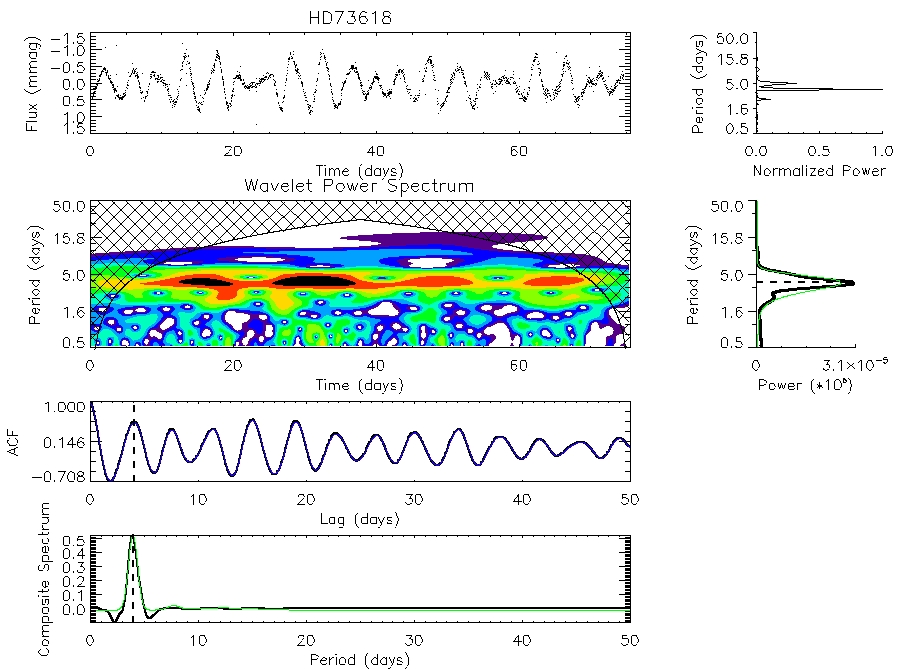}
  \caption{The wavelet map of HD\,73618 based on {\it K2} C05 data. The description follows that for HD\,73045, see Fig.\,\ref{wavel1}.}
  \label{wavel18c18}
\end{figure*}

\begin{figure*}
\centering
 \includegraphics[width=0.85\textwidth]{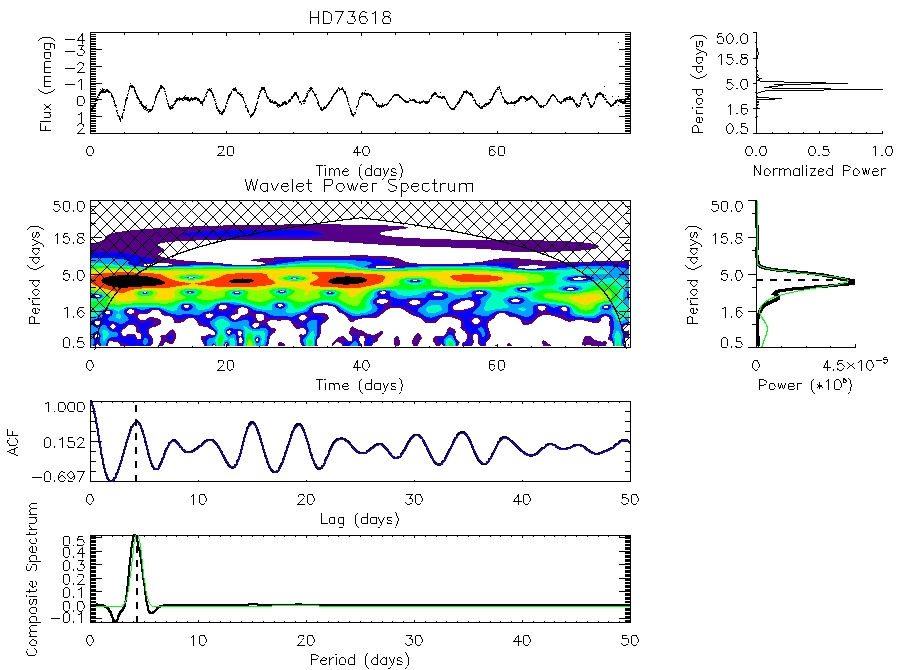}
  \caption{The wavelet map of HD\,73618 based on {\it K2} C16 data. The description follows that for HD\,73045, see
  Fig.\,\ref{wavel1}.}
  \label{wavel18c16}
\end{figure*}

\begin{figure}
\centering
 \includegraphics[width=0.85\textwidth]{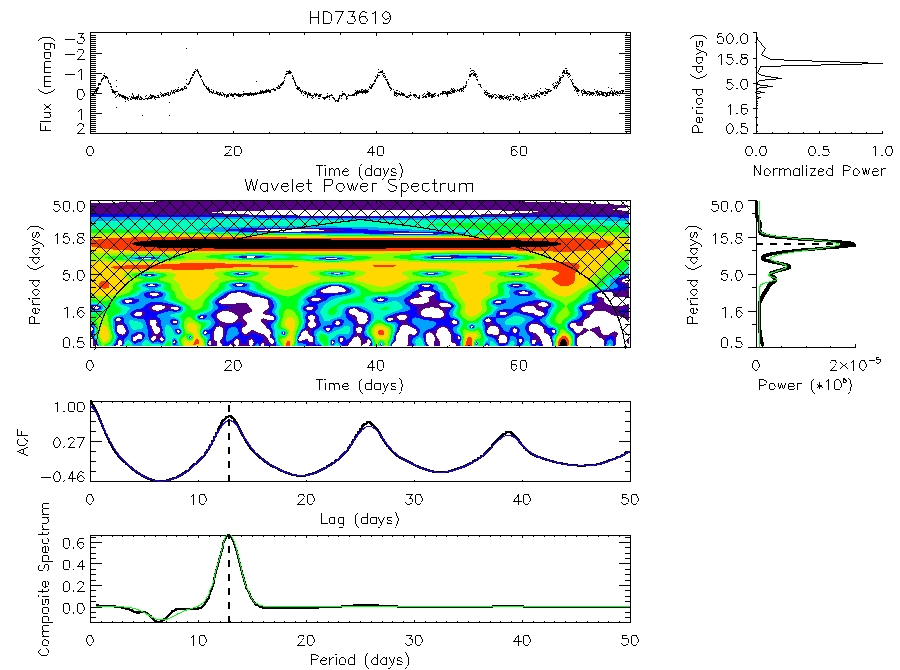}
  \caption{The wavelet map of HD\,73619 based on {\it K2} C05 data. The description follows that for HD\,73045, see
  Fig.\,\ref{wavel1}.}
  \label{wavel19c16}
\end{figure}

%%%%%%
%%%%%%

\begin{figure*}
\centering
 \includegraphics[width=0.85\textwidth]{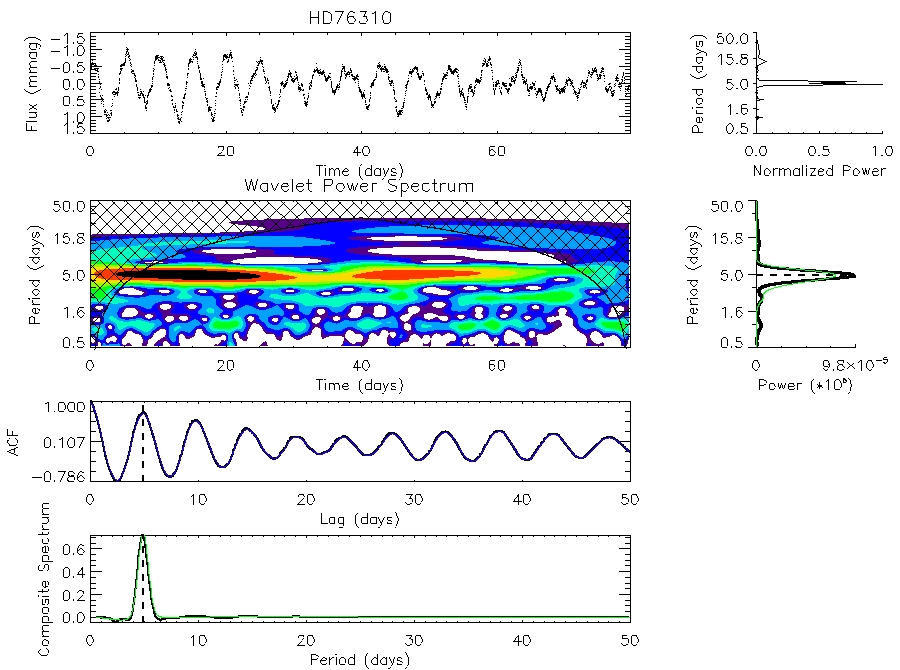}
  \caption{The wavelet map of HD\,73610 based on {\it K2} C16 data. The description follows that for HD\,73045, see
  Fig.\,\ref{wavel1}.}
  \label{wavel10c16}
\end{figure*}

\begin{figure*}
\centering
 \includegraphics[width=0.85\textwidth]{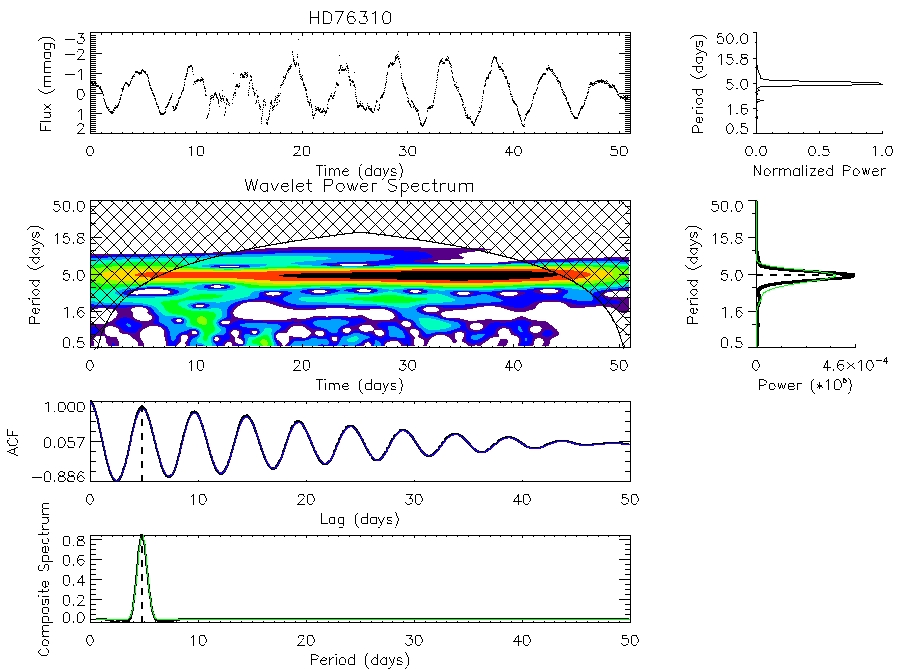}
  \caption{The wavelet map of HD\,73610 based on {\it K2} C18 data. The description follows that for HD\,73045, see
  Fig.\,\ref{wavel1}.}
  \label{wavel10c18}
\end{figure*}

%\twocolumn
\begin{figure}
\centering
 \includegraphics[width=0.7\textwidth]{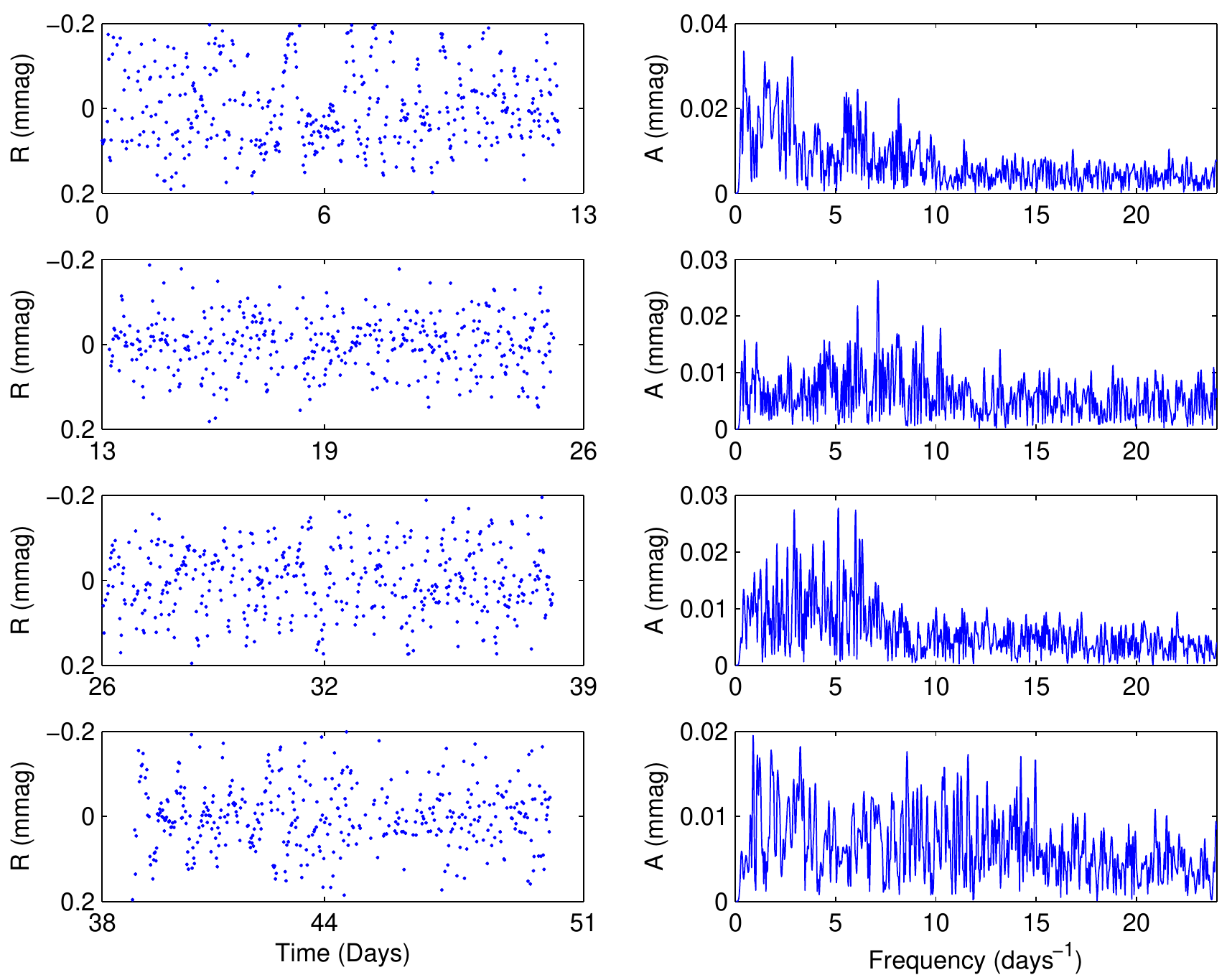}
  \includegraphics[width=0.7\textwidth]{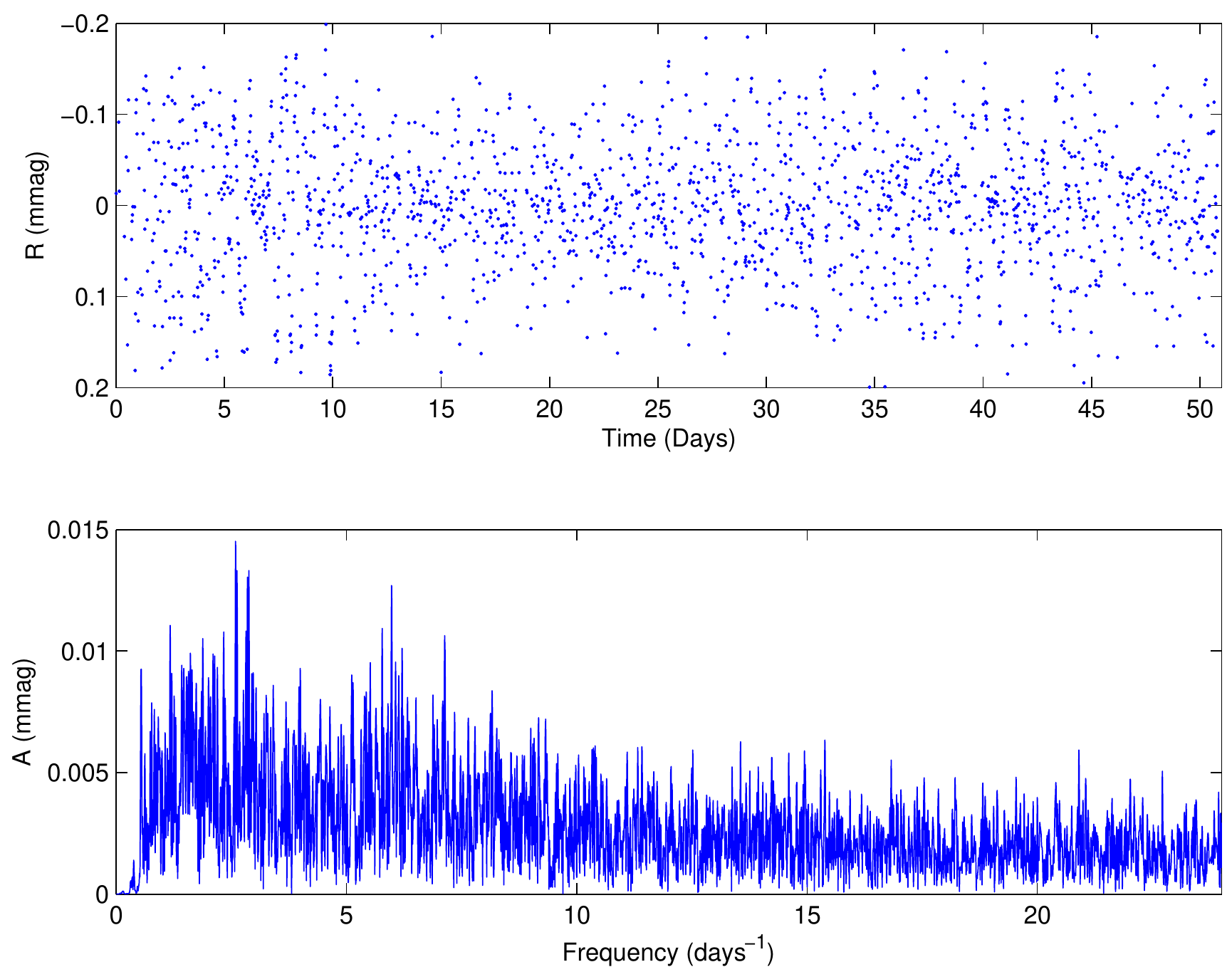}
  \caption{As Fig.\,\ref{pulsation2_c05}, but for C18. Here the light curve was split into four segments. No evidence for pulsational signals is identified based on either method of analysis.}
  \label{pulsation2_c18}
\end{figure}

\bsp    % typesetting comment
\label{lastpage}
\end{document}